\documentclass[11pt]{report}
\usepackage{amsmath,amscd,amsfonts,amsthm,bm}
\usepackage{amssymb}
\usepackage{setspace}
\usepackage{epsfig}
\usepackage{graphicx}
\usepackage[english]{babel}
\usepackage{fancyhdr}
\newcommand\Sym{{\mathbb S}}
\newcommand\R{\mathbb R}
\newcommand\T{\mathbb T}

\newcommand\x{\times}
\DeclareMathOperator{\ubnabla}{\ub{\nabla}}
\DeclareMathOperator{\ubbnabla}{\ubb{\nabla}}
\let\div\undefined
\DeclareMathOperator{\div}{div} 
\DeclareMathOperator{\ubdiv}{\ub{\operatorname{div}}}
\DeclareMathOperator{\skw}{\ubb{\operatorname{skw}}}
\DeclareMathOperator{\Skw}{\ubb{\operatorname{Skw}}}
\newcommand\tr{\operatorname{tr}}
\newcommand\Span{\operatorname{Span}}
\newcommand\Ker{\operatorname{Ker}}
\newcommand\ubbeps{\operatorname{\ubb\epsilon}}

\newcommand\ubblap{\operatorname{\ubb\Delta}}
\DeclareMathOperator{\ubbcurlc}{\ubb{\operatorname{curl}}_{\text{c}}}
\DeclareMathOperator{\ubbcurlr}{\ubb{\operatorname{curl}}_{\text{r}}}
\DeclareMathOperator{\ubbcurls}{\ubb{\operatorname{curl}}_{\text{s}}}
\DeclareMathOperator{\ubcurl}{\ub{\operatorname{curl}}}

\renewcommand\H{\mathcal H}
\newtheorem{thm}{Theorem}
\newtheorem{prop}[thm]{Proposition}
\newtheorem{lem}[thm]{Lemma}
\newtheorem{cor}[thm]{Corollary}
\newtheorem{remark}{Remark}
\usepackage{amssymb}  
\usepackage{graphics}
\def\underput#1#2#3{
\mathchoice
{\vtop{\ialign{##\crcr\hfil$#2$\vrule width0pt height0pt depth#3\hfil\crcr
\noalign{\nointerlineskip}\hfil$\scriptstyle#1$\hfil\crcr}}}
{\vtop{\ialign{##\crcr\hfil$#2$\vrule width0pt height0pt depth#3\hfil\crcr
\noalign{\nointerlineskip}\hfil$\scriptstyle#1$\hfil\crcr}}}
{\vtop{\ialign{##\crcr\hfil$\scriptstyle#2$\vrule width0pt height0pt 
depth#3\hfil\crcr
\noalign{\nointerlineskip}\hfil$\scriptscriptstyle#1$\hfil\crcr}}}
{\vtop{\ialign{##\crcr\hfil$\scriptscriptstyle#2$\vrule width0pt height0pt
depth#3\hfil\crcr
\noalign{\nointerlineskip}\hfil$\scriptscriptstyle#1$\hfil\crcr}}}}
\def\stack#1#2#3{\rlap{#1}\lower#3\hbox{#2}}

\def\strikespace{1truept}
\def\strikedist{3pt}

\def\dstrike{\vrule width7pt height0pt depth.4pt}
\def\tstrike{\vrule width7pt height0pt depth.4pt}
\def\sstrike{\hbox{\vrule width5pt height0pt depth.4pt}}
\def\ssstrike{\hbox{\vrule width3pt height0pt depth.4pt}}
\def\strike{{\mathchoice{\dstrike}{\tstrike}{\sstrike}{\ssstrike}}}

\def\doubledstrike{\stack{$\dstrike$}{$\dstrike$}{\strikespace}}
\def\doubletstrike{\stack{$\tstrike$}{$\tstrike$}{\strikespace}}
\def\doublesstrike{\stack{$\sstrike$}{$\sstrike$}{\strikespace}}
\def\doublessstrike{\stack{$\ssstrike$}{$\ssstrike$}{\strikespace}}
\def\doublestrike{{\mathchoice{\doubledstrike}{\doubletstrike}%
 {\doublesstrike}{\doublessstrike}}}

\def\ub#1{\underput\strike{#1}{\strikedist}}
\def\ubb#1{\underput\doublestrike{#1}{\strikedist}}

\def\ipd #1#2{\partial #1/\partial #2}

\begin{document}
\pagenumbering{roman}
\pagestyle{plain}
\pagestyle{empty}

\begin{center}
\begin{large}
UNIVERSITY OF MINNESOTA
\vskip 0.75in
This is to certify that I have examined this copy of a doctoral thesis
by
\vskip 0.5in
NICOLAE TARFULEA
\vskip 0.5in
and have found that is is complete and satisfactory in all respects,\linebreak
and that any and all revisions required by the final\linebreak
examining committee have been made. 
\vskip 0.75in
DR. DOUGLAS N. ARNOLD
\vskip 0.0in
\hrulefill\linebreak
Name of Faculty Advisor
\vskip 0.75in
\hrulefill\linebreak
Signature of Faculty Advisor
\vskip 0.75in
\hrulefill\linebreak
Date
\vskip 0.75in
GRADUATE SCHOOL
\end{large}
\end{center}
\pagebreak
\begin{center}
\begin{large}
Constraint Preserving Boundary Conditions for\linebreak
Hyperbolic Formulations of Einstein's Equations\linebreak
\vskip 1.0in
A THESIS\linebreak
SUBMITTED TO THE GRADUATE SCHOOL\linebreak
OF THE UNIVERSITY OF MINNESOTA\linebreak
BY\linebreak
\vskip 1.0in
Nicolae Tarfulea\linebreak
\vskip 1.0in
IN PARTIAL FULFILLMENT OF THE REQUIREMENTS\linebreak
FOR THE DEGREE OF\linebreak
DOCTOR OF PHILOSOPHY\linebreak
\vskip 0.5in
Dr.\ Douglas N. Arnold, Advisor\linebreak
\vskip 0.3in
July 2004
\end{large}
\end{center}
\pagebreak
\vspace*{7.0in}
\centerline{\large{\copyright \phantom{i}Nicolae Tarfulea 2004}}

\pagebreak
\pagestyle{plain}
\pagenumbering{roman}

\addcontentsline{toc}{section}{Acknowledgments} 
\vspace*{1.0in}
\noindent {\bf \Huge Acknowledgments}
\vspace*{0.5in}

I am most grateful and indebted to my thesis advisor, Prof. Douglas N. Arnold,
for the large doses of guidance, patience, and encouragement he has
shown me during my time at University of Minnesota and Penn State University.

I wish to thank my other thesis committee members 
Professors Carme Calderer, Bernardo Cockburn, and Fernando Reitich 
for their support and insightful commentaries on my work.

A substantial part of this thesis was written while I was supported by the
University of Minnesota Doctoral Dissertation Fellowship, which is gratefully 
acknowledged.
\newpage
\addcontentsline{toc}{section}{Abstract}
\vspace*{1.0in}
\noindent{\bf \Huge Abstract}
\vspace*{.5in}

Einstein's system of equations in the ADM decomposition involves
two subsystems of equations: evolution equations and constraint
equations. 
For numerical relativity, one typically solves the constraint equations only on the
initial time slice, and then uses the evolution equations to advance the solution
in time. Our interest is in the case 
when the spatial domain is bounded and appropriate boundary conditions are imposed.
A key difficulty, which we address in this thesis, is what
boundary conditions to place at the artificial boundary that lead to long time stable 
numerical solutions.
We develop an effective technique for finding well-posed
constraint preserving boundary conditions for constrained first order symmetric 
hyperbolic systems. 
By using this technique, we study the preservation of constraints by some
first order symmetric hyperbolic formulations of Einstein's equations derived from
the ADM decomposition linearized around Minkowski spacetime with arbitrary lapse and shift
perturbations, and the closely related question of their 
equivalence with the linearized ADM system. 
Our main result is the finding of well-posed maximal nonnegative
constraint preserving boundary conditions for each of the first order symmetric
hyperbolic formulations under investigation, 
for which the unique solution of the corresponding 
initial boundary value problem provides a solution to the linearized ADM system on polyhedral domains.

We indicate how to transform first order symmetric hyperbolic systems with constraints 
into equivalent unconstrained first order symmetric hyperbolic systems (extended 
systems) by building-in the constraints. We analyze and prove the equivalence 
between the original and extended systems in both the case of pure Cauchy 
problem and initial boundary value problem.  
These results seem to be very useful for transforming constrained numerical simulations
into unconstrained ones. As applications, we derive the extended systems
corresponding to the very same hyperbolic formulations of Einstein's equations for which
boundary conditions consistent with the constraints have been found.
Boundary conditions for these extended systems that make them equivalent to the 
original constrained systems are provided.

\setstretch{1.5}
\tableofcontents
\pagebreak
\addcontentsline{toc}{section}{List of Figures}
\listoffigures

\pagebreak
\pagestyle{plain}
\oddsidemargin=0.5in

\pagenumbering{arabic}
\chapter{Introduction}
\section{Background and Motivations}

\parskip 10pt plus2pt minus1pt
\parindent0pt

In a nutshell, general relativity says
that ``matter tells space how to curve, while the curvature of space tells matter
how to move,'' in a now famous phrase that the physicist John Wheeler once said.
So, general relativity serves both as a theory of space and time and
as a theory of gravitation.
Einstein began with two basic but subtle and powerful ideas: gravity and acceleration 
are indistinguishable and matter in free-fall always takes the shortest possible path
in curved spacetime. 
One of Einstein's most important discoveries was a system of  equations 
which relates spacetime and matter.

While the theory of general relativity has tremendous 
philosophical implications and has given rise to exotic new physical concepts like
black holes and dark matter, it is also crucial in some areas of modern technology such
as global positioning systems.
All these predictions and applications make general relativity a spectacularly
successful theory. However, though a number of its major predictions have been carefully
verified by experiments and observations, there are other key predictions, as the existence
of gravitational waves, that remain to be fully tested.
Einstein's equations possess solutions describing wavelike undulations in the 
spacetime. These gravitational waves correspond to ripples in spacetime itself, they
are not waves of any substance or medium. Like electromagnetic waves, gravitational
waves move at the speed of light and carry energy. In spite of carrying enormous 
amounts of energy from some of the most violent events in the universe, gravitational
waves are almost unobservable. For detecting and analyzing them, state--of--the--art 
detectors have been built in the United States and overseas.
The development of these observatories, just coming online now, is one of the
grandest scientific undertakings of our time, and the most expensive project ever funded
by the National Science Foundation. With a network of gravitational waves detectors, 
mankind could open up a whole new window on the cosmic space. The study of the universe 
using gravitational waves would 
not be just a simple extension of the optical and electromagnetic possibilities, it would
be the exploitation of an entirely new spectrum that could unveil parts and aspects
of the universe inaccessible so far. 

This enormous technological effort to build ultra-sensitive detectors has been followed
by an intense quest for developing computer methods to solve Einstein's 
equations. Having invested so much to detect gravitational waves, it is crucial
that we be able to interpret the waveforms detected. Most recent investigations in 
numerical relativity have been based on first order hyperbolic formulations derived 
from the Arnowitt, Deser, and Misner, or ADM, decomposition \cite{ADM} (see also \cite{Y})
of Einstein's equations and some results have been obtained 
in spherical symmetry and axisymmetry. However, in the general three spatial 
dimensions case, which is needed for the simulation of realistic astrophysical systems, 
it has not been possible to
obtain long term stable and accurate evolutions. One might argue that present day
computational resources are still insufficient to carry out high enough resolution
three dimensional simulations. However, the difficulty is likely to be more fundamental
than that. It seems that there is insufficient understanding of the structure
of Einstein's equations and there are too many unsolved questions related to how to
approach them numerically. 

Einstein's system of equations can be decomposed into two subsystems 
of equations (ADM decomposition): evolution equations and 
constraint equations (Hamiltonian constraint and momentum constraints).
For numerical relativity, one typically solves the constraint
equations only on the initial time slice, and then uses the evolution equations to 
advance the solution in time. A very difficult task is to derive good boundary 
conditions, and this problem is crucial if one takes into account that it seems 
impossible to have in the near future the computational power to put the boundaries 
far away from sources, far enough that they would not affect the region of numerical 
spacetime being looked at. 
Traditionally, most numerical relativity treatments have been careful to impose initial
data that satisfies the constraints. However, very rarely boundary conditions that lead
to well-posedness are used and much less frequently are they consistent with the 
constraints. Stewart \cite{S} has addressed this subject within Frittelli-Reula
formulation \cite{FR} linearized around flat space with unit lapse 
and zero shift in the quarter plane. Both main system and constraints propagate as
first order strongly hyperbolic systems. This implies that vanishing values of
the constraints at $t=0$ will propagate along characteristics. One wants the 
values of the incoming constraints at the boundary to vanish. 
However, one can not just impose them to vanish along the 
boundaries since the constraints involve derivatives of the fields across the boundary, 
not just the values of the fields themselves. If the Laplace-Fourier transforms
are used, the linearity of the differential equations gives algebraic equations
for the transforms of the fields.  
Stewart deduces boundary conditions for the main system in terms of Laplace-Fourier 
transforms that preserve the constraints by imposing the incoming modes for
the system of constraints to vanish and translating these conditions in terms
of Laplace-Fourier transforms of the main system variables. 
The idea of imposing the vanishing of the incoming constraints as boundary conditions
is pursued further in \cite{CLT} within Einstein-Christoffel formulation \cite{AY}
in the simple case of spherical symmetry. The radial derivative is eliminated in favor
of time derivative in the expression of the incoming constraints by using the main
evolution system. In \cite{CPSTR}, these techniques are refined and employed for the
linearized generalized Einstein-Christoffel formulation \cite{KST} around flat 
spacetime with vanishing lapse and shift perturbations on 
a cubic box. By considering well posed boundary conditions for the constraint 
system and trading normal derivatives for time and tangential ones, face systems 
are obtained that need to be solved first together with the compatibility conditions at 
the edges of the faces. The solutions of the face systems are used to impose
well posed constraint preserving boundary conditions for the main system. 
A construction with several points in common with the one just described can be also 
found in \cite{SW}. These two papers, \cite{CPSTR} and \cite{SW}, are the closest
to our work.  A different approach can be found in \cite{FG2}, \cite{FG3}, where the 
authors stray away from the general trend of seeking to impose the constraints 
along the boundary.
Their method consists in making the four components of the 
Einstein tensor projection along the normal to the boundary vanish. 
In the case of Einstein-Christoffel formulation restricted to spherical symmetry, 
the same boundary conditions as in \cite{CLT} are obtained.

Before we end this brief review, it should also be mentioned here the work done on 
boundary conditions for 
Einstein's equations in harmonic coordinates \cite{SSW}, \cite{SW}, when Einstein's 
equations become a system of second order hyperbolic equations for the metric components. The
question of the constraints preservation does not appear here, as it is hidden in the
gauge choice (the constraints have to be satisfied only at the initial surface, the harmonic
gauge guarantees their preservation in time).

What follows next is a summary of the contents of this dissertation, with emphasis on the
ideas that connect the different parts.

\section{Thesis Organization}
This dissertation is divided into three main parts. The first part is represented
by Chapter 2 and mainly describes our results concerning first order symmetric 
hyperbolic (FOSH) systems of partial differential equations. A special attention 
is being placed on FOSH systems with constraints and their well-posedness with or 
without boundary conditions. This first part represents a portion of the background
theory needed for the rest of the dissertation. The second part, Chapter 3,
is focused on  the ADM decomposition of Einstein's equations due to Arnowitt, Deser, 
and Misner \cite{ADM} and some important first order hyperbolic formulations derived
from it. A novelty in this part is represented by the introduction and analysis in
Subsection 3.4.4 of a new first order symmetric hyperbolic formulation of the
linearized ADM decomposition due to Arnold \cite{A}.
The third part, represented by Chapter 4, is the most important part of this
thesis. Here, we address a key difficulty in numerical relativity, the derivation
of boundary conditions that lead to well posedness and consistent with the constraints.
In the beginning of Chapter 4 we introduce and analyze a simpler model problem which
gives good insight for the more complex case of Einstein's equations. The core
of Chapter 4 consists of the analysis of three important first order symmetric hyperbolic
formulations of Einstein's equations for which we provide well-posed constraint-preserving
boundary conditions.

In the remainder of this introduction, we will describe the principal results
of this dissertation.

\section{Principal Results}
We have developed an effective and general technique for finding well-posed
constraint preserving boundary conditions for constrained first order symmetric 
hyperbolic systems. The key point of this technique is the matching of the general forms
of maximal nonnegative boundary conditions for the main system and the system
of constraints.

By applying this technique, we study the preservation of constraints by the linearized 
Einstein-Christoffel system around Minkowski spacetime with arbitrary lapse and shift
perturbations, and the closely related question of the 
equivalence of that system and the linearized ADM system. Our interest is in the case 
when the spatial domain is bounded and appropriate boundary conditions are imposed. 
However, we also consider the pure Cauchy problem with the result that
the linearized Einstein-Christoffel and ADM systems are equivalent.
Our main result is the finding of two distinct sets of well-posed maximal nonnegative
constraint preserving boundary conditions for which the unique solution of the 
corresponding linearized Einstein-Christoffel initial boundary value problem 
provides a solution to the linearized ADM system on polyhedral domains.

We have also obtained similar results for a very recent symmetric
hyperbolic formulation of Einstein's equations introduced by Alekseenko and 
Arnold in \cite{AA}. A new first order symmetric hyperbolic formulation of 
linearized Einstein equations due to Arnold \cite{A} is analyzed. Again, the main 
result is the finding of well-posed constraint preserving boundary conditions. 
In fact, same ideas should be applicable to some other formulations and/or in different
contexts, as, for example, linearization about some other backgrounds. 
However, while the strategy of finding adequate boundary conditions is similar, the
technical apparatus employed depends very much on the formulation under investigation. 

Returning to the more general framework of constrained 
first order symmetric hyperbolic systems, we indicate how to 
transform such systems into equivalent unconstrained first order symmetric
hyperbolic systems (extended systems) by building in the constraints. 
We also analyze and prove the equivalence 
between the original and extended systems in both the case of pure Cauchy 
problem and initial boundary value problem.
As applications, we derive extended systems corresponding to the (EC), (AA),
and (A) formulations respectively and boundary conditions that make them equivalent 
to the original constrained systems.
These results seem to be useful for transforming constrained numerical simulations
into unconstrained ones. 

\chapter{Symmetric Hyperbolic Systems}
\section{Introduction}

In this chapter some basic results on first order symmetric hyperbolic
(or FOSH) systems of partial differential equations are briefly reviewed, with
special attention being given to systems with constraints and boundary conditions. 
All these results represent background material relevant to the discussions
of the hyperbolic formulations of the Einstein equations which follow in the next
chapters. Much more information on hyperbolic systems can be found in the
books by John \cite{J}, Kreiss and Lorenz \cite{KL}, Gustafsson, Kreiss and Oliger 
\cite{GKO}, and Evans \cite{E}, among many others. 

The second section of this chapter is intended to enlist the basic definition 
of FOSH systems and some relevant existence and uniqueness results. The third section
is dedicated to the analysis of constrained initial value problems in a more
abstract framework and in the case of FOSH systems of partial differential
equations. The emphasis is on the equivalence between a given
system subject to constraints and a corresponding extended unconstrained
system. The fourth section deals with boundary conditions for FOSH systems
and the connections between the initial boundary value problem for a given
FOSH system with constraints and that for the extended system.
Section 2.3 and a substantial
part of Section 2.4 represent our contribution to the subject.

\section{Initial Value Problems}
In this section we will be concerned with a linear first order system
of equations for a column vector $u=u(x,t)=u(x_1,\ldots ,x_N,t)$ with
$m$ components $u_1,\ldots ,u_m$. Such a system can be written as
\begin{equation}\label{hs1}
Lu=\partial_t u+
\sum_{i=1}^{N}A_i(x,t)\partial_i u=f(x,t)\mbox{ in }\R^N\times (0,T],
\end{equation}
where $T>0$.
Here $A_1,\ldots , A_N$ are given $m\times m$ matrix functions, and $f$ is a 
given $m$-vector field. We will further assume that $A_i$ are of class $C^2$, with
bounded derivatives over $\R^N\times [0,T]$, and $f\in H^1(\R^N\times (0,T);\R^m)$. 

As initial data we prescribe the values of $u$ on the hyperplane $t=0$
\begin{equation}\label{id}
u=u_0\mbox{ on }\R^N\times\{ t=0\},
\end{equation}
with $u_0\in H^1(\R^N;\R^m)$.
For each $w\in \R^N$, define
\begin{equation}
A(w)(x,t)=\sum_{i=1}^Nw_iA_i(x,t)\ \ (x\in\R^N,\, t\geq 0).
\end{equation}
The system \eqref{hs1} is called symmetric hyperbolic if $A_i(x,t)$ is a symmetric
$m\times m$ matrix for each $x\in\R^N$, $t\geq 0$ ($i=1,\ldots, N$). Thus, the 
$m\times m$ matrix $A(w)(x,t)$ has only real eigenvalues and the corresponding 
eigenvectors
form a basis of $\R^N$ for each $w\in \R^N$, $x\in\R^N$, and $t\geq 0$.

\begin{remark}
More general systems having the form
\begin{equation}
A_0\partial_t u+\sum_{i=1}^{N}A_i(x,t)\partial_i u+B(x,t)u=f(x,t)
\end{equation}
are also called {\rm symmetric}, provided the matrix functions $A_i$ are symmetric
for $i=0,\ldots ,N$, and $A_0$ is positive definite. The results set forth below can
be easily extended to such systems.
\end{remark}
As in \cite{E}, Section 7.3.2., let us define the bilinear form
$$A[u,v;t]:=\int_{\R^N}\sum_{i=1}^{N}(A_i(\,\cdot\, ,\, t\, ) \partial_iu)\cdot v\, dx$$
for $0\leq t\leq T$, $u$, $v\in H^1(\R^N;\R^m)$.

{\bf Definition.}
We say
\[ u\in L^2((0,T);\,  H^1(\R^N;\R^m)),\mbox{ with } u'\in L^2((0,T);\,  L^2(\R^N;\R^m)),\]
is a weak solution of the initial value problem \eqref{hs1}, \eqref{id} provided \newline
(i) $(u',v)+A[u,v;t]=(f,v)$ for each $v\in H^1(\R^N;\R^m)$ and a.e. $0\leq t\leq T$, and
\newline
(ii) $u(0)=u_0$.\newline
Here $(\, \cdot\, ,\,\cdot\, )$ denotes the inner product in $L^2(\R^N;\R^m)$.

By using energy methods and the vanishing viscosity technique (see \cite{E}, Section 7.3.2.), 
the following existence and uniqueness result can be proven:
\begin{thm}\label{thm:weak}
The initial value problem \eqref{hs1}, \eqref{id} has a unique weak solution.
\end{thm}
In what follows, we will be more interested in first order symmetric hyperbolic systems with
constant coefficients. For such systems, a more general result (including regularity) is valid.
\begin{thm}\label{thm:evans}\cite{E}, Section 7.3.3. (also \cite{T}, Section 16.1.)
Assume $u_0\in H^s(\R^N;\R^m)$, with $s>N/2+m$. Then there is a unique solution $u\in C^1(\R^N
\times [0,\infty);\R^m)$ of the initial value problem \eqref{hs1}, \eqref{id}.
\end{thm}
The main tool used for proving this theorem is the Fourier transform. The
unique $C^1$ solution is given by:
\begin{equation}
\label{nhs}
u(x,t)=\frac{1}{(2\pi )^{N/2}}\int_{\R^N}e^{ix\cdot w}[e^{itA(w)}\hat{u}_0(w)
+\int_0^te^{i(t-s)A(w)}\hat{f}(w,s)ds]\, dw.
\end{equation}

\section{Constrained Initial Value Problems}
\subsection{Abstract Framework}
We introduce in this subsection an extended system corresponding to a given constrained
system defined on Hilbert spaces and investigate the equivalence of these two systems.

Let us consider the following system subject to constraints:
\begin{gather}\label{a1}
\dot y=Ay+f,\\ \label{a2}
By=0,\\ \label{a3}
y(0)=y_0,
\end{gather}
where $A:D(A)\subseteq Y\to Y,\ B:D(B)\subseteq Y\to X$ are densely defined 
closed linear operators on the Hilbert spaces $(X,\, \langle\, \cdot\, ,\,\cdot\, \rangle_X)$ 
and $(Y,\, \langle\,\cdot\, ,\,\cdot\, \rangle_Y)$, and $f:[0,\infty)\to Y$. 
Moreover, suppose $A$ is skew-symmetric and
\begin{equation}\label{cond}
A(\mbox{Ker}\, B)\subseteq \mbox{Ker}\, B. 
\end{equation}
Of course, we assume that the
compatibility condition $By_0=0$ is satisfied. Moreover, another more subtle 
compatibility condition must hold: $Bf=0$. This is because, for any fixed 
$\overline{t}\in [0,\infty )$, $B([y(t)-y(\overline{t})]/(t-\overline{t}))=0$, for all $t>0$, and
passing to the limit as $t\to \overline{t}$, it turns out that 
$B(\dot y(\overline{t}))=0,\ \forall\overline{t}\in [0,\infty )$ 
(since $B$ is a closed operator). By operating on \eqref{a1} with $B$, it follows that
$$0=B(\dot y(\overline{t}))=BAy(\overline{t})+Bf(\overline{t})=Bf(\overline{t}),\ 
\forall\overline{t}\in [0,\infty ),$$
where the last equality comes from \eqref{cond}.

\begin{remark}
If $f=0$, then the energy of the solution is preserved:
$$E(t)=\frac{1}{2}\| y(t)\|^2=\frac{1}{2}\| y_0\|^2, \ \forall t\geq 0.$$
\end{remark}

\begin{thm}\label{thm:ivp_equiv} 
The pair $(y,z)^T$ is a solution of the associated unconstrained system
\begin{gather}\label{ae1}
\left(\begin{array}{cc}
\dot y \\
\dot z
\end{array}\right)=
\left(\begin{array}{cc}
A&-B^*\\
B&0
\end{array}\right)
\left(\begin{array}{cc}
y \\
z
\end{array}\right) + \left(\begin{array}{cc}
f \\
0
\end{array}\right),\\ \label{ae2}
y(0)=y_0,\ z(0)=0,
\end{gather}
if and only if $z\equiv 0$ and $y$ satisfies the constrained system
\eqref{a1}--\eqref{a3}.
\end{thm}
\begin{proof}
If $z\equiv 0$ and $y$ satisfies \eqref{a1}--\eqref{a3}, then it is
clear that $(y,0)$ satisfies \eqref{ae1}--\eqref{ae2}.

Now suppose $(y,z)$ satisfies \eqref{ae1}--\eqref{ae2}. Observe
that we can split $Y$ as a direct sum as following
\begin{equation}
Y=\mbox{Ker}\, B\oplus (\mbox{Ker}\, B)^{\bot}=\mbox{Ker}\, B\oplus\overline{
\mbox{Im}\, B^*}.
\end{equation}

According to this decomposition of $Y$,
\begin{equation}\label{split}
y(t)=y_1(t)+y_2(t),
\end{equation}
where $y_1(t)\in\mbox{Ker}\, B$, and $y_2(t)\in \overline{\mbox{Im}\, B^*}$.

Since both $\mbox{Ker}\, B$ and $\overline{\mbox{Im}\, B^*}$ are closed and the
corresponding projections are continuous,
\begin{equation}
\dot y(t)=\dot y_1(t)+\dot y_2(t),
\end{equation}
with $\dot y_1(t)\in\mbox{Ker}\, B$, and $\dot y_2(t)\in \overline{\mbox{Im}\, B^*}$.

From \eqref{cond}, \eqref{ae1}, and \eqref{ae2}, we obtain that
\begin{gather}
\dot y_1=Ay_1+f,\\
y_1(0)=y_0,
\end{gather}
and
\begin{gather}
\dot y_2=Ay_2-B^*z,\\
\dot z=By_2,\\
y_2(0)=0,\ z(0)=0.
\end{gather}
Observe that
$$\frac{1}{2}(\| z\|^2)^.=\langle\dot z,z\rangle_X=\langle By_2,z\rangle_X=
\langle y_2,B^*z\rangle_Y =\langle y_2,
Ay_2-\dot y_2\rangle_Y
=-\langle y_2,\dot y_2\rangle_Y =-\frac{1}{2}(\| y_2\|^2)^{.}.$$
Therefore,
$$\| z(t)\|^2+\| y_2(t)\|^2=\| z(0)\|^2+\| y_2(0)\|^2=0,$$
which implies $z\equiv 0$ and $y_2\equiv 0$. Thus, $y=y_1$ and
\eqref{a1}--\eqref{a3} are satisfied.
\end{proof}

\subsection{Constrained First Order Symmetric Hyperbolic Problems}
In this subsection, we will prove a result similar to Theorem~\ref{thm:ivp_equiv} for
the initial value problem 
\begin{gather}\label{a1h}
\partial_t u=Au+f,\\ \label{a2h}
Bu=0,\\ \label{a3h}
u(x,0)=u_0,
\end{gather}
where $A=\sum_{i=1}^NA_i\partial_i$, with $A_i\in \R^{m\times m}$ constant
symmetric matrices, and
$B=\sum_{i=1}^NB_i\partial_i$, with $B_i\in\R^{p\x m}$ constant
matrices. Of course, we assume that \eqref{cond} and  the compatibility conditions 
$Bu_0=0$, $Bf(\cdot ,\, t)=0,\ \forall t\geq 0$, hold.
\begin{thm} (equivalence for classical solutions) 
If $u_0\in H^s(\R^N;\, \R^m)$ and $f(\cdot ,t)\in H^s(\R^N;\, \R^m),$ $\forall t\geq 0$,
for $s>N/2+m$, then the pair $(u,z)^T\in C^1(\R^N\x[0,\infty);\R^{m+p})$ is a solution of the associated unconstrained system
\begin{gather}\label{un1}
\frac{\partial }{\partial t}\left(\begin{array}{cc}
u \\
z
\end{array}\right)=
\left(\begin{array}{cc}
A&-B^*\\
B&0
\end{array}\right)
\left(\begin{array}{cc}
u \\
z
\end{array}\right) + \left(\begin{array}{cc}
f \\
0
\end{array}\right),\\ \label{un2}
u(x,0)=u_0,\ z(x,0)=0,
\end{gather}
if and only if $z\equiv 0$, and $u \in C^1(\R^N\x[0,\infty);\R^{m})$ 
satisfies the constrained system
\eqref{a1h}--\eqref{a3h}.
\end{thm}
\begin{proof}
If $z\equiv 0$ and $u$ satisfies \eqref{a1h}--\eqref{a3h}, then it is
clear that $(y,0)^T$ satisfies \eqref{un1}--\eqref{un2}. 

Now, let us prove the converse. Denote by 
$$\overline{A}=\left(\begin{array}{cc}
A&-B^*\\
B&0
\end{array}\right) =\sum_{j=1}^N\overline{A}^j\frac{\partial }{\partial x_j}.$$
From \eqref{nhs}, we know that the solution of \eqref{un1}--\eqref{un2}
has the following expression
\begin{equation}\label{uz}
\begin{array}{ll}
\left(\begin{array}{c}
u\\
z
\end{array}
\right) (x,t)&=\frac{1}{(2\pi )^{N/2}}\int_{\R^N}e^{ix\cdot y}[e^{-it\overline{A}(y)}
\left(\begin{array}{c}
\hat{u}_0\\
0
\end{array}
\right) (y)\\
 &+\int_0^te^{-i(t-s)\overline{A}(y)}
\left(\begin{array}{c}
\hat{f}\\
0
\end{array}\right) (y,s)ds]\, dy.
\end{array}
\end{equation}

The next step in the proof is to show that
\begin{equation}
\label{Any}
\overline{A}^n(y)\left(\begin{array}{c}
\hat{u}_0\\
0
\end{array}\right)=\left(\begin{array}{c}
A^n(y)\hat{u}_0\\
0
\end{array}\right)
\end{equation}
for all positive integer $n$.

We are going to prove \eqref{Any} by induction.

For $n=1$, we have
$$\overline{A}(y)\left(\begin{array}{c}
\hat{u}_0\\
0
\end{array}\right)=\left(\begin{array}{c}
A(y)\hat{u}_0\\
B(y)\hat{u}_0
\end{array}\right). $$

But, since $Bu_0=0$, it follows that $B(y)\hat{u}_0=0$ by taking the Fourier
transform. So
$$\overline{A}(y)\left(\begin{array}{c}
\hat{u}_0\\
0
\end{array}\right)=\left(\begin{array}{c}
A(y)\hat{u}_0\\
0
\end{array}\right). $$

Assume that \eqref{Any} is true for $n=k-1$ and let us prove it for
$n=k$.

\[ \overline{A}^k(y)\left(\begin{array}{c}
\hat{u}_0\\
0
\end{array}\right)=\overline{A}(y)\left(\begin{array}{c}
A^{k-1}(y)\hat{u}_0\\
0
\end{array}\right)=\left(\begin{array}{c}
A^k(y)\hat{u}_0\\
B(y)A^{k-1}(y)\hat{u}_0
\end{array}\right).\]

Since $u_0\in\Ker B$, from \eqref{cond}, we can see that
\begin{equation}\label{BA}
BA^{k-1}u_0=0.
\end{equation}
Applying the Fourier transform to \eqref{BA}, we get
$$B(y)A^{k-1}(y)\hat{u}_0=0.$$
Thus, 
\[ \overline{A}^k(y)\left(\begin{array}{c}
\hat{u}_0\\
0
\end{array}\right)=\left(\begin{array}{c}
A^k(y)\hat{u}_0\\
0
\end{array}\right)\]
and the proof of \eqref{Any} is complete. 

From \eqref{Any}, observe that
\begin{equation}\label{iA1}
e^{-it\overline{A}(y)}
\left(\begin{array}{c}
\hat{u_0}\\
0
\end{array}
\right)(y)=\sum_{n=0}^{\infty}\frac{(-it)^n}{n!}\overline{A}^n(y)
\left(\begin{array}{c}
\hat{u}_0\\
0
\end{array}\right)=\left(\begin{array}{c}
e^{-itA(y)}\hat{u}_0\\
0
\end{array}\right) .
\end{equation}

Same arguments show that
\begin{equation}\label{iA2}
e^{-i(t-s)\overline{A}(y)}
\left(\begin{array}{c}
\hat{f}\\
0
\end{array}
\right)(y)=\sum_{n=0}^{\infty}\frac{[-i(t-s)]^n}{n!}\overline{A}^n(y)
\left(\begin{array}{c}
\hat{f}\\
0
\end{array}\right)=\left(\begin{array}{c}
e^{-i(t-s)A(y)}\hat{f}\\
0
\end{array}\right) .
\end{equation}

From \eqref{iA1} and \eqref{iA2}, it follows that
\begin{equation*}
\left(\begin{array}{c}
u \\
z
\end{array}\right)=\left(\begin{array}{c}
u \\
0
\end{array}\right) ,
\end{equation*}
with
$$u(x,\, t)=\frac{1}{(2\pi )^{N/2}}\int_{\R^N}e^{ix\cdot y}[e^{-itA(y)}
\hat{u}_0(y)
+\int_0^te^{-i(t-s)A(y)}
\hat{f}(y,s)ds]\, dy $$
the unique $C^1$ solution of \eqref{a1h}--\eqref{a3h}.
\end{proof}

\section{Boundary Conditions}
In general, one has to be careful when choosing boundary conditions for
a hyperbolic equation (or system). This can be seen even in the simple
case of a first order equation in one space dimension (the transport equation).
It seems that any acceptable boundary conditions should give the incoming
modes into the spatial domain, but they must not try to change the behavior
of the outgoing modes. In several dimensions the situation is much more
complicated since there is not easy to identify the incoming and outgoing
modes. Worse, there may also be waves moving tangent to the boundary, and
it is not very clear how these modes could be casted into the boundary
conditions.

A few approaches to these questions have been proposed. Some answers
have been given by Friedrichs \cite{F} via the ``energy method'' (see also the
work done by Courant and Hilbert). This method provides criteria which are sufficient
for constructing boundary conditions that lead to a well-posed problem.
Other sufficient conditions have been pointed out by Lax and Phillips in their
very interesting work \cite{LP}.
A necessary and sufficient condition for having a well-posed initial boundary
condition has been proved by Hersh \cite{H}, but his result was only for systems
with constant coefficients and defined on a half-space with non-characteristic
boundary conditions. Using Fourier and Laplace
transforms, he constructed solutions and derived a necessary and sufficient
condition for well-posedness. 
Later on (in the 1970s and 1980s), more technical approaches came up. Kreiss 
\cite{K}, 
Majda and Osher \cite{MO}, among others, proved pretty complicated algebraic results
concerning boundary conditions. Remarkably, Kreiss \cite{K} gave a criteria that 
determine
whether a boundary condition is admissible or not. The main point of his approach
was the possibility to solve for incoming modes in terms of outgoing modes and
boundary conditions. Majda and Osher \cite{MO} generalized Kreiss' theory to the case
of uniformly characteristic boundary. Other significant contributions to this
subject have been made by Rauch \cite{R}, Higdon \cite{Hi}, Secchi \cite{Se1}--\cite{Se5},
among many others.

\subsection{Maximal Non-Negative Boundary Conditions}\label{2.4.1} 
In this subsection we prove a well-posedness result for first order symmetric hyperbolic
initial boundary value problems that closely follows the ideas of \cite{LP}, \cite{R}, and \cite{F}.
Moreover, we give an algebraic characterization of maximal non-negative boundary conditions
which will be used later for determining constraint preserving boundary conditions for
some constrained first order symmetric hyperbolic systems.       
  
Consider the symmetric hyperbolic system of equations \eqref{hs1} on
$\Omega_T=\Omega\times (0,T)$, where $\Omega\subset \R^N$ is a bounded
domain with a  smooth boundary $\partial\Omega$, and $f\in L^2(\Omega_T,\R^m)$.

Set $n(x)=(n_1,\ldots ,n_N)$ be the outer normal to $\Omega$ at $x\in\partial\Omega$,
 and denote by $A_n(x,t)$ the boundary matrix
\begin{equation}
\label{eq:boundarymatrix}
A_n(x,t)=\sum_{i=1}^Nn_iA^i.
\end{equation}

We supplement \eqref{hs1} with the initial condition \eqref{id},
with $u_0\in H^1(\Omega)$,
and with linear boundary conditions of the following form
\begin{equation}\label{bc}
E(x,t)u(x,t)=0,\ \mbox{ on }\partial\Omega\times (0,T).
\end{equation}

Of course, we suppose that the compatibility condition $E(x,0)u_0(x)=0$
holds on $\partial\Omega$.

In fact, by choosing a function of $H^1(\Omega_T,\R ^m)$ satisfying
\eqref{id} and \eqref{bc}, and changing the variable, we may
assume that $u_0=0$; hereafter, we stick with this choice of $u_0$.

Also, the boundary condition \eqref{bc} may be regarded as
\begin{equation}\label{bc1}
u(x,t)\in N(x,t)=\Ker E(x,t),\ \forall (x,t)\in\partial\Omega\x (0,T).
\end{equation}

Denote the formal adjoint of $L$ by $L^*$
$$L^*u=-\partial_t u-
\sum_{i=1}^{N}\partial_i (A^iu).$$

Associated to \eqref{hs1}, \eqref{id}, and \eqref{bc} (or \eqref{bc1}),
we consider the adjoint problem
\begin{gather}\label{noua}
L^*v=f \ \mbox{ in }\Omega_T,\\ \label{zece}
v(x,t)\in (A_n(x,t)N(x,t))^\bot, \ \forall (x,t)\in\partial\Omega\x (0,T),\\ \label{unspe}
v(x,T)=0 \ \ \mbox{ in }\Omega.
\end{gather}

Next, define the admissible spaces of solutions for both the original problem
and the adjoint problem:
$$\H =\{ u\in H^1(\Omega_T,\R^m):\ u\mbox{ satisfies \eqref{id} and
\eqref{bc} (or \eqref{bc1})} \}$$
and
$$\H_* =\{ v\in H^1(\Omega_T,\R^m):\ v\mbox{ satisfies \eqref{zece} and
\eqref{unspe} } \}.$$

Observe that, if $u\in\H$ and $v\in\H_*$, then from Green's formula
\begin{equation*}
(v,Lu)-(u,L^*v)=\int_{\partial\Omega\x (0,T)}v^TA_nu\, d\sigma ,
\end{equation*}
it follows that
\begin{equation*}
(v,Lu)=(u,L^*v).
\end{equation*}

{\bf Definition.} The function $u\in L^2(\Omega_T,\R^m)$ is said 
to be a weak solution of \eqref{hs1}, \eqref{id}, and \eqref{bc} if 
\begin{equation*}
(v,f)-(L^*v,u)=0,\ \forall v\in\H_*.
\end{equation*}

{\bf Definition.} The function $u\in L^2(\Omega_T,\R^m)$ is said to be 
a strong solution of
\eqref{hs1}, \eqref{id}, and \eqref{bc}, if the pair $(u,f)$
belongs to the closure of the graph of $L$; in other words, if $u$ is
the limit in the $L^2$ norm of a sequence of functions $\{ u_k\}\subset\H$
 such that $f_k=Lu_k\to f$ in the $L^2$ norm.

\begin{remark}
If $u$ solves \eqref{hs1}, \eqref{id}, and \eqref{bc} in the
strong sense, then it also solves the problem in the weak sense.
\end{remark}

{\bf Definition.} The boundary condition \eqref{bc} (or \eqref{bc1})
is called non-negative if the matrix $A_n(x,t)$ is non-negative over $N(x,t)$
\begin{equation}\label{pos}
u^TA_nu\geq 0,\ \forall u\in N(x,t),\ (x,t)\in\partial\Omega\x (0,T).
\end{equation}

\begin{thm}(\cite{LP}, \cite{R}) \label{thm:existence}
If the boundary condition \eqref{bc} is maximal non-negative at each
point $(x,t)\in \partial\Omega\x (0,T)$ (meaning that \eqref{pos} holds
and there is no other larger subspace containing $N(x,t)$ and having the same
property), then \eqref{hs1}, \eqref{id}, and \eqref{bc} has a unique
strong solution for any given integrable function $f$.
\end{thm}

\medskip

In order to prove this theorem, we need a couple of intermediate results.

\begin{lem}\label{lem:ineq}
For all $u\in\H$ satisfying the condition \eqref{pos}, there exists
a positive constant $C>0$ that does not depend on $u$ such that
\begin{equation}\label{in}
\| u\|_{L^2(\Omega_T,\R^m)}\leq C\| Lu\|_{L^2(\Omega_T,\R^m)}.
\end{equation}
\end{lem}
\begin{proof}
Symmetric operators with smooth $A^i$
satisfy the following identity
\begin{equation}\label{uLu}
u^TLu=\frac{1}{2}\frac{\partial }{\partial t}(u^Tu)+\frac{1}{2}\sum_{i=1}^N
\frac{\partial }{\partial x_i}(u^TA^iu)+u^TKu,
\end{equation}
where
\begin{equation}
K=-\frac{1}{2}\sum_{i=1}^N\frac{\partial A^i}{\partial x_i}.
\end{equation}

If we integrate \eqref{uLu} over $\Omega_s=\Omega\x (0,s)$, we get
\begin{equation}\label{iulu}
\int_{\Omega_s}u^TLu=\frac{1}{2}\int_{\Omega\x \{ s\}}u^Tu+\frac{1}{2}
\int_{\partial\Omega\x (0, s)}u^TA_nu+\int_{\Omega_s}u^TKu.
\end{equation}

From \eqref{pos} and \eqref{iulu}, it is easy to see that
\begin{equation}\label{iuluu}
\int_{\Omega\x \{ s\}}u^Tu\leq 2\int_{\Omega_s}u^TLu+2\| K\| 
\int_{\Omega_s}u^Tu\leq (1+2\| K\|)\int_{\Omega_s}u^Tu+\int_{\Omega_T}(Lu)^T(Lu).
\end{equation}

Denote by
$$\phi (s)=\int_{\Omega_s}u^Tu.$$

Then \eqref{iuluu} recasts into
\begin{equation}\label{fp}
\phi '(s)\leq (1+2\| K\|)\phi (s)+\| Lu\|_{L^2(\Omega_T,\R^m)}^2.
\end{equation}

Applying the Gronwall's lemma to \eqref{fp}, it follows that
$$\phi (s)\leq \frac{e^{s(1+2\| K\| )}}{1+2\| K\| }\| Lu\|_{L^2(\Omega_T,\R^m)}^2.$$

Hence,
$$\phi (T)=\| u\|_{L^2(\Omega_T,\R^m)}^2\leq
\frac{e^{T(1+2\| K\|)}}{1+2\| K\|} \| Lu\|_{L^2(\Omega_T,\R^m)}^2.$$

Then, the inequality \eqref{in} holds for $$C=\sqrt{ \frac{e^{T(1+2\| K\|)}}
{ 1+2\| K\| }}.$$
\end{proof}
A simple consequence of this lemma is stated next.

\begin{cor}
For any square integrable function $f$ there is at most one solution $u\in \H$
satisfying non-negative boundary conditions to the problem
\eqref{hs1}, \eqref{id}, and \eqref{bc}.
\end{cor} 

\begin{lem}
If the boundary conditions \eqref{bc1} are maximal non-negative for $L$, then the adjoint boundary
conditions \eqref{zece} are non-negative for $L^*$.
\end{lem}
\begin{proof}
According to the expression of $L^*$, the boundary matrix for $L^*$ is equal
to $-A_n$. Hence it suffices to show that $v^TA_nv$ is non-positive for each
$v$ satisfying the adjoint boundary condition. 

Arguing by contradiction, suppose that there is a $v$ such that $v^TA_nv>0$.

Consider the linear space $N(x,t)\oplus v$. The elements in this space have the form
$u+av$, where $u\in N(x,t)$ and $a$ is a real number.

Observe that
$$(u+av)^TA_n(u+av)=u^TA_nu+2av^TA_nu+a^2v^TA_nv=u^TA_nu+a^2v^TA_nv\geq 0,$$
which is in contradiction with the maximality of $N(x,t)$.
\end{proof}

\begin{proof} (of Theorem~\ref{thm:existence}) {\it Existence.}
We claim that $L(\H )$ is dense in $L^2(\Omega_T,\R^m)$. Arguing by contradiction,
let us suppose the contrary. Then there exists a non-trivial function $v$
orthogonal to $L(\H )$, i.e.
$$(Lu,v)=0,\ \forall u\in\H .$$
Therefore, $v$ is a weak solution to the adjoint problem corresponding to $f=0$.

Now, let us prove that $v=0$. 
From \cite{LP}, Theorem 1.1, we get that $v$ is a strong solution; therefore there exists
a sequence $v_n\to v$ in $L^2$, such that
$f_n=Lv_n\to 0$ in $L^2$ and $v_n$ satisfies the
(non-negative) adjoint boundary conditions, which implies
$$\| v_n\|\leq C\| f_n\|\to 0.$$
So, $v=0$.
In conclusion, if $f$ is any element of $L^2(\Omega_T,\R^m)$, 
then we can find a sequence  $\{ u_n\}\subset \H$ so that
$Lu_n\to f,\mbox{ as }n\to \infty, \mbox{ in } L^2.$

From Lemma~\ref{lem:ineq}, it follows that $\{ u_n\}$ is a Cauchy sequence
in $L^2$. Thus, there is $u\in\ L^2(\Omega_T,\R^m$  such that
$u_n\to u\mbox{ and }Lu_n\to f\mbox{ in }L^2.$

This implies that $u$ is a strong solution of \eqref{hs1},
\eqref{id}, and\eqref{bc} for the given $f\in L^2(\Omega_T,\R^m)$.

\smallskip

{\it Uniqueness and Continuous Dependence.} Follow from Lemma~\ref{lem:ineq}. 
\end{proof}

There are results concerning the regularity of the solution of \eqref{hs1},
\eqref{id}, and \eqref{bc}. If $f\in L^1((0,T); L^2(\Omega ))$ and $u_0\in L^2(\Omega)$
then the solution $u$ belongs to $C((0,T);L^2(\Omega))$ (see \cite{R}). In fact,
this regularity result can be improved by imposing more regularity for $f$ and $u_0$
and, in addition, compatibility conditions at the corner $\{ t=0\}\x\partial\Omega$.
These compatibility conditions are computed in the following fashion. Denote by
$\pi (x)$ the orthogonal projection onto $N(x)^{\perp}$. The compatibility condition
of order $j$ comes from expressing $\partial_t^j(\pi u)$ at $\{ t=0\}\x\partial\Omega$
in terms of $u_0$ and $f$ and requiring that the resulting expression vanishes.
For example, for $j=0$, we find $\pi u_0=0$ on $\partial\Omega$, or $u_0\in N(x)$
for all $x\in\partial\Omega$. For $j=1$, we have to impose: $\pi u_0=0$ and
$\pi (f(0,\cdot)-\sum_{i=1}^NA_i\partial_iu_0)=0$ on $\partial\Omega$.

{\bf Definition.} (\cite{R}) A smooth vector field $\ub{\gamma}$ on
$\overline{\Omega}$ is called {\it tangential} if and only if, for every $x\in
\partial\Omega$, $\langle\ub{\gamma}(x),\ub{n}(x)\rangle =0$. For $s$ a positive
integer, the space $H^s_{{\rm tan}}(\Omega )$ consists of those $u\in L^2(\Omega)$
with the property that for any $l\leq s$ and tangential fields $\{\ub\gamma_i\}_{i=1}^l$,
$\ub\gamma_1\ub\gamma_2\,\cdots\,\ub\gamma_l u\in L^2(\Omega)$ (see \cite{BR} for more
on $H^s_{{\rm tan}}$ spaces).
  
\begin{thm}(\cite{R})
Suppose $s\geq 1$ is an integer and $A_i,\, N,\, \partial\Omega$ are of class
$C^{s,1}$. Suppose the data $u_0\in H^s$ and $\partial_t^jf\in L^1((0,T);
H_{{\rm tan}}^{s-j}(\Omega))$ for $0\leq j\leq s$ and in addition there is a 
$0<T'\leq T$ such that $\partial_t^jf\in L^1((0,T');
H^{s-j}(\Omega))$, $0\leq j\leq s$. If the data satisfy the compatibility conditions
up to order $s-1$, then the solution $u$ of \eqref{hs1},
\eqref{id}, and \eqref{bc} lies in $C^j([0,T];H_{{\rm tan}}^{s-j}(\Omega))$ for
$0\leq j\leq s$.
\end{thm}
We close this subsection by giving an algebraic characterization of maximal
non-negative boundary conditions.

Suppose that the boundary matrix $A_n(x,t)$ has $l_0$ 0--eigenvalues 
$\lambda_1(x,t)$,$\,\ldots\,$ ,$\lambda_{l_0}(x,t)$, $l_{-}$ negative eigenvalues
$\lambda_{l_0+1}(x,t)$,$\,\ldots\,$ ,$\lambda_{l_0+l_{-}}(x,t)$, and $l_+$ positive
eigenvalues $\lambda_{l_0+l_-+1}(x,t)$, $\,\ldots\, $, $\lambda_{m}(x,t)$. Let
$\ub{e}_1(x,t)$,$\,\ldots\,$, $\ub{e}_{l_0}$, $\ub{e}_{l_0+1}$,$\,\ldots\,$,  
 $\ub{e}_{l_0+l_-}$, $\ub{e}_{l_0+l_-+1}$,$\,\ldots\,$,
$\ub{e}_m(x,t)$ be the corresponding eigenvectors.
Naturally, at $(x,t)\in \partial\Omega\x (0,T)$,
any vector of $N(x,t)$ has the form $v=\sum_{i=1}^m\alpha_i\ub{e}_i(x,t)$.
The non-negative condition \eqref{pos} implies:
$$v^TA_n(x,t)v=\sum_{i=1}^m\lambda_i\alpha_i^2\geq 0,$$
or
\begin{equation}\label{la}
\sum_{\lambda_i>0}\lambda_i\alpha_i^2\geq -\sum_{\lambda_j<0}\lambda_j\alpha_j^2.
\end{equation}
Observe that the dimension of $N(x,t)$ must be equal to the number of
positive and null eigenvalues counted with their multiplicities.

Now, any $v\in N(x,t)$ can be written as
$v=Q\alpha$,
where $\alpha =(\alpha_1,\ldots ,\alpha_m)^T$, and
$Q=(\ub{e}_1(x,t),\ldots ,\ub{e}_m(x,t))$.

Since $N(x,t)$ is a subspace of codimension $l_-$,
there exists a $l_-\x m$ matrix $E(x,t)$ such that 
$$N(x,t)=\{ v\, :\ E(x,t)v=0\}.$$
So,
$EQ \alpha =0$, or
\begin{equation}\label{0np}
S_0\alpha_0+S_-\alpha_- -S_+\alpha_+=0,
\end{equation}
where $\alpha_0=(\alpha_1$,$\,\ldots\,$,$\alpha_{l_0})$,
$\alpha_-=(\alpha_{l_0+1}$,$\,\ldots\,$,$\alpha_{l_0+l_-})$,
$\alpha_+=(\alpha_{l_0+l_-+1}$,$\,\ldots\,$,$\alpha_{m})$, and
$S_0=EQ_0,\ S_-=EQ_-,\ S_+=-EQ_+$, with
$Q_0=(\ub{e}_1(x,t)$,$\,\ldots\,$,$\ub{e}_{l_0}(x,t))$, 
$Q_-=(\ub{e}_{l_0+1}(x,t)$,$\,\ldots\,$,
$\ub{e}_{l_0+l_-}(x,t))$, and $Q_+=(\ub{e}_{l_0+l_-+1}(x,t)$,$\,\ldots\,$,$\ub{e}_m(x,t))$.

Since $N(x,t)$ is maximal non-negative, it follows that $\Ker{A_n(x,t)}\subseteq
N(x,t)$, and so,
$$S_0=EQ_0=0_{l_-\x l_0}.$$

Therefore, \eqref{0np} reads
\begin{equation}\label{np}
S_-\alpha_- -S_+\alpha_+=0.
\end{equation}

Observe that the $l_-\x l_-$ matrix $S_-$ is invertible. If not, 
there exists $\alpha_-\neq 0$ so that $S_-\alpha_-=0$, and so,
$$v=\sum_{i=l_0+1}^{l_0+l_-}\alpha_i\ub{e}_i(x,t)$$
belongs to $N(x,t)$. But this is in contradiction with \eqref{la}. Hence,
$S_-$ is invertible.

So, \eqref{np} recasts into
\begin{equation*}
\alpha_-=S_-^{-1}S_+\alpha_+.
\end{equation*}

Now, we must have \eqref{la} satisfied, which leads to 
\begin{equation*}
\| \left(\begin{array}{ccc}
\sqrt{|\lambda_{l_0+1}|}&\ldots &0\\
\vdots &  & \vdots\\
0&\ldots & \sqrt{|\lambda_{l_0+l_-}|}
\end{array}\right) S_- ^{-1}S_+
\left(\begin{array}{ccc}
1/\sqrt{\lambda_{l_0+l_-+1}}&\ldots &0\\
\vdots &  &\vdots\\
0&\ldots & 1/\sqrt{\lambda_{m}}
\end{array}\right)\| \leq 1.
\end{equation*}

Let us state what we have just proved into the following
proposition.

\begin{prop}\label{MaxNNeg}
The boundary condition \eqref{bc} is maximal non-negative
if and only if there exists a $l_-\x l_+$ matrix $M(x,t)$ such that
\begin{equation}\label{mnn}
E(x,t)=\left(\begin{array}{c}
\ub{e}_{l_0+1}^T(x,t)\\
\vdots \\
\ub{e}_{l_0+l_-}^T(x,t)
\end{array}\right)-M(x,t)
\left(\begin{array}{c}
\ub{e}_{l_0+l_-+1}^T(x,t)\\
\vdots \\
\ub{e}_{m}^T(x,t)
\end{array}\right),
\end{equation}
with
\begin{equation*}
\| \left(\begin{array}{ccc}
\sqrt{|\lambda_{l_0+1}|}&\ldots &0\\
\vdots &  &\vdots\\
0&\ldots & \sqrt{|\lambda_{l_0+l_-}|}
\end{array}\right) M
\left(\begin{array}{ccc}
1/\sqrt{\lambda_{l_0+l_-+1}}&\ldots &0\\
\vdots &  &\vdots\\
0&\ldots & 1/\sqrt{\lambda_{m}}
\end{array}\right)\| \leq 1.
\end{equation*}
\end{prop}
The following corollary is an immediate consequence of Proposition~\ref{MaxNNeg}.
\begin{cor}
A sufficient condition for having a maximal non-negative boundary
condition of the  form \eqref{bc}, with $E(x,t)$ given by \eqref{mnn},
is
 $$\| M\| \leq \frac{\min_{\lambda_j>0}{\sqrt{\lambda_j}}}{
\max_{\lambda_j<0}{\sqrt{| \lambda_j |}}}.$$
\end{cor}

\subsection{Boundary Conditions for Constrained Systems}

In this subsection we establish some connections between the problems
\eqref{a1h}--\eqref{a3h} and \eqref{un1}--\eqref{un2} restricted
to a bounded smooth domain $\Omega$. More precisely, we analyze
the links between the boundary subspaces and boundary matrices associated
to the two problems.

First of all, it is easy to see that the boundary matrix associated to 
\eqref{un1}--\eqref{un2} is given by
\begin{equation*}
\overline{A}_n(x)=\left(\begin{array}{cc}
A_n(x)&B_n^T(x)\\
B_n(x)&0
\end{array}\right) ,
\end{equation*}
where, as in Subsection~\ref{2.4.1}, $A_n(x)=-\sum_{j=1}^Nn_jA^j$, and 
$B_n(x)=-\sum_{j=1}^Nn_jB^j$,
with $n(x)=(n_1,\ldots ,n_N)$ the outer normal to $\Omega$ at $x\in\partial\Omega$. 

\begin{lem} 
\[ A_n(x)(\Ker B_n(x))\subset\Ker B_n(x), \] for all $x\in\partial\Omega $.
\end{lem}
\begin{proof}
Let $v\in \Ker B_n(x)$ and $u\in\Ker B$ such that $v=\hat{u}(n)$.
Applying the Fourier transform to \eqref{cond}, it follows that
$B_n(x)A_n(x)\hat{u}(n)=0$, or $B_n(x)A_n(x)v=0$. So, 
$A_n(x)(\Ker B_n(x))\subset\Ker B_n(x)$, for all $x\in\partial\Omega $.
\end{proof}

\begin{lem}
\begin{equation}\label{equal}
\Ker{\overline{A}_n(x)}=(\Ker A_n(x)\cap\Ker B_n(x))\x \Ker B_n^T(x),
\end{equation}
for all $x\in\partial\Omega $.
\end{lem}
\begin{proof}
Clearly,
\begin{equation}\label{incl1}
(\Ker A_n(x)\cap\Ker B_n(x))\x \Ker B_n^T(x)\subset\Ker{\overline{A}_n(x)},
\end{equation}
for all $x\in\partial\Omega $.

Let $\left(\begin{array}{c} u\\ z \end{array}\right) \in\Ker{\overline{A}_n(x)}$.
Then
\begin{gather}\label{Anu}
A_nu+B_n^Tz=0,\\ \label{Bnu}
B_nu=0.
\end{gather}
From \eqref{Bnu}, $u\in\Ker B_n$, and so, $A_nu\in\Ker B_n$, from the previous lemma. 
Applying $B_n$ to
\eqref{Anu}, it follows that $B_nB_n^Tz=0$, which implies $\| B_n^Tz\| =0$,
by multiplying it with $z^T$. Thus $z\in \Ker B_n^T$. Returning to \eqref{Anu},
observe that $u\in\Ker A_n$. Putting together all this information, we have that
$$\left(\begin{array}{c} u\\ z \end{array}\right) \in 
(\Ker A_n\cap\Ker B_n)\x \Ker B_n^T.$$
Hence,
\begin{equation}\label{incl2}
\Ker{\overline{A}_n(x)}\subset (\Ker A_n(x)\cap\Ker B_n(x))\x \Ker B_n^T(x),
\end{equation}
for all $x\in\partial\Omega $.

From \eqref{incl1} and \eqref{incl2}, we get \eqref{equal}.
\end{proof}
\begin{lem}
If $u$ is a non-negative vector for $A_n$ and $z\bot B_nu$, then 
$\left(\begin{array}{c} u\\ z \end{array}\right)$ is non-negative for
$\overline{A}_n$.
\end{lem}
\begin{proof}
Under the given hypotheses, the conclusion is a simple consequence of the 
following equalities
\begin{equation*}
(u^T,z^T)\overline{A}_n\left(\begin{array}{c} u\\ z \end{array}\right) =
u^TA_nu+2z^TB_nu=u^TA_nu.
\end{equation*}
\end{proof}
An immediate corollary of this lemma is 
\begin{cor}\label{cor:nonneg}
The subspace $\overline{N}=N\x [B_n(N)]^{\bot }$ is non-negative for
$\overline{A}_n$ if and only if $N$ is non-negative for $A_n$.
\end{cor}
\subsubsection{Inhomogeneous Boundary Conditions}
In this part, we consider the problem of finding $u(x,t)\in\R^m$, 
$x\in\Omega\subset\R^N$, $t>0$, for \eqref{a1h}, subject to
initial condition \eqref{a3h}, constraints \eqref{a2h}, and linear
inhomogeneous boundary conditions
\begin{equation}\label{Ei}
E(x,t)u(x,t)=g(x,t)\mbox{ on }\partial\Omega\x (0,T).
\end{equation}
Assume $g$ is a vector function defined on $\partial\Omega$ for all
time $t>0$ such that there exists $\tilde{g}$ satisfying the
constraint equation \eqref{a2h} and $g=E\tilde{g}$ on the
boundary $\partial\Omega$ for all time $t>0$. Then, by substituting
$\tilde{u}=u-\tilde{g}$, we arrive to the constrained initial
{\it homogeneous} boundary value problem
\begin{gather}\label{eq2}
\dot{\tilde{u}}=A\tilde{u}+A\tilde{g}-\dot{\tilde{g}}+f,\\
\tilde{u}(x,0)=u_0(x)-\tilde{g},\\
B\tilde{u}=0,\\ \label{B2}
E\tilde{u}=0.
\end{gather}
It is easy to see that the compatibility conditions for this last
problem are satisfied and, in fact, \eqref{eq2}--\eqref{B2} is
equivalent to the original constrained initial {\it inhomogeneous} boundary value
problem \eqref{a1h}--\eqref{a3h} and \eqref{Ei}. Therefore, in this way and for a
restricted set of boundary inhomogeneities, the treatment of
inhomogeneous boundary conditions reduces to the treatment of 
the corresponding homogeneous ones.

\chapter{Hyperbolic Formulations of Einstein Equations}
\label{ch:main}
\section{Introduction}
Einstein's equations can be viewed as equations for geometries, that is,
its solutions are equivalent classes under spacetime diffeomorphisms of
metric tensors. To break this diffeomorphism invariance, Einstein's 
equations must be first transformed into a system having a well-posed Cauchy
problem. The initial method to solve this problem has been by ``fixing the
gauge'' \cite{FB}, \cite{CB2}, or in other words, by imposing some conditions on
the metric components which would select only one representative from each
equivalent class of Einstein's solutions. By using an ingenious choice of
gauge fixing, the so called ``harmonic gauge,'' Einstein's equations can be 
converted to a set of coupled wave equations, one for each metric component.
Thus, by prescribing at an initial hypersurface values for the variables and 
their normal derivatives, one gets unique solutions to this system of wave
equations. In Appendix B, we explain this approach, restricted to the linearized
case for the sake of simplicity. For a more detailed discussion on this topic,
see \cite{Re}, \cite{FH}, and references therein. 

Another way to deal with the gauge freedom of Einstein's equations is by
prescribing the time foliation along evolution, that is, by prescribing a 
lapse-shift pair along evolution \cite{ADM}. Einstein's equations are
then decomposed into evolution equations and constraint equations on the
foliation hypersurfaces. An analogous decomposition occurs for Maxwell's
equations, which are canonically split into constraint (divergence) equations
and evolution (curl) equations.
Both the constraints and the evolution equations are not uniquely determined
by this procedure: by taking combinations of the constraints or/and by adding any combination of
constraints to any of the evolution equations one gets an equivalent decomposition.  

There have been numerical schemes to solve Einstein's equations based on the
ADM decomposition \cite{ADM}, but they have had only a very limited success.
The instabilities observed in the ADM based schemes might be at least in part 
caused by the weakly hyperbolicity of the first order differential reduction
of the ADM evolution equations, as argued in \cite{KST}. In fact, it is
known that some of the stability problems of numerical schemes are due to 
properties of the equations themselves. By rewriting the equations in a different
form, one can obtain better stability of computations for the very same
numerical methods.

There have been a large number of first order hyperbolic formulations derived
from the ADM decomposition in recent years \cite{As1}, \cite{As2}, 
\cite{FR}, \cite{AY}, \cite{BM}, \cite{BMSS}, \cite{KST}, \cite{AA}, among others. 
To give a survey of all of them appears to be a very difficult and extensive task, 
which is out of the scope of this dissertation (see \cite{Hern} for a more
comprehensive review). All such formulations
must be equivalent since they describe the same physical phenomenon. However, they
can admit different kinds of unphysical solutions which can grow rapidly
in time and overwhelm the physical solution in numerical computations. This is
one reason (signaled in \cite{KST}) why some formulations of Einstein's
equations behave numerically better than others. From this point of view, first
order symmetric hyperbolic formulations of Einstein's equations present a special
attraction for a couple of reasons. First of all, there is a large body of
experience and numerical codes that are stable for numerical simulations
of FOSH systems derived from various applications (transport equations, 
wave equations, electromagnetism, etc.). Also, the symmetric hyperbolicity
of the system ensures well--posedness and gives bounds on the solution growth.
In fact, hyperbolicity refers to algebraic conditions on the principal part of the
equations which imply well--posedness, that is, if appropriate initial data
is given on an appropriate hypersurface, then a unique solution can be found
in a neighborhood of that hypersurface, and that solution depends continuously, 
with respect to an appropriate norm, on the initial data.

In this chapter, we discuss some FOSH formulations of Einstein's equations introduced
in recent years. Moreover, a new formulation due to Arnold \cite{A} is presented and
analyzed.

\section{Einstein Equations}

In general relativity, spacetime is a 4-dimensional manifold of events endowed with
a pseudo-Riemannian metric
\begin{equation}
ds^2=g_{\alpha \beta}dx^{\alpha}dx^{\beta}.
\end{equation}
This metric determines curvature on the manifold, and Einstein's
equations relate the curvature at a point of spacetime to the
mass-energy there:
\begin{equation} 
G_{\alpha \beta}=8\pi T_{\alpha \beta},
\end{equation}
where $G_{\alpha \beta }$ is the {\it Einstein tensor} and $T_{\alpha \beta }$
is the {\it energy-momentum tensor}.

The Einstein tensor $G$ is a second order tensor built from
the given metric $g_{\alpha\beta}$ as follows:
$$G_{\alpha\beta}=R_{\alpha\beta}-\frac{1}{2}Rg_{\alpha\beta}.$$
Here $R_{\alpha\beta}$ is the {\it Ricci tensor: }$$R_{\alpha\beta}=R_{\alpha\delta\beta}^{\delta},$$
where $$R_{\alpha\beta\gamma}^\delta=\partial_{\alpha}\Gamma ^\delta_{\beta\gamma}-
\partial_{\beta}\Gamma ^\delta_{\alpha\gamma}+\Gamma^\epsilon _{\beta\gamma}
\Gamma^\delta _{\epsilon\alpha}-\Gamma^\epsilon _{\alpha\gamma}
\Gamma^\delta _{\epsilon\beta}$$ is the {\it Riemann curvature tensor}.

The {\it Christoffel symbols} are defined as
$$\Gamma^\alpha_{\beta\delta}=\frac{1}{2}g^{\alpha\lambda}(
\partial_{\delta} g_{\beta\lambda}+
\partial_{\beta} g_{\lambda\delta}-
\partial_{\lambda} g_{\beta\delta}).$$

By $R$ we denote the {\it scalar curvature}  $$R=g^{\alpha\beta}R_{\alpha\beta}.$$ 

The energy-momentum tensor $T_{\alpha\beta}$ can be better understood by looking
at two of the simplest energy-momentum tensors in general
relativity, namely, the energy-momentum tensors for incoherent matter or dust
and for a perfect fluid.

{\bf a)} {\it Incoherent matter (non-interacting incoherent matter or dust)}

Such a field may be characterized by two quantities, the {\it 4-velocity} 
vector field of flow $$ u^a=\frac{dx^a}{d\tau},$$ where $\tau$ is the proper
time along the world-line of a dust particle and a scalar field
$$\rho_0=\rho_0(x)$$ 
describing the {\it proper density} of the flow, that is, the density which would be measured by an observer moving with the flow (a co-moving observer).

The simplest second-rank tensor we can construct from these two quantities is
$$T^{ab}=\rho_0 u^au^b$$ and this turn out to be the {\it energy-momentum tensor} for the matter field.

Now, let us investigate this tensor in special relativity in Minkowski coordinates. In this case
$$u^a=\frac{dx^a}{d\tau}=\gamma (1, u_x, u_y, u_z),$$ where $\gamma =(1-u^2)^{-1/2}$ and $\tau$ is the proper time defined by
$$d\tau ^2=ds^2=dt^2-dx^2-dy^2-dz^2=dt^2(1-u^2)=\gamma ^{-2}dt^2.$$

Then, the $T^{00}$ component of $T^{ab}$ is 
$$T^{00}=\rho_0\frac{dx^0}{d\tau}\frac{dx^0}{d\tau}=\rho_0\frac{dt^2}{d\tau^2}=
\gamma ^2\rho_0 .$$

This quantity has a simple physical interpretation. First of all, in special
relativity, the mass of a body in motion is greater than its rest mass by a 
factor $\gamma$ ($m=\gamma m_0$). In addition, if we consider a moving three-dimensional volume element, then its volume decreases by a factor $\gamma$
through the Lorentz contraction. Thus, from the point of view of a fixed
as opposed to a comoving observer, the density increases by a factor $\gamma^2$.
Hence, if a field of material of proper density $\rho_0$ flows past a fixed
observer with velocity $\ub u$, then the observer will measure a density 
$\rho =\gamma ^2\rho_0$.

The component $T^{00}$ may therefore be interpreted as the {\it relativistic
energy density} of the matter field since the only contribution to the energy of the field is from the motion of the matter.

The components of $T^{ab}$ are
\[T^{ab}=\rho \left( \begin{array}{cccc}
1& u_x& u_y& u_z\\
u_x& u_{x}^{2}& u_xu_y& u_xu_z\\
u_y& u_xu_y& u_{y}^2& u_yu_z\\
u_z& u_xu_z& u_yu_z& u^2_{z}
\end{array}\right) .\]

Next, we will show that the equations governing the force-free motion of a matter field of dust can be written in the following very succinct way
$$\partial _bT^{ab}=0.$$

When $a=0$, this equation becomes exactly the classical {\it equation of continuity} 
$$\frac{\partial\rho}{\partial t}+ div\, (\rho\ub u)=0,$$ which 
expresses the conservation of matter with density $\rho$ moving with velocity
$\ub u$. Since matter is the same as energy, it follows that the conservation of
energy equation for dust is $$\partial_bT^{0b}=0.$$

Writing the equations corresponding to $a\in \{1,2,3\}$, we get
$$\frac{\partial }{\partial t}(\rho\ub u)+\frac{\partial }{\partial x}(\rho u_x
\ub u)+\frac{\partial }{\partial y}(\rho u_y\ub u)+
\frac{\partial }{\partial z}(\rho u_z\ub u)=0.$$

Combining this with the equation of continuity, we obtain
$$\rho [\frac{\partial \ub u}{\partial t}+(\ub u\cdot\nabla)\ub u ]=0,$$ 
which is the Euler equation of motion
for a perfect fluid in classical fluid dynamics (in the absence of pressure
and external forces).

We have seen that the requirement that the energy-momentum tensor has zero
divergence in special relativity is equivalent to demanding conservation of energy and conservation of momentum in the matter field (hence the name {\it 
energy-momentum tensor}). 

If we use a non-flat metric, then the {\it conservation law} 
$$\partial_bT^{ab}=0$$ is replaced by its covariant counterpart
$$\nabla_bT^{ab}=0.$$ 

\bigskip

{\bf b)} {\it Perfect Fluid}

\medskip

A {\it perfect fluid} is characterized by three quantities
\begin{enumerate}
\item{ a {\it 4-velocity} $u^a=dx^a/d\tau$,}
\item{ a {\it proper density} $\rho_0=\rho_0(x)$,}
\item{ a {\it scalar pressure} $p=p(x)$.}
\end{enumerate}

Observe that, if $p$ vanishes, a perfect fluid reduces to incoherent matter.
This suggests that we take the energy-momentum tensor for a perfect fluid to be of the form
$$T^{ab}=\rho_0u^au^b+pS^{ab},$$ for some symmetric tensor $S^{ab}$. Since this 
tensor depends on the velocity and the metric, the simplest assumption we can
make is
$$ S^{ab}=\lambda u^au^b+\mu g^{ab},$$ where $\lambda$ and $\mu$ are constants.

Considering the conservation law $$\partial_bT^{ab}=0$$ in special relativity in Minkowski coordinates and demanding that it reduces in an appropriate limit to the continuity equation and the Euler equation in the absence of body forces, we obtain that $\lambda =1$ and $\mu =-1$. Therefore, the energy-momentum tensor of a perfect fluid is
$$T^{ab}=(\rho_0+p)u^au^b-pg^{ab}.$$

In the full theory, we again take the covariant form $\nabla_bT^{ab}=0$ for
the conservation law. In addition, $p$ and $\rho$ are related by an {\it equation
of state} which, in general, is an equation of the form $p=p(\rho , T)$, where
$T$ is the absolute temperature. Usually, $T$ is constant and so, that
equation of state reduces to $p=p(\rho )$.

\newcommand{\N}{\mathbb N}

The Einstein equations can be viewed in three different ways: 
\vspace{1mm}

1. The field equations are differential equations for determining
the metric tensor $g_{ab}$ from a {\it given} energy-momentum tensor $T_{ab}$.
An important case of the equations is when $T_{ab}=0$, in which case
we are concerned with finding {\it vacuum} solutions.

\vspace{1mm}

2. The field equations are equations from which the energy-momentum
tensor can be read off corresponding to a {\it given} metric tensor $g_{ab}$.
In fact, this rarely turns out to be very useful in practice because the
resulting $T_{ab}$ are usually physically unrealistic. In particular, it 
frequently turns out that the energy density goes negative in some region,
which is rejected as unphysical.

\vspace{1mm}

3. The field equations consist of {\it ten equations} connecting
{\it twenty quantities} (the ten components of $g_{ab}$ and the ten
components of $T_{ab}$). In this way, the field equations are viewed as
{\it constraints} on the simultaneous choice of $g_{ab}$ and $T_{ab}$.
This point of view is useful when one can partly specify the geometry
and the energy-momentum tensor from physical considerations and then the
equations are used to determine both quantities completely. 

\begin{remark}

\begin{enumerate}
\item{ Each of the 10 equations $G_{\alpha\beta}=8\pi T_{\alpha\beta}$ is a 2nd order PDE in 4 independent variables and 10 unknowns. These 10 equations involve lots of terms.}
\item{The equations are not independent, due to the Bianchi identities
$\nabla_\alpha G^{\alpha\beta}=0$. The energy-momentum tensor also must
satisfy these identities.}
\item{Solutions are not unique, because of gauge freedom (any diffeomorphism of
the manifold gives a reparametrization, and hence another solution).}
\end{enumerate}
\end{remark}

A subtle consequence of Einstein's equations is that relatively accelerating
bodies emit gravitational waves. These gravitational waves are very slight
variations in the spacetime metric tensor, which propagate at the speed of
light.
It is generally believed that extremely violent movements of huge masses, such
as collisions of black holes, should generate detectable gravitational waves.
However, because of their tiny amplitude, gravitational waves have eluded detection
until now. When gravitational waves will be detected, we will must determine the
cosmological event that could have caused them. This is in fact an inverse problem
and, as usual, we need the solution of the direct problem, which in this case
is the numerical solution of Einstein's equations.

\section{ADM (3+1) Decomposition}
\label{sec:ADM}
\begin{figure} 
 \centering 
 \includegraphics[width=10cm,height=9cm]{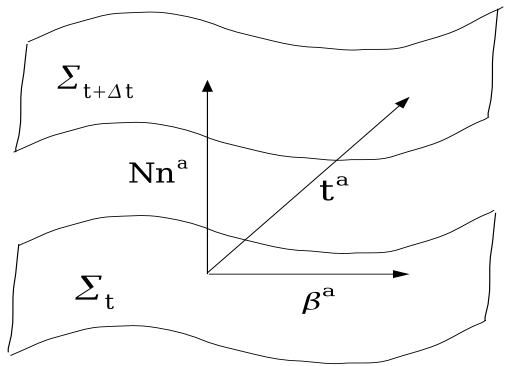} 
 \caption{A spacetime diagram for the ADM decomposition illustrating the definition
of the lapse function, $N$, and shift vector $\beta^a$.} 
 \label{ADM_figure} 
\end{figure}

In numerical relativity, the Einstein equations are usually solved as an
initial-boundary value problem. In other words, the spacetime is foliated
and each slice $\Sigma_t$ is characterized by its intrinsic geometry $\gamma_{ij}$ and 
extrinsic curvature $K_{ij}$. Subsequent slices are connected via the 
lapse function $N$ and shift vector $\beta^i$ (see Figure~\ref{ADM_figure}). 
The ADM decomposition \cite{ADM} (also \cite{Y}) 
of the line element
$$ds^2=-N^2dt^2+\gamma_{ij}(dx^i+\beta^idt)(dx^j+\beta^jdt),\ i,j=1,2,3,$$
allows one to express six of the ten components of Einstein's equations 
in vacuum as a system of evolution equations for the metric $\gamma_{ij}$ and the 
extrinsic curvature $K_{ij}$:
\begin{gather}\label{ev1}
\dot{\gamma}_{ij} = -2NK_{ij} + 2 \nabla_{(i}\beta_{j)},
\\ \label{ev2}
\dot{K}_{ij} =
N[R_{ij}+(K_l^l)K_{ij}-2K_{il}K_j^l]+\beta^l\nabla_l K_{ij}
+K_{il}\nabla_j\beta^l + K_{lj}\nabla_i\beta^l-\nabla_i\nabla_j N,
\\ \label{Ham}
R_i^i + (K_i^i)^2-K_{ij}K^{ij}=0,
\\ \label{Mom}
\nabla^jK_{ij}-\nabla_iK_j^j=0,
\end{gather}
where we use a dot to denote time differentiation.
The spatial Ricci tensor $R$ has components given
by second order spatial partial differential
operators applied to the spatial metric components $\gamma_{ij}$. Indices are raised
and traces taken with respect to the spatial metric, and parenthesized
indices are used to denote the symmetric part of a tensor.

The system of equations for $\gamma_{ij}$ and $K_{ij}$ is first order in time and second
order in space. It is not hyperbolic in any usual sense, and direct numerical
approaches have been unsuccessful. Therefore, many authors have considered reformulations
of \eqref{ev1}, \eqref{ev2} into more standard first order hyperbolic
systems. Typically, these approaches involve introducing other variables, like
the first spatial derivatives of the spatial metric components $\gamma_{ij}$ (or
quantities closely related to them). In the rest of this chapter we present some
of the most important first order hyperbolic formulations derived from the ADM
decomposition, as well as their linearizations around the Minkowski spacetime.
In the last section of this chapter we introduce a new first order symmetric hyperbolic
formulation of the linearized ADM equations around Minkowski's spacetime that 
has surprising resemblances with Maxwell's equations. This motivates the introduction
of the linearized ADM equations around the flat spacetime in the following
subsection.

\subsection{Linearized ADM Decomposition}
In this subsection, we derive the linearized ADM decomposition around
the Minkowski spacetime.

A trivial solution to the ADM system \eqref{ev1}--\eqref{Mom} is 
the Minkowski spacetime in Cartesian coordinates, given by 
$\gamma_{ij}=\delta_{ij}$, $K_{ij}=0$, $\beta^i=0$,
$N=1$. To derive the linearization of \eqref{ev1}--\eqref{Mom} about this solution,
we write
$\gamma_{ij}=\delta_{ij}+\bar g_{ij}$, $K_{ij}=\bar K_{ij}$,
$\beta^i=\bar\beta^i$, $N=1+\bar N$, where the bars indicate
perturbations, assumed to be small.  If we substitute these expressions
into \eqref{ev1}--\eqref{Mom} and ignore terms which
are at least quadratic in the perturbations and their derivatives, then
we obtain a linear system for the perturbations. Dropping the bars, the
system is
\begin{gather}
\label{LG}
\dot g_{ij}=-2K_{ij}+2\partial_{(i}\beta_{j)},
\\ \label{LKADM}
\dot K_{ij}=\partial^l\partial_{(j}g_{i)l}-\frac12\partial^l\partial_l
g_{ij}-\frac12\partial_i\partial_jg_l^l-\partial_i\partial_j N,
\\ \label{LC}
C:=\partial^j(\partial^lg_{lj}-\partial_jg_l^l)=0,
\\ \label{LCj}
C_j:=\partial^lK_{lj}-\partial_jK_l^l=0.
\end{gather}
The usual approach to solving the system \eqref{LG}--\eqref{LCj} is to
begin with initial data $g_{ij}(0)$ and $K_{ij}(0)$ defined on $\R^3$
and satisfying the constraint equations \eqref{LC}, \eqref{LCj}, and to
define $g_{ij}$ and $K_{ij}$ for $t>0$ via the Cauchy problem for the
evolution equations \eqref{LG}, \eqref{LKADM}.  It can be easily shown
that the constraints are then satisfied for all times.  Indeed, if we
apply the Hamiltonian constraint operator defined in \eqref{LC} to the
evolution equation \eqref{LG} and apply the momentum constraint operator
defined in \eqref{LCj} to the evolution equation \eqref{LKADM}, we obtain
the first order symmetric hyperbolic system \begin{equation*} \dot C =
-2\partial^j C_j, \quad \dot C_j = -\frac12 \partial_j C.
\end{equation*} Thus if $C$ and $C_j$ vanish at $t=0$, they vanish for
all time.

\section{Hyperbolic Formulations}
In this section, we present a number of popular hyperbolic formulations of
Einstein's equations derived from the ADM decomposition. In the last section,
we present and analyze a new first order symmetric hyperbolic formulation of Einstein's
equations in the linearized case due to Arnold \cite{A}.

\subsection{Kidder--Scheel--Teukolsky (KST) Family}
In this subsection, we present a many-parameter family of hyperbolic
representations  of Einstein's equations intoduced by Kidder, Scheel, and
Teukolsky \cite{KST}. 

In order to write the evolution equations \eqref{ev1} and \eqref{ev2}
in first-order form, we have to eliminate the second order derivatives of
the spatial metric. For this purpose, we introduce new variables
\begin{equation}\label{d}
d_{kij}=\partial_k\gamma_{ij}.
\end{equation}
Since $$R_{ij}=\frac{1}{2}\gamma^{ab}(\partial_{(i}\partial_{a}\gamma_{bj)}+
\partial_a\partial_{(i}\gamma_{j)b}-\partial_a\partial_b\gamma_{ij}-
\partial_{(i}\partial_{j)}\gamma_{ab})+\mbox{lower order},$$
the evolution system \eqref{ev1}, \eqref{ev2}, together with \eqref{d}
give
\begin{gather}\label{s1}
\dot{\gamma}_{ij}=-2NK_{ij}+\nabla_i\beta_j+\nabla_j\beta_i,\\
\dot{K}_{ij}= \frac{1}{2}N\gamma^{ab}(\partial_{(i}d_{abj)}+
\partial_ad_{(ij)b}-\partial_ad_{bij}-
\partial_{(i}d_{j)ab})-\partial_i\partial_jN+\ldots ,\\ \label{s3}
\dot{d}_{kij}=-2N\partial_kK_{ij}-2K_{ij}\partial_kN+\partial_k(\nabla_i\beta_j+
\nabla_j\beta_i).
\end{gather}
Since we have introduced a new variable that we will evolve independently of the
metric, we get additional constraints,
\begin{gather}
C_{kij}:=d_{kij}-\partial_k\gamma_{ij}=0,\\
C_{klij}:=\partial_{[k}d_{l]ij}=0.
\end{gather}

The system \eqref{s1}--\eqref{s3} has been proven to be only weakly
hyperbolic. Its characteristic matrix has eigenvalues $\{ 0,\, \pm 1\}$, 
but does not have a complete set of eigenvectors. Fortunately, the hyperbolicity
of the system can be changed by {\it densitizing} the lapse and adding constraints to
the evolution equations.

We densitize the lapse by writing: $N=\gamma^{\sigma}e^Q$, where $\gamma
=\det{(\gamma_{ij})}$, $\sigma$ is the densitization parameter, which is an 
arbitrary constant, and $Q$ is the lapse density, which will be chosen independent 
of the dynamical fields.

By adding terms proportional to the constraints, we can modify the evolution system
\eqref{s1}--\eqref{s3} without affecting the physical solution:
\begin{gather}\label{ss2}
\dot{K}_{ij}=(\ldots )+\theta N\gamma_{ij}C+\zeta N\gamma^{ab}C_{a(ij)b},\\ 
\label{ss3}
\dot{d}_{kij}=(\ldots )+\eta N\gamma_{k(i}C_{j)}+\chi N\gamma_{ij}C_k,
\end{gather}
where $(\ldots )$ represents the same thing that was before, $C$ is the Hamiltonian
constraint, $C_i$ are the momentum constraints, and $\{\theta ,\, \zeta,\, \eta,\, \chi
\}$ are arbitrary parameters.

By carrying out the computations, the evolution system, up to the principal part, 
is now given by ({\bf System 1} in \cite{KST}):
\begin{equation}\label{S1}
\begin{gathered}
\dot{\gamma}_{ij}=0,\\
\dot{K}_{ij}=\frac{1}{2}N\gamma^{ab}[\partial_ad_{bij}-(1+\zeta )\partial_ad_{(ij)b}
-(1-\zeta )\partial_{(i}d_{abj)}+\\ 
(1+2\sigma )\partial_{(i}d_{j)ab}-\theta\gamma_{ij} \gamma^{cd}\partial_ad_{cdb}+
\theta\gamma_{ij}\gamma^{cd}\partial_ad_{bcd}],\\
\dot{d}_{kij}=-2N\partial_kK_{ij}+N\gamma^{ab}(\eta \gamma_{k(i}\partial_aK_{bj)}
+\chi \gamma_{ij}\partial_aK_{bk}-\eta\gamma_{k(i}\partial_{j)}K_{ab}-\\
-\chi\gamma_{ij}\partial_kK_{ab}).
\end{gathered}
\end{equation}
The eigenvalues of the characteristic matrix of this system are $\{ 0,\pm 1,\pm c_1,
\pm c_2,\pm c_3\}$, where 
$c_1=\sqrt{2\sigma}$, $c_2=2^{-3/2}\sqrt{\eta-4\eta\sigma-2\chi-12\sigma\chi-3
\eta\zeta}$, and \newline
$c_3=2^{-1/2}\sqrt{2+4\theta-\eta-2\theta\eta+2\chi+4\theta\chi-\eta\zeta}$.

If all $c_i$ are real, then it can be proven that the system is strongly hyperbolic
unless one of the following conditions occurs:
\begin{gather}
c_i=0,\\
c_1=c_3\neq 1,\\
c_1=c_3=1\neq c_2.
\end{gather}
By differentiating the constraints in time, we obtain the following system for
the evolution of the constraints (up to the principal part)
\begin{equation}\label{c1}
\dot{C}= -\frac{1}{2}(2-\eta+2\chi )N\gamma^{pq}\partial_pC_q,
\end{equation}
\begin{equation}
\dot{C}_i=-(1+2\theta)N\partial_iC+\frac{1}{2}N\gamma^{pq}\gamma^{rs}
[(1-\zeta )\partial_qC_{prsi}+(1+\zeta)\partial_pC_{siqr}-(1+2\sigma)\partial_p
C_{qirs}],
\end{equation}
\begin{equation}
\dot{C}_{kij}=0,
\end{equation}
\begin{equation}\label{c4}
\dot{C}_{klij}=\frac{1}{2}\eta N(\gamma_{j[l}\partial_{k]}C_i+\gamma_{i[l}
\partial_{k]}C_j)+\chi N\gamma_{ij}\partial_{[k}C_{l]}.
\end{equation}
The eigenvalues for the characteristic matrix of this last system is a subset
of the eigenvalues of the evolution equations $\{ 0,\pm c_2,\pm c_3\}$.
Moreover, the constraint evolution system is strongly hyperbolic whenever
the regular evolution system is strongly hyperbolic.
 
We define two new variables: the generalized extrinsic curvature $P_{ij}$
and the generalized derivative of the metric $M_{kij}$
\begin{gather}\label{Pij}
P_{ij}=K_{ij}+\hat{z}\gamma_{ij}K,  \mbox{ so } K_{ij}=P_{ij}-\frac{\hat{z}}{1+3\hat{z}}
\gamma_{ij}P, \\
M_{kij}=\frac{1}{2}[\hat{k}d_{kij}+\hat{e}d_{(ij)k}+\gamma_{ij}(\hat{a}d_k+\hat{b}b_k)
+\gamma_{k(i}(\hat{c}d_{j)}+\hat{d}b_{j)})],
\end{gather}
where we introduce seven additional parameters $\{\hat{a},\, \hat{b},\,\hat{c},\,
\hat{d},\,\hat{e},\,\hat{k},\,\hat{z}\}$, and $d_j=\gamma^{ab}d_{jab}$, $b_j=\gamma^{ab}
d_{abj}$.

The System 1, \eqref{S1}, rewritten for these new variables, up
to the principal parts, becomes ({\bf System 2} in \cite{KST}):
\begin{equation}\label{S2}
\begin{gathered}
\dot{\gamma}_{ij}=0,\\
\dot{P}_{ij}=-N\gamma^{ab}(\mu_1\partial_aM_{bij}+\mu_2\partial_aM_{(ij)b}
+\mu_3\partial_{(i}M_{abj)}+\mu_4\partial_{(i}M_{j)ab}+\\
+\mu_5\gamma_{ij}\gamma^{cd}
\partial_aM_{cdb}+\mu_6\gamma_{ij}\gamma^{cd}\partial_aM_{bcd}),\\
\dot{M}_{kij}=-N(\nu_1\partial_kP_{ij}+\nu_2\partial_{(i}P_{j)k}+
\nu_3\gamma^{ab}g_{k(i}\partial_aP_{bj)}+\nu_4\gamma_{ij}\gamma^{ab}\partial_aP_{bk}
+\\
+\nu_5\gamma^{ab}\gamma_{k(i}\partial_{j)}P_{ab}+\nu_6\gamma_{ij}\gamma^{ab}
\partial_kP_{ab}),
\end{gathered}
\end{equation}
where $\mu_i=\mu_i(\sigma,\,\theta,\,\zeta,\, \eta,\,\chi,\,\hat{a},\, \hat{b},\,
\hat{c},\,\hat{d},\,\hat{e},\,\hat{k},\,\hat{z})$ and 
$\nu_i=\nu_i(\sigma,\,\theta,\,\zeta,\, \eta,\,\chi,\,\hat{a},\, \hat{b},\,\hat{c},\,
\hat{d},\,\hat{e},\,\hat{k},\,\hat{z})$.

The System 2, \eqref{S2}, is also strongly hyperbolic. 
Moreover, it has the same eigenvalues as System 1, \eqref{S1}, but the eigenvectors
are different.

\bigskip

{\bf Conclusion: } By densitizing the lapse, adding constraints to the evolution
equations, and changing variables, Kidder, Scheel, and Teukolsky \cite{KST} got a 
twelve-parameter family of strongly hyperbolic formulations of Einstein's equations. 
They have observed that the choice of parameters can have a
huge impact on the amount of time and accuracy of numerical simulations.
Unfortunately, nobody knows why one particular parameter choice behaves much better than
others.

Some well-known formulations can be recovered by making appropriate choices for 
parameters. For example, we can recover the Fritteli--Reula (FR) system \cite{FR} 
if the following choice of parameters is taken in \eqref{S2}:
$$\{ \sigma,\,\theta,\,\zeta,\, \eta,\,\chi,\,\hat{z},\,\hat{k}, \, \hat{a},\, \hat{b},\,\hat{c},\,
\hat{d},\,\hat{e}\}=\{ 1/2,\, -1,\, -1,\, 4,\, -2,\, -1,\, 1,\, -1,\, 0,\, 0, \,
0,\, 0\}.$$
In the next subsection, we present another popular hyperbolic formulation which 
can be recovered from \eqref{S2}. 

\subsection{Einstein--Christoffel (EC) formulation}
The EC formulation was originally derived  directly from the ADM system \eqref{ev1}--\eqref{Mom}
by Anderson and York \cite{AY} in 1999. It can be also recovered if we make the following 
choice of parameters in \eqref{S2}:
$$\{ \sigma,\,\theta,\,\zeta,\, \eta,\,\chi,\,\hat{z},\,\hat{k}, \, \hat{a},\, \hat{b},\,\hat{c},\,
\hat{d},\,\hat{e}\}=\{ 1/2,\, 0,\, -1,\, 4,\, 0,\, 0,\, 1,\, 0,\, 0,\, 2,\, -2,\, 0\}.$$
In this case, the system is (up to the principal parts):
\begin{equation}\label{ECg}
\begin{gathered}
\partial_t\gamma_{ij}=0,\\
\partial_tP_{ij}=-N\gamma^{ab}\partial_aM_{bij},\\ 
\partial_tM_{kij}=-N\partial_kP_{ij}.
\end{gathered}
\end{equation}
Essentially the coupled part of this system, i.e. the last two equations, is a set of six coupled quasilinear scalar 
wave equations with nonlinear source terms.

\subsubsection{Linearized Einstein--Christoffel Formulation}
First, we replace the lapse $N$ in \eqref{ev1}--\eqref{Mom} with $\alpha\sqrt \gamma$ 
where $\alpha$ denotes the lapse density. A trivial solution to this system is Minkowski spacetime 
in Cartesian coordinates, given by $\gamma_{ij}=\delta_{ij}$, $K_{ij}=0$, $\beta^i=0$,
$\alpha=1$. To derive the linearization, we write
$\gamma_{ij}=\delta_{ij}+\bar g_{ij}$, $K_{ij}=\bar K_{ij}$,
$\beta^i=\bar\beta^i$, $\alpha=1+\bar\alpha$, where the bars indicate
perturbations, assumed to be small.  If we substitute these expressions
into  \eqref{ev1}--\eqref{Mom} (with $N=\alpha\sqrt \gamma$), and ignore terms which
are at least quadratic in the perturbations and their derivatives, then
we obtain a linear system for the perturbations. Dropping the bars, the
system is
\begin{gather}
\label{G}
\dot g_{ij}=-2K_{ij}+2\partial_{(i}\beta_{j)},
\\ \label{KADM}
\dot K_{ij}=\partial^l\partial_{(j}g_{i)l}-\frac12\partial^l\partial_l
g_{ij}-\partial_i\partial_jg_l^l-\partial_i\partial_j\alpha,
\\ \label{C}
C:=\partial^j(\partial^lg_{lj}-\partial_jg_l^l)=0,
\\ \label{Cj}
C_j:=\partial^lK_{lj}-\partial_jK_l^l=0,
\end{gather}
where we use a dot to denote time differentiation.
\begin{remark}
For the linear system the effect of densitizing the lapse is to change
the coefficient of the term $\partial_i\partial_j g_l^l$ in
\eqref{KADM}.  Had we not densitized, the coefficient would have
been $-1/2$ instead of $-1$, and the derivation of the linearized
EC formulation below would not be possible.
\end{remark}

The usual approach to solving the system \eqref{G}--\eqref{Cj} is to
begin with initial data $g_{ij}(0)$ and $K_{ij}(0)$ defined on $\R^3$
and satisfying the constraint equations \eqref{C}, \eqref{Cj}, and to
define $g_{ij}$ and $K_{ij}$ for $t>0$ via the Cauchy problem for the
evolution equations \eqref{G}, \eqref{KADM}.  It can be easily shown
that the constraints are then satisfied for all times.  Indeed, if we
apply the Hamiltonian constraint operator defined in \eqref{C} to the
evolution equation \eqref{G} and apply the momentum constraint operator
defined in \eqref{Cj} to the evolution equation \eqref{KADM}, we obtain
the first order symmetric hyperbolic system \begin{equation*} \dot C =
-2\partial^j C_j, \quad \dot C_j = -\frac12 \partial_j C.
\end{equation*} Thus if $C$ and $C_j$ vanish at $t=0$, they vanish for
all time.

The linearized EC formulation provides an
alternate approach to obtaining a solution of \eqref{G}--\eqref{Cj}
with the given initial data, based on solving a system with better
hyperbolicity properties.  If $g_{ij}$, $K_{ij}$ solve
\eqref{G}--\eqref{Cj}, define
\begin{equation}\label{deff}
f_{kij}= \frac12 [\partial_k g_{ij}-
(\partial^l g_{li}-\partial_i g_l^l)\delta_{jk}-
(\partial^l g_{lj}-\partial_j g_l^l)\delta_{ik}].
\end{equation}
Then $-\partial^kf_{kij}$ coincides with the first three terms of the
right-hand side of \eqref{KADM}, so
\begin{equation}\label{K}
\dot K_{ij}=-\partial^kf_{kij}-\partial_i\partial_j\alpha.
\end{equation}
Differentiating \eqref{deff} in time, substituting \eqref{G},
and using the constraint equation \eqref{Cj}, we obtain
\begin{equation}\label{F}
\dot f_{kij}=-\partial_kK_{ij}+L_{kij},
\end{equation}
where
\begin{equation}\label{defL}
L_{kij}=\partial_k\partial_{(i}\beta_{j)}-
\partial^l\partial_{[l}\beta_{i]}\delta_{jk}-
\partial^l\partial_{[l}\beta_{j]}\delta_{ik}.
\end{equation}
The evolution equations \eqref{K} and \eqref{F} for $K_{ij}$
and $f_{kij}$,
together with the evolution equation \eqref{G} for $g_{ij}$,
form the linearized EC system.
As initial data for this system we use the given initial
values of $g_{ij}$ and $K_{ij}$ and derive the initial
values for $f_{kij}$ from those of $g_{ij}$ based on \eqref{deff}:
\begin{equation}\label{finit}
f_{kij}(0)= \frac12 \{\partial_k g_{ij}(0)-
[\partial^l g_{li}(0)-\partial_i g_l^l(0)]\delta_{jk}-
[\partial^l g_{lj}(0)-\partial_j g_l^l(0)]\delta_{ik}\}.
\end{equation}

A purpose of this dissertation is to study the preservation of constraints by the
linearized EC  system and the closely related question of the equivalence
of that system and the linearized ADM system. Our interest is in the
case when the spatial domain is bounded and appropriate boundary
conditions are imposed, but first we consider the result for the pure
Cauchy problem in the remainder of this subsection.

Suppose that $K_{ij}$ and $f_{kij}$ satisfy the evolution equations
\eqref{K} and \eqref{F} (which decouple from \eqref{G}).
If $K_{ij}$ satisfies the momentum constraint \eqref{Cj} for all
time, then from \eqref{K} we obtain a constraint which must be
satisfied by $f_{kij}$:
\begin{equation}\label{Fc}
\partial^k(\partial^lf_{klj}-\partial_jf_{kl}^{\hphantom{kl}l})=0.
\end{equation}
The following theorem shows that the pair of constraints \eqref{Cj},
\eqref{Fc} is preserved by the linearized EC evolution.
\begin{thm}\label{thm:constraint0}
Let initial data $K_{ij}(0)$, $f_{kij}(0)$ be given satisfying the
constraints \eqref{Cj} and \eqref{Fc}.  Then the unique solution
of the evolution equations \eqref{K}, \eqref{F} satisfy
\eqref{Cj} and \eqref{Fc} for all time.
\end{thm}
\begin{proof}
It is immediate from the evolution equations
that each component $K_{ij}$ satisfies the
inhomogeneous wave equation
\begin{equation*}
\ddot K_{ij}=\partial^k\partial_k K_{ij}-\partial^kL_{kij}-
\partial_i\partial_j\dot\alpha .
\end{equation*}
Applying the momentum constraint operator defined in \eqref{Cj}, we
see that each component $C_j$ satisfies the homogeneous wave equation
\begin{equation}\label{w}
\ddot C_j =\partial^k\partial_k C_j.
\end{equation}
Now $C_j=0$ at the initial time by assumption, so if we can show
that $\dot C_j=0$ at the initial time, we can conclude that $C_j$
vanishes for all time.  But, from \eqref{K} and the definition of $C_j$,
\begin{equation}\label{Cjd}
\dot C_j = -\partial^k(\partial^lf_{klj} -\partial_j
f_{kl}^{\hphantom{kl}l}),
\end{equation}
which vanishes at the initial time by assumption.  Thus we have shown
$C_j$ vanishes for all time, i.e., \eqref{Cj} holds.  In view
of \eqref{Cjd}, \eqref{Fc} holds as well.
\end{proof}

In view of this theorem it is straightforward to establish the
key result that for given initial data satisfying the
constraints, the unique solution of the linearized EC evolution
equations satisfies the linearized ADM system, and so the linearized
ADM system and the linearized EC system are equivalent.

\begin{thm}\label{thm:equiv0}
Suppose that initial data $g_{ij}(0)$ and $K_{ij}(0)$ are given
satisfying the Hamiltonian constraint \eqref{C} and momentum constraint
\eqref{Cj}, respectively, and that initial data $f_{kij}(0)$ is defined
by \eqref{finit}.  Then the unique solution of the linearized EC
evolution equations \eqref{G}, \eqref{K}, \eqref{F} satisfies the
linearized ADM system \eqref{G}--\eqref{Cj}.
\end{thm}
\begin{proof}
First we show that the initial data $f_{kij}(0)$ defined in
\eqref{deff} satisfies the constraint \eqref{Fc}.  Applying the
constraint operator in \eqref{Fc} to \eqref{deff} we find
\begin{equation*}
\partial^k(\partial^lf_{klj} -\partial_j
f_{kl}^{\hphantom{kl}l})=
\frac12\partial_j(\partial^l\partial^kg_{kl}-
\partial^k\partial_kg_l^l)=\frac12\partial_jC,
\end{equation*}
which vanishes at time $0$ by \eqref{C}.  From
Theorem~\ref{thm:constraint0}, we conclude that $C_j=0$
for all time, i.e., \eqref{Cj} holds.
Then from \eqref{G} and \eqref{Cj} we see that $\dot C=-2\partial^j
C_j=0$, and, since $C$ vanishes at initial time by assumption,
$C$ vanishes for all time, i.e., \eqref{C} holds as well.

It remains to verify \eqref{KADM}.  From \eqref{F} and \eqref{G}
we have 
\begin{equation*}
\dot f_{kij}=\frac12 \partial_k\dot g_{ij}-
\partial^l\partial_{[l}\beta_{i]}\delta_{jk}-
\partial^l\partial_{[l}\beta_{j]}\delta_{ik}.
\end{equation*}
Applying the momentum constraint operator to
\eqref{G} and using \eqref{Cj}, it follows that
\begin{equation*}
\frac12 (\partial^l\dot g_{li}-
\partial_i\dot g_l^l)=\partial^l\partial_{[l}\beta_{i]},
\end{equation*}
so
$f_{kij}-[\partial_k g_{ij}-(\partial^lg_{li}-
\partial_ig_l^l)\delta_{kj}-(\partial^lg_{lj}-
\partial_jg_l^l)\delta_{ki}]/2$ does not depend on time.  
In view of \eqref{finit}, we have \eqref{deff}.

Substituting \eqref{deff} in \eqref{K} gives \eqref{KADM},
as desired.
\end{proof}

\subsection{Alekseenko--Arnold (AA) Formulation}
In this subsection we present a first order symmetric hyperbolic formulation of the
full nonlinear ADM system \eqref{ev1}--\eqref{Mom} which involves fewer unknowns 
than other hyperbolic formulations and does not require any arbitrary parameters.
The hyperbolic systems involves 14 unknowns, namely the components
of the extrinsic curvature and eight particular combinations of the first derivatives
of the spatial metric. In order to derive the AA formulation, we introduce
the notations $\Sym $ for the 6-dimensional space of symmetric matrices and $\T$ for
the 8-dimensional space of triply-indexed arrays $(w_{ijk})$ which are skew-symmetric
in the first two indices and satisfy the cyclic property
\begin{equation}\label{cyclic}
w_{ijk}+w_{jki}+w_{kij}=0.
\end{equation}
Define the operators $M:C^{\infty}(\R^3,\Sym )\to C^{\infty}(\R^3,\R^3 )$, 
$L:C^{\infty}(\R^3,\Sym )\to C^{\infty}(\R^3,\T )$, and 
$L^*:C^{\infty}(\R^3,\T )\to C^{\infty}(\R^3,\Sym )$ by
\begin{gather}
(Mu)_i=2\gamma^{pq}\partial_{[p}u_{i]q},\\
(Lu)_{ijk}=\partial_{[i}u_{j]k},\\
(L^*v)_{ij}=-\gamma_{qi}\gamma_{rj}\partial_pv^{p(qr)},
\end{gather}
respectively. Observe that the operators $L$ and $L^*$ are formal adjoints to
each other with respect to the scalar products $\langle\, u,w\,\rangle : =\int u_{ij}w^{ij}dx$ and
$\langle\, v,z\,\rangle :=\int v_{ijk}z^{ijk}dx$ on the spaces $C^{\infty}(\R^3,\Sym )$ and
$C^{\infty}(\R^3,\T )$ respectively.

Next, we introduce new variables
\begin{equation}
f_{ijk}=-\frac{1}{\sqrt{2}}[(L\gamma)_{ijk}+(M\gamma)_{[i}\gamma_{j]k}].
\end{equation}
Observe that $(f_{ijk})$ belongs to $\T$ and so, it has eight independent components.

Now we are able to write down the new formulation derived in \cite{AA} for the
ADM system \eqref{ev1}--\eqref{Mom}.
\begin{equation}\label{aa}
\begin{gathered}
\partial_0\gamma_{ij}=-2NK_{ij}+2\gamma_{k(i}\partial_{j)}\beta^k,\\
\frac{1}{\sqrt{2}}\partial_0K_{ij}=-N(L^*f)_{ij}+\ldots =
N\gamma_{mi}\gamma_{nj}\partial_lf^{l(mn)}+\ldots ,\\
\frac{1}{\sqrt{2}}\partial_0f_{ijk}=[L(NK)]_{ijk}+\ldots =
\partial_{[i}(NK)_{j]k}+\ldots .
\end{gathered}
\end{equation}
Here $\partial_0:=\partial_t-\beta^i\partial_i$ is the convective derivative and the
omitted terms are algebraic expressions involving  $\gamma_{ij}$, their spatial 
derivatives $\partial_k\gamma_{ij}$, $K_{ij}$, the lapse $N$, and the shift $\beta$. 

\subsubsection{Linearized AA Formulation}

The linearized AA formulation provides an
alternate approach to obtaining a solution of \eqref{LG}--\eqref{LCj}
with the given initial data, based on solving a system with better
hyperbolicity properties.  If $g_{ij}:=\gamma_{ij}$ and $K_{ij}:=\kappa_{ij}$ solve
\eqref{LG}--\eqref{LCj}, define
\begin{equation}\label{AAdeff}
\lambda_{kij}=-\frac{1}{\sqrt{2}}[\partial_{[k}\gamma_{i]j}+(M\gamma)_{[k}\delta_{i]j}],
\end{equation}
where $(M\gamma)_i=\partial^l\gamma_{il}-\partial_i\gamma_l^l$.

Then, proceeding as in \cite{AA}, we obtain an evolution system for 
$\kappa_{ij}$ and $\lambda_{kij}$
\begin{gather}\label{AAK}
\frac{1}{\sqrt{2}}\dot\kappa_{ij}=\partial^k\lambda_{k(ij)}-
\frac{1}{\sqrt{2}}\partial_i\partial_j\alpha ,\\ \label{AAF}
\frac{1}{\sqrt{2}}\dot\lambda_{kij}=\partial_{[k}\kappa_{i]j}+\eta_{kij},
\end{gather}
where
\begin{equation}\label{AAdefL}
\eta_{kij}=-\frac12(\partial_j\partial_{[k}\beta_{i]}+\partial^m\partial_{[m}
\beta_{k]}\delta_{ij}-\partial^m\partial_{[m}
\beta_{i]}\delta_{kj}).
\end{equation}
The evolution equations \eqref{AAK} and \eqref{AAF} for $\kappa_{ij}$
and $\lambda_{kij}$ form the linearized AA system.
As initial data for this system we use the given initial
values of $\gamma_{ij}$ and $\kappa_{ij}$ and derive the initial
values for $\lambda_{kij}$ from those of $\gamma_{ij}$ based on \eqref{AAdeff}:
\begin{equation}\label{AAfinit}
\lambda_{kij}(0)= -\frac{1}{\sqrt{2}}
[\partial_{[k}\gamma (0)_{i]j}+(M\gamma (0))_{[k}\delta_{i]j}].
\end{equation}
As shown in \cite{AA}, if $\gamma$ and $\kappa$ satisfy
the ADM system and $\lambda$ is defined by \eqref{AAdeff}, then $\kappa$
and $\lambda$ satisfy the symmetric hyperbolic system \eqref{AAK},
\eqref{AAF}.
Conversely, to recover the solution to the ADM system from \eqref{AAK},
\eqref{AAF}, the same $\kappa$ should be taken at time $0$, and
$\lambda$ should be given by \eqref{AAfinit}. Having $\kappa$
determined, the metric perturbation $\gamma$ is defined as follows
from \eqref{LG}
\begin{equation}\label{AAgamma}
\gamma_{ij}=\gamma_{ij}(0)-2\int_0^t(\kappa_{ij}-\partial_{(i}\beta_{j)})(s)\, ds.
\end{equation}

\subsection{Arnold (A) Formulation}
In this subsection, we present and analyze the linearized case of a 
new first order symmetric hyperbolic formulation of Einstein's equations due to Arnold \cite{A}.

\subsubsection{Notations and Identities}
We begin by introducing a number of notations and listing a couple
of useful identities. Let $\ubb\gamma$ be a symmetric $3\times 3$--matrix function and 
$\ub v$ a vector field (we use underbars for 3-vectors, double underbars 
for $3\times3$ matrices).

$\skw\ubb\gamma = (\ubb\gamma-\ubb\gamma^T)/2$

$\Skw\ub v = \left(\begin{matrix} 0 & -v_3 & v_2 \\ v_3 & 0 & -v_1 \\ -v_2 & v_1 &
0\end{matrix}\right)$

$\ubcurl\ub v=\left(\begin{matrix} \ipd{u_2}z-\ipd{u_3}y,  & \ipd{u_3}x-\ipd{u_1}z, &  
\ipd{u_1}y-\ipd{u_2}x
\end{matrix}\right) $

$\ubbcurlr\ubb\gamma$ is the matrix whose rows are the curls of the rows
of $\ubb\gamma$

$\ubbcurlc\ubb\gamma$ is the matrix whose columns are the curls of the
columns of $\ubb\gamma$

$\ubbcurls\ubb\gamma=(\ubbcurlr\ubb\gamma+\ubbcurlc\ubb\gamma)/2$

$\ubbnabla\ub v$ is the matrix whose rows are the gradients of the
entries of $\ub v$

$\ubdiv\ubb\gamma$ is the vector whose components are the divergences
of the rows of $\ubb\gamma$.

$\ubbeps\ub v = [\ubbnabla\ub v +(\ubbnabla\ub v)^T]/2$

$\ubb R\ubb\gamma=\ubbeps\ubdiv\ubb\gamma-1/2\ubblap\ubb\gamma
-1/2\ubbnabla\ubnabla\tr\ubb\gamma$

$\ub M \ubb\gamma = \ubdiv\ubb\gamma-\ubnabla\tr\ubb\gamma$

$\ubb M^*\ub v = -\ubbeps\ub v+(\div\ub v)\ubb\delta$ ($\ubb\delta$ being the 
$3\times 3$ unit matrix)

For two vector fields $\ub a$ and $\ub b$, denote by $\ub a\odot\ub b$ the
following symmetric matrix function
$\ub a\odot\ub b=(\ub a\;\ub b^T+\ub b\;\ub a^T)/2$.

The proofs of the following identities imply only elementary computations
and we leave them to the reader:
\begin{equation}\label{curlr*}
\ubbcurlr^*=\ubbcurlr
\end{equation}
\begin{equation}\label{curlc*}
\ubbcurlc^*=\ubbcurlc
\end{equation}
\begin{equation}\label{curls*}
\ubbcurls^*=\ubbcurls
\end{equation}
\begin{equation}\label{curlrcurlc}
\ubbcurlc\ubb\gamma = (\ubbcurlr\ubb\gamma)^T 
\ \mbox{(therefore $\ubbcurls\ubb\gamma$ is symmetric)}
\end{equation}
\begin{equation}\label{tr}
\tr\ubbcurlr\ubb\gamma =0
\end{equation}
\begin{equation}\label{Mcurlr}
\ub M\ubbcurlr\ubb\gamma =0
\end{equation}
\begin{equation}\label{Mcurls}
\ub M\ubbcurls\gamma =1/2\ubcurl \ub M \ubb \gamma =
-\frac12\ub M \Skw(\ub M\ubb\gamma) 
\end{equation}
\begin{equation}\label{curlsrc}
\ubbcurls\ubb\gamma =\ubbcurlr\ubb\gamma -\frac12\Skw(\ub M\ubb\gamma)
=\ubbcurlc\ubb\gamma +\frac12\Skw(\ub M\ubb\gamma)
\end{equation}
\begin{equation}\label{skewM}
\skw\ubbcurlr\ubb\gamma = -\frac12\Skw(\ub M\ubb\gamma)
\end{equation}
\begin{equation}\label{R}
\ubb R\ubb \gamma = \frac12\ubbcurlc\ubbcurlr\ubb\gamma +
\frac12 (\div\ub M\ubb\gamma )\ubb\delta
\end{equation}
\begin{equation}\label{Meps}
\div\ub M\ubbeps\ub v =0
\end{equation}
\begin{equation}\label{Mnn}
\ub M\ubbnabla\ubnabla N =0 \mbox{ for any function $N$}
\end{equation}
\begin{equation}\label{MM*}
\ub M\ubb M^*\ub v =-\ubdiv\ubbeps\ub v-\ubnabla\div\ub v
\end{equation}

\subsubsection{A New FOSH Formulation of Linearized ADM}

If we choose lapse $N$, shift $\ub\beta$, and linearize the ADM system
about the Minkowski's flat space, the perturbations $\ubb\gamma$, $\ubb\kappa$
of the metric and the extrinsic curvature, respectively, are symmetric
matrix fields satisfying
\begin{gather}\label{dotgamma}
\dot{\ubb\gamma}=-2\ubb\kappa +2\ubbeps\ub\beta,
\\ \label{dotkappa}
\dot{\ubb\kappa}=\ubb R\ubb\gamma-\ubbnabla\ubnabla N,
\\ \label{ham}
\div\ub M\ubb\gamma=0,
\\\label{mom}
\ub M\ubb\kappa=0.
\end{gather}
Of course, \eqref{dotgamma}--\eqref{mom} is exactly the linearized ADM system
\eqref{LG}--\eqref{LCj} written in vector--matrix notation. 

This system should be supplemented with initial conditions
\begin{equation}\label{ic}
\ubb\gamma (0)=\ubb\gamma_0,\quad\ubb\kappa (0)=\ubb\kappa_0,
\end{equation}
(that satisfy the constraints \eqref{ham} and \eqref{mom}) and boundary conditions 
if the domain has frontier.  

\begin{lem}\label{Mdkem}
\begin{equation}\label{Mkd}
\ub M\dot{\ubb\kappa} = -\frac12\ubnabla (\div\ub M\ubb\gamma).
\end{equation}
\end{lem}
\begin{proof} It follows from \eqref{dotkappa}, \eqref{ham}, and the
identity:$$\ubdiv\ubbcurlc\ubb\gamma =\ubcurl\ubdiv\ubb\gamma .$$
\end{proof}

\begin{thm}
If the initial conditions  $\ubb\gamma_0$ and $\ubb\kappa_0$ satisfy the Hamiltonian constraint \eqref{ham}
and the momentum constraint \eqref{mom} respectively, then the constraints are
automatically satisfied for all time by any solution of the pure Cauchy problem \eqref{dotgamma}, \eqref{dotkappa},
and \eqref{ic}.
\end{thm}
\begin{proof} From \eqref{Meps}, \eqref{dotgamma}, and \eqref{Mkd}, observe that
\[ \partial_t^2(\div\ub M\ubb\gamma )=-2\div\ub M\dot{\ubb\kappa}=\Delta (\div
\ub M\ubb\gamma ).\]

Denote by $\phi = \div\ub M\ubb\gamma $. Then $\partial_t^2\phi =\Delta\phi ,\ 
\phi (0)=0 $, and $\dot{\phi } (0)=0$. Therefore, $$\phi =\div\ub M\ubb\gamma\equiv 0.$$ 

Moreover, from Lemma~\ref{Mdkem}, $$\ub M\dot{\ubb\kappa}=-\frac12\ubnabla(\div\ub M\ubb\gamma) = 0,$$
and from $\ub M\ubb\kappa_0=0$, we get $\ub M\ubb\kappa=0$ for all time.
\end{proof}

From \eqref{dotgamma} and \eqref{mom}, it follows that
\begin{gather}\label{K2}
\partial_t^2\ubb\kappa= -2 \ubb R \ubb\kappa -\ubbnabla\ubnabla\dot N,\\ \label{MK}
\ub M\ubb\kappa =0.
\end{gather}

Taking into account \eqref{curlsrc}, \eqref{R}, and \eqref{ham}, the equation
 \eqref{K2} transforms into 
\begin{equation}
\label{KC}
\partial_t^2\ubb\kappa = -\ubbcurls\ubbcurls\ubb\kappa -
\ubbnabla\ubnabla\dot N.
\end{equation}

Introduce $\ubb\nu =\dot{\ubb\kappa}$, and $\ubb\mu =\ubbcurls\ubb\kappa$. Then the 
equations \eqref{KC}, \eqref{MK} induce the following first order symmetric hyperbolic
system with constraints
\begin{gather}\label{eq:n}
\dot{\ubb\nu} = -\ubbcurls\ubb\mu -\ubbnabla\ubnabla\dot N,\\ \label{eq:m}
\dot{\ubb\mu} = \ubbcurls\ubb\nu ,\\ 
\ub M\ubb\nu =0,\\ \label{eq:Mn}
\ub M\ubb\mu =0.
\end{gather}

Here, the initial data is 
\begin{equation}\label{eq:id}
\ubb\nu (0)=\ubb R \ubb\gamma (0)-
\ubbnabla\ubnabla N(0),\ \ubb\mu (0)=\ubbcurls\ubb\kappa (0).
\end{equation}

If $\ubb{\gamma}_0$ and $\ubb\kappa_0$ satisfy the Hamiltonian constraint \eqref{ham}
and the momentum constraint \eqref{mom} respectively, then
it is not hard to see that the compatibility conditions $\ub M\ubb\nu (0)=0$ 
and $\ub M\ubb\mu (0)=0$ are satisfied.

\begin{prop}
Suppose that $\ubb{\gamma}_0$ and $\ubb{\kappa}_0$ satisfy 
the Hamiltonian constraint \eqref{ham} and the momentum constraint \eqref{mom} respectively. 
Then, the problems \eqref{dotgamma}--\eqref{ic} and \eqref{eq:n}--\eqref{eq:id}
are equivalent. 
\end{prop}
\begin{proof}
From construction, it is obvious that once we have a solution for 
\eqref{dotgamma}--\eqref{ic}, then  \eqref{eq:n}--\eqref{eq:id} has solution. 

Let us prove that the converse is also valid. Assume we can solve 
\eqref{eq:n}--\eqref{eq:id} and define
\begin{gather}\label{eq:k}
\ubb{\kappa}=\ubb{\kappa}_0+\int_0^t\ubb{\nu}(s)\, ds,\\ \label{eq:g}
\ubb{\gamma}=\ubb{\gamma}_0+\int_0^t[-2\kappa (s)+2\ubb{\epsilon}\ub{\beta}(s)]\, ds.
\end{gather}
Obviously, $\ubb{\gamma} (0)=\ubb{\gamma}_0$ and $\ubb{\kappa} (0)=\ubb{\kappa}_0$ 
and so, \eqref{ic} is verified.
Now, we prove that \eqref{ham} is satisfied. From \eqref{Mcurls} it follows that
\begin{equation}
\partial_t(\ub{M}\ubb{\nu})=-\frac{1}{2}\ubcurl\ub{M}\ubb{\mu},
\end{equation}
and from here
\begin{equation}
\partial_t(\div\ub{M}\ubb{\nu})=0.
\end{equation}
Since $\div\ub{M}\ubb{\nu}_0=0$, we get $\div\ub{M}\ubb{\nu}=0$ for all time.

Next, we show that $\ubb{\kappa}$ satisfies \eqref{mom}. From \eqref{eq:m}, 
\eqref{eq:id}, and \eqref{eq:k} it follows that 
\begin{equation}\label{eq:mk}
\ubb{\mu}=\ubbcurls\ubb{\kappa}
\end{equation}
for all time. By using \eqref{eq:n}--\eqref{eq:m}, \eqref{eq:k}, and \eqref{eq:mk}
we can see that $\ub{M}\ubb{\kappa}$ satisfies the initial value problem
\begin{gather}
\partial^2_t(\ub{M}\ubb{\kappa})=\frac{1}{4}\ub{\Delta}{\ub{M}\ubb{\kappa}},\\
\ub{M}\ubb{\kappa}(0)=0,\ \partial_t(\ub{M}\ubb{\kappa})(0)=0,
\end{gather}
whose only solution is the trivial solution. So, $\ub{M}\ubb{\kappa}=0$ for all time.

Finally, let us prove that \eqref{dotgamma} and \eqref{dotkappa} are verified.
Observe that \eqref{dotgamma} is trivially satisfied by taking the derivative
of \eqref{eq:g} with respect to time. The identity \eqref{dotkappa} is a little
bit more delicate, but it follows from \eqref{R}, \eqref{eq:n}, 
\eqref{eq:k}, and \eqref{eq:mk}
\begin{equation}
\partial_t\ubb{\nu}=\partial_t^2\ubb{\kappa}=-\ubbcurls\ubbcurls\ubb{\kappa}-
\ubb{\nabla}\ub{\nabla}\dot{N}=-2\ubb{R}\ubb{\kappa}-\ubb{\nabla}\ub{\nabla}\dot{N}=
\partial_t(\ubb{R}\ubb{\gamma}-\ubb{\nabla}\ub{\nabla}N),
\end{equation}
and from \eqref{eq:id}, we get \eqref{dotkappa}.
\end{proof}

\subsubsection{Equivalent Unconstrained Initial Value Problem}

Observe that the constrained problem 
\eqref{eq:n}--\eqref{eq:id} can be put into the form 
\eqref{ae1}--\eqref{ae2} for
\[ A=\left(\begin{array}{cc}
0&-\ubbcurls\\
\ubbcurls&0
\end{array}\right),\ B=\left(\begin{array}{cc}
\ub M &0\\
0&\ub M
\end{array}\right),\ f=\left(\begin{array}{c}
-\ubb\nabla\, \ub\nabla\dot N\\
0
\end{array}\right), \mbox{ and } u_0=\left(\begin{array}{c}
\ubb\nu (0)\\
\ubb\mu (0) \end{array}\right).\]
 
Moreover, the condition \eqref{cond} is satisfied since
\[ Bx=0 \Leftrightarrow \ub M\ubb\mu =\ub M\ubb\nu =0 \Leftrightarrow \ubbcurls\ubb
\mu =\ubbcurlr\ubb\mu
\mbox{ and } \ubbcurls\ubb\nu =\ubbcurlr\ubb\nu \Leftrightarrow \]
\[ \ub M\ubbcurls\ubb\mu =\ub M\ubbcurls\ubb\nu =0 \Leftrightarrow BAx=0.\]

Therefore, we can transform our system into a $18\times 18$ symmetric hyperbolic
(unconstrained) system

\begin{equation}\label{A_ext_system}
\left( \begin{array}{c}
\dot{\ubb\nu}\\
\dot{\ubb\mu}\\
\dot{\underline{p}}\\
\dot{\underline{q}}
\end{array}\right) =\left( \begin{array}{cccc}
0&-\ubbcurls &-\ubb M^* &0\\
\ubbcurls&0&0&-\ubb M^* \\
\ub M &0&0&0\\
0&\ub M &0&0
\end{array}\right) 
\left( \begin{array}{c}
\ubb\nu\\
\ubb\mu\\
\underline{p}\\
\underline{q}
\end{array}\right) +
\left( \begin{array}{c}
-\ubb\nabla\, \ub\nabla\dot N \\
0\\
0\\
0
\end{array}\right) .
\end{equation}

The initial data for the above system reads

\begin{equation}\label{A_ext_id}
\left( \begin{array}{c}
\ubb\nu \\
\ubb\mu\\
\underline{p}\\
\underline{q}
\end{array}\right) (0)=
\left(\begin{array}{c}
\ubb R\ubb \gamma_0-\ubb\nabla\,\ub\nabla N(0)\\
\ubbcurls\ubb\kappa_0\\
0\\
0
\end{array}\right) .
\end{equation}

\chapter{Boundary Conditions for Einstein's Equations}

\section{Introduction}
A very difficult situation is encountered if initial {\it boundary} value problems
are considered for Einstein's equations. It has become clear in the numerical relativity
community that in order for constraints to be preserved during evolution, the boundary
conditions have to be chosen in an appropriate way. This problem is widely open in its
full generality for Einstein's equations. The main difficulty comes from the necessity
to deal with unphysical instabilities, which will always be triggered in numerical
simulations by round-off errors and lead to constraint violating numerical solutions.
In other words, the numerical solution of the initial boundary value problem with
constrained initial data fails to solve the constraints soon after the initial time.
This serious drawback motivates the quest for boundary conditions that preserve the 
constraints. Only in the last few years work along this line has been pursued; see \cite{S},
\cite{CLT}, \cite{CPSTR}, \cite{FG2}, \cite{FG3}, \cite{Holst}, among others. In view of our work, 
of  particular interest is the recent paper by Calabrese et al. \cite{CPSTR} on the generalized \cite{KST} 
Einstein--Christoffel formulation \cite{AY} linearized around Minkowski spacetime, which
employs techniques with some points in common with the ones used in our approach of the
problem in this chapter. It is here where the differential equations satisfied by the
constraints are very important. In particular, if they form a hyperbolic system, then
the study of its well--posedness is proven to be a very good starting point for what
boundary conditions we must force upon the evolution system for the dynamical variables.
By exploiting this idea, we have been able to develop a technique that provides maximal
nonnegative constraint preserving boundary conditions for various hyperbolic formulations 
of Einstein's equations in the linearized case. This entire chapter represents our
original contribution to the subject.

\section{Model Problem}
In this section we consider a constrained first order symmetric hyperbolic system, which, as we
shall prove, admits well-posed constraint preserving boundary conditions.
The analysis of this model problem gives a good deal of insight for finding
same type of  boundary conditions for various hyperbolic formulations of Einstein's equations.
Although the problem is simple, the techniques used here reveal our basic strategy
to tackle the more complex case of Einstein's equations.  

For $t>0$ and $x$ in $\R^3$, we are interested in finding a solution $(w_i,\, v_i,\, u_{ij})$ 
for the following first order symmetric hyperbolic system
\begin{equation}\label{hyps}
\dot{w}_i=v_i, \quad
\dot{v}_i=\partial^ju_{ij}+f_i, \quad
\dot{u}_{ij}=\partial_jv_i,
\end{equation}
with the initial data
\begin{equation}\label{idwaves}
w_i(0)=w_i^0, \quad
v_i(0)=v_i^0,\quad 
u_{ij}(0)=u_{ij}^0,
\end{equation}
and subject to the constraint
\begin{equation}\label{hypsC}
C:=\partial^iv_i=0.
\end{equation}
Assume that the initial conditions \eqref{idwaves} are
compatible with the constraint \eqref{hypsC}
\begin{equation}\label{nhypscomp1}
\partial^iv_i^0=0, \quad
\partial^iu_{ij}^0=0.
\end{equation}
Also, assume that the forcing terms satisfy the compatibility condition
\begin{equation}\label{nhypscomp2}
\partial^if_i=0, \quad  \mbox{ for all time }t\geq 0.
\end{equation}
Then, for the pure Cauchy problem \eqref{hyps}, \eqref{idwaves}, it can be easily shown that the
constraint \eqref{hypsC} is satisfied for all time. Indeed, if we differentiate twice in time
the constraint \eqref{hypsC} and use the main system \eqref{hyps}, it follows that
\begin{equation}\label{twiceC}
\ddot{C}=\partial^i\ddot{v}_i=\partial_t[\partial^i(\partial^ju_{ij}+f_i)]=\partial^i\partial^j\dot{u}_{ij}+
\partial_t(\partial^if_i)=\partial^j\partial_j\partial^iv_i+\partial_t(\partial^if_i)=\partial^j\partial_j C+
\partial_t(\partial^if_i).
\end{equation}
Since $\partial^if_i=0$ for all time, it follows that $C$ satisfies the wave equation on $\R_+\times\R^3$.
\begin{equation}\label{twiceCdel}
\ddot{C}=\Delta C.
\end{equation}
Moreover, from the compatibility conditions \eqref{nhypscomp1}
\begin{equation}
\partial^iv_i(0)=\partial^iv_i^0=0.
\end{equation}
So,
\begin{equation}\label{Cofzero}
C(0)=0.
\end{equation}
Also,
\begin{equation}
\partial^i\dot{v}_i(0)=\partial^j\partial^iu_{ij}(0)+\partial^i\dot{f}_i(0).
\end{equation}
The first term on the right side vanishes from the second compatibility condition in \eqref{nhypscomp1}.
The second term vanishes because of \eqref{nhypscomp2}. Thus, 
\begin{equation}\label{Cdotofzero}
\dot{C}(0)=0.
\end{equation}
From \eqref{twiceCdel}, \eqref{Cofzero}, and \eqref{Cdotofzero}, we conclude that $C=0$ for all time.

\subsection{Constraint-Preserving Boundary Conditions for the Model Problem}
In this subsection, we are interested in finding suitable boundary conditions for \eqref{hyps} on the
boundary $\partial\Omega$ of  
a {\it bounded} spatial domain $\Omega$ such that the solution of the resulting initial-{\it boundary} value
problem satisfies the constraint \eqref{hypsC} for all time.
\subsubsection{On Polyhedral Domains}
First, we investigate the existence of constraint-preserving boundary conditions on polyhedral domains, with
the result that there exists a set of such boundary conditions. Moreover, these boundary conditions are also
maximal nonnegative, and so, the corresponding initial-boundary value problem is well-posed.

\begin{figure} 
\centering 
\includegraphics[width=6cm,height=6cm]{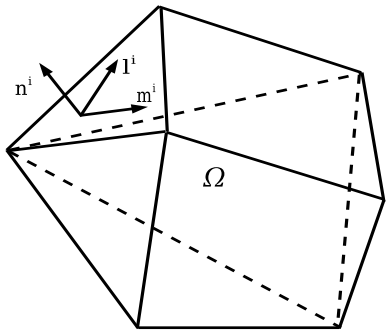} 
\caption{A polyhedral domain.} 
\label{polyhedral_domain} 
\end{figure}

Let $\Omega$ be a polyhedral domain in $\R^3$.
Consider an arbitrary face of $\partial\Omega$ and let $n^i$ denote its
exterior unit normal.  Denote by $m^i$ and $l^i$ two additional vectors
which together $n^i$ form an orthonormal basis (see Figure~\ref{polyhedral_domain}). 
The projection operator orthogonal to $n^i$ is then given by $\tau_i^j:=m_im^j+l_il^j$
(and does not depend on the particular choice of these tangential
vectors).  Note that
\begin{equation}\label{delid}
\delta_i^j=n_in^j + \tau_i^j, \quad \tau_i^j\tau_j^k=\tau_i^k.
\end{equation}
Consequently,
\begin{equation}\label{dotid}
a_lb^l = n^ja_j n_ib^i + \tau_l^j a_j \tau_i^lb^i \text{\quad for all
vectors $a_l$, $b^l$}. 
\end{equation}
We shall prove that the following set of boundary conditions
\begin{equation}\label{modelbc}
n^in^ju_{ij}=0,\quad 
m^iv_i=0, \quad
l^iv_i=0,
\end{equation}
is constraint-preserving and, together with \eqref{hyps} and \eqref{idwaves}, leads to
a well-posed initial-boundary value problem. Observe that these boundary conditions can
be written as well:
\begin{equation}\label{modelbc2}
n^in^ju_{ij}=0,\quad 
\tau^i_jv_i=0. \quad
\end{equation}
Therefore, they do not depend on the particular choice of the tangential vectors
$m^i$ and $l^i$.
\begin{thm}\label{thm:mpconstraint}
Given initial conditions  $w_i(0)$, $v_i(0)$, $u_{ij}(0)$
and forcing terms $f_i$ satisfying the compatibility conditions \eqref{nhypscomp1} and \eqref{nhypscomp2}
respectively, define $w_i$, $v_i$, and $u_{ij}$ for positive time by the evolution equations
\eqref{hyps} and the boundary conditions \eqref{modelbc}, or \eqref{modelbc2}.
Then, the constraint \eqref{hypsC} is satisfied for all time.
\end{thm}
\begin{proof} First we prove that $C=0$ on any boundary face of $\Omega$.
From the first identity of \eqref{delid}, the constraint $C$ can be decomposed as
follows
\begin{equation}\label{Cninj1}
C=\partial^iv_i=n^in^j\partial_jv_i+\tau^{ij}\partial_jv_i.
\end{equation}
By the third equation in \eqref{hyps} and the second identity of \eqref{delid},
\eqref{Cninj1} reads
\begin{equation}\label{Cninj2}
C=n^in^j\dot{u}_{ij}+\tau^{ik}\tau^j_k\partial_jv_i.
\end{equation}
From the boundary conditions \eqref{modelbc2}, we know that $n^in^ju_{ij}=0$ for all time,
and so the first term on the right-hand side vanishes. Similarly, we know that
$\tau^{ik}v_i=0$ on the boundary face, and so the second term vanishes as well (since
the differential operator $\tau^j_k\partial_j$ is purely tangential).
We have established that $C=0$ holds on any boundary face of $\Omega$.

Finally, since $C$ also satisfies the wave equation \eqref{twiceCdel}, together
with the trivial initial conditions \eqref{Cofzero} and \eqref{Cdotofzero},
we conclude that the constraint $C$ vanishes for all time.
\end{proof}

\begin{thm}\label{thm:mpwp}
The differential system \eqref{hyps}, together with initial data \eqref{idwaves} and
boundary conditions \eqref{modelbc2} is well-posed.
Furthermore, if the initial conditions $w_i(0)$, $v_i(0)$, $u_{ij}(0)$ and the forcing terms
$f_i$ are given satisfying the compatibility conditions \eqref{nhypscomp1} and \eqref{nhypscomp2}
respectively, then the  solution satisfies the constraint \eqref{hypsC} for all time.
\end{thm}
\begin{proof}
We begin by showing that the boundary conditions \eqref{modelbc2} are maximal nonnegative for
the hyperbolic system \eqref{hyps}, and so, according to the classical theory of \cite{F} and \cite{LP}
(see also \cite{R}), the initial-boundary value problem is well-posed. We recall the definition
of maximal nonnegative boundary conditions. Let $V:=\R^3\times\R^3\times\R^{3\times 3}$. Obviously,
$\dim{V}=15$. The boundary operator $A_n$ associated to the hyperbolic system \eqref{hyps} is the
symmetric linear operator $V\to V$ which assigns to any element $(w_i,\, v_i,\, u_{ij})$  of $V$ the
element $(\tilde{w}_i,\, \tilde{v}_i,\, \tilde{u}_{ij})$ defined by
\begin{equation}\label{tildewvu}
\tilde{w}_i=0, \quad
\tilde{v}_i=-n^ju_{ij}, \quad
\tilde{u}_{ij}=-n_jv_i.
\end{equation}
A subspace $N$ of $V$ is called nonnegative for $A_n$ if
\begin{equation}\label{witwi}
w^i\tilde{w}_i+v^i\tilde{v}_i+u^{ij}\tilde{u}_{ij}\geq 0,
\end{equation}
whenever $(w_i,\, v_i,\, u_{ij})\in N$ and $(\tilde{w}_i,\, \tilde{v}_i,\, \tilde{u}_{ij})$ defined by
\eqref{tildewvu}. We claim that the subspace $N$ defined by the boundary conditions \eqref{modelbc2} is maximal nonnegative.
The dimension of $N$ is clearly 12. Since $A_n$ has three positive, nine zero, and three negative eigenvalues,
a nonnegative subspace is {\it maximal} nonnegative if and only if it has dimension 12, and so, the
proof reduces to the verification of nonnegativity condition \eqref{witwi}. Since
\begin{equation}
w^i\tilde{w}_i+v^i\tilde{v}_i+u^{ij}\tilde{u}_{ij}=-2v^in^ju_{ij},
\end{equation}
the verification of \eqref{witwi} reduces to showing that $v^in^ju_{ij}\leq 0$ whenever \eqref{modelbc2}
holds. In fact, we prove that $v^in^ju_{ij}=0$ if the boundary conditions \eqref{modelbc2} are verified. For this, we use the 
orthogonal decomposition of $v^i=n^kv_kn^i+\tau^{ki}v_k$. From the second boundary condition in
\eqref{modelbc2}, the tangential part $\tau^{ki}v_k$ of $v^i$ vanishes. Thus, $v^in^ju_{ij}=n^in^ju_{ij}n^kv_k$.
From the first boundary condition in \eqref{modelbc2}, the right-hand side vanishes, and so, 
$v^in^ju_{ij}=0$. This ends the proof of the maximal nonnegativity of the boundary conditions
\eqref{modelbc2}.

The fact that the solution satisfies the constraint for all time follows from 
Theorem~\ref{thm:mpconstraint}.
\end{proof}

\subsubsection{On Piecewise Regular Domains}\label{secmodpoly}
Here we investigate the existence of constraint-preserving boundary conditions on more general
domains, namely on piecewise regular domains. We shall provide a set of such boundary conditions
which generalize the boundary conditions \eqref{modelbc2}. Although these boundary conditions are not maximal nonnegative
in general, and so, the classical theory of \cite{F} and \cite{LP} can not be
employed, we have been able to prove an energy inequality which is a key ingredient in 
proving well-posedness.

\begin{figure} 
 \centering 
 \includegraphics[width=5cm,height=6cm]{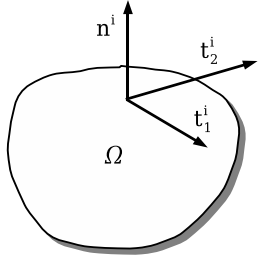} 
 \caption{A 3D domain.} 
 \label{3D_domain} 
\end{figure}

Let $\Omega$ be a bounded domain in $\R^3$. Assume that the boundary of $\Omega$, denoted
by $\partial\Omega$ is an almost everywhere regular surface in $\R^3$, that is, for almost
every point $p\in \partial\Omega$, there exists an open set $V$ in $\R^3$ and a difeomorphism
$\phi :U\to V\cap\partial\Omega$, $(x^1,\, x^2)\to \phi (x^1,\, x^2)$, of an open set $U\subset\R^2$ 
onto $V\cap\partial\Omega\subset\R^3$.
The mapping $\phi$ is called a {\it parameterization} or a system of (local) coordinates in a neighborhood
of $p$. 
From \cite{DoC}, Corollary 2, p. 183, we know that for all
$p\in\partial\Omega$ there exists a parameterization $\phi$ in a
neighborhood $V$ of $p$ such that the coordinate curves intersect orthogonally 
for each $q\in V$ (such a $\phi$ is called {\it orthogonal} parameterization). Since working in an orthogonal parameterization
turns out to simplify our computations, we assume that $\phi$ is an orthogonal parameterization.
At each point $p$ of $V\cap\partial\Omega$ define the tangent vectors $\ub{t}_i=\partial\phi/\partial x^i$,
$i=1,\, 2$, and the unit normal to the boundary (see Figure~\ref{3D_domain})
\begin{equation}
\ub n=\frac{\ub t_1\wedge \ub t_2}{|\ub t_1\wedge \ub t_2|}.
\end{equation}
For the given parameterization $\phi$, the derivatives of $\ub n$ belong to the tangential plane and
are given by
\begin{equation}
\begin{gathered}
\frac{\partial\ub n}{\partial x^1}=a_{11}\ub t_1+a_{21}\ub t_2,\\
\frac{\partial\ub n}{\partial x^2}=a_{12}\ub t_1+a_{22}\ub t_2.
\end{gathered}
\end{equation}
The trace of the matrix $(a_{ij})_{1\leq i,\, j\leq 2}$ does not depend on the choice of
parameterization. Hence, the {\it mean curvature} $H$, which is one half of
the trace of $(a_{ij})_{1\leq i,\, j\leq 2}$, is parameterization independent. 
Since the principal curvatures $k_1$ and $k_2$ are the roots of the quadratic equation
\[\det\left(\begin{array}{cc}
a_{11}-k&a_{12}\\
a_{21}&a_{22}-k
\end{array}\right) =0,\]
we can write
\begin{equation}
H=\frac12(a_{11}+a_{22})=\frac12(k_1+k_2).
\end{equation}

We shall prove that the following set of boundary conditions
\begin{equation}\label{modelbcH}
n^in^ju_{ij}+2Hn^iw_i=0,\quad
\tau^{ij}v_i=0,
\end{equation}
is constraint-preserving and, together with \eqref{hyps} and \eqref{idwaves}, leads to well-posedness.

First we prove the equivalent of Theorem~\ref{thm:mpconstraint} under the new circumstances.
\begin{thm}\label{thm:mpconstraintH}
Given initial conditions  $w_i(0)$, $v_i(0)$, $u_{ij}(0)$
and forcing terms $f_i$ satisfying the compatibility conditions \eqref{nhypscomp1} and \eqref{nhypscomp2}
respectively, define $w_i$, $v_i$, and $u_{ij}$ for positive time by the evolution equations
\eqref{hyps} and the boundary conditions \eqref{modelbcH}.
Then, the constraint \eqref{hypsC} is satisfied for all time.
\end{thm}
\begin{proof}
Let $E=\langle\, \ub t_1,\, \ub t_1\,\rangle$, $F=\langle\, \ub t_1,\, \ub t_2\,\rangle =0$, and
$G=\langle\, \ub t_2,\, \ub t_2\,\rangle$ be the coefficients of the first fundamental form in the
basis $\{\, \ub t_1,\, \ub t_2\, \}$. 
The derivatives 
of the vectors
$\ub t_1$ and $\ub t_2$ in the basis $\{\, \ub t_1,\,\ub t_2,\,\ub n\,\}$ are (see \cite{DoC}, p. 232):
\begin{equation}\label{secderiv}
\begin{gathered}
\frac{\partial\ub t_1}{\partial x^1}=\Gamma^1_{11}\ub t_1+\Gamma^2_{11}\ub t_2+L_1\ub n,\\
\frac{\partial\ub t_1}{\partial x^2}=\frac{\partial\ub t_2}{\partial x^1}=
\Gamma^1_{12}\ub t_1+\Gamma^2_{12}\ub t_2+L_2\ub n,\\
\frac{\partial\ub t_2}{\partial x^2}=\Gamma^1_{22}\ub t_1+\Gamma^2_{22}\ub t_2+L_3\ub n , 
\end{gathered}
\end{equation}
where $\Gamma^k_{ij}$ are the Christoffel symbols and $L_1=-a_{11}E$,
$L_2=-a_{12}E=-a_{21}G$, and $L_3=-a_{22}G$.

Since the constraint $C$ satisfies the wave equation \eqref{twiceCdel} with zero initial conditions \eqref{Cofzero},
\eqref{Cdotofzero}, the proof of this theorem reduces to prove that $C=0\ a.e.$ on the boundary $\partial\Omega$ for
all time. From the first identity of \eqref{delid},
observe that on the boundary the constraint $C$ is $\partial^i v_i=\delta^{ij}\partial_jv_i=n^in^j\partial_jv_i+
\tau^{ij}\partial_jv_i$. From the main system \eqref{hyps}, the first term of the right-hand side is 
equal to $n^in^j\dot{u}_{ij}$. Since the parameterization is orthogonal, 
the tangential projection of $\partial_jv_i$ is 
$$\tau^{ij}\partial_jv_i=\frac{1}{E}t_1^it_1^j\partial_jv_i+\frac{1}{G}t_1^it_1^j\partial_jv_i,$$
where $t_i^j$, $1\leq j\leq 3$, represent the components of the tangent vector $\ub t_i$, $i=1,2$ . 
By the Leibnitz rule of differentiation, the right-hand side of this identity can be written as follows
\begin{equation}  
\tau^{ij}\partial_jv_i=\frac{1}{E}t_1^j\partial_j(t_1^iv_i)+\frac{1}{G}t_2^j\partial_j(t_2^iv_i)
-\frac{1}{E}v_it_1^j\partial_jt_1^i-\frac{1}{G}v_it_2^j\partial_jt_2^i.
\end{equation}
The first two terms of the right-hand side vanish due to the second boundary condition in \eqref{modelbcH} 
combined with the tangentiality of the differential operators $t_i^j\partial_j$, $i=1,2$. Furthermore, by the
chain rule, $t_k^j\partial_jt_k^i=\partial t_k^i/\partial x^k$, $k=1,2$. Thus,
\begin{equation}\label{tauijpvi}
\tau^{ij}\partial_jv_i=-\frac{1}{E}v_i\frac{\partial t_1^i}{\partial x^1}-\frac{1}{G}v_i\frac{\partial t_2^i}{\partial x^2}.
\end{equation}
By the first identity in \eqref{secderiv}, the first term of the right-hand side of \eqref{tauijpvi} is
\begin{equation}
-\frac{1}{E}v_i\frac{\partial t_1^i}{\partial x^1}=-\frac{1}{E}\Gamma^1_{11}(t_1^iv_i)-\frac{1}{E}\Gamma^2_{11}(t_2^iv_i)+a_{11}n^iv_i.
\end{equation}
The first two terms of the right side of the last identity vanish due to the second boundary condition in \eqref{modelbcH}, and so
\begin{equation}
-\frac{1}{E}v_i\frac{\partial t_1^i}{\partial x^1}=a_{11}n^iv_i.
\end{equation}
Similarly,
\begin{equation}
-\frac{1}{G}v_i\frac{\partial t_2^i}{\partial x^2}=a_{22}n^iv_i.
\end{equation}
Thus,
\begin{equation}\label{tauijpvi2}
\tau^{ij}\partial_jv_i=(a_{11}+a_{22})n^iv_i.
\end{equation}
Therefore, $\partial^i v_i=n^in^j\dot{u}_{ij}+(a_{11}+a_{22})n^iv_i$. By the first equation in \eqref{hyps}, $\dot{w}_i=v_i$, combined
with the definition of the mean curvature, $H=(a_{11}+a_{22})/2$, it follows that $\partial^i v_i=(n^in^ju_{ij}+2Hn^iw_i)^.$.
Finally, from the first boundary condition in \eqref{modelbcH}, $C=\partial^i v_i=0$ $a.e.$ on $\partial\Omega$ for all time.
This ends the proof of this theorem. 
\end{proof}

\begin{thm}\label{gd_cp}
Let $\Omega$ be a bounded domain in $\R^3$ with (a.e.) regular boundary and the mean curvature $H\geq 0$ almost everywhere.
Given the initial conditions $w_i^0$, $v_i^0$,
$u_{ij}^0$ and the forcing terms $f_i$ satisfying the compatibility conditions \eqref{nhypscomp1} and \eqref{nhypscomp2}
respectively, the system of differential equations \eqref{hyps}, together with the boundary conditions \eqref{modelbcH} and
the initial conditions \eqref{idwaves}, is well-posed. Moreover, its solution satisfies the constraint \eqref{hypsC} for all
time.
\end{thm} 
\begin{proof}
For any solution $w_i$, $v_i$, $u_{ij}$ of \eqref{hyps} satisfying the boundary conditions \eqref{modelbcH},
define the energy 
\begin{equation*}
E(t)=\frac12 [\int_{\Omega}(w_iw^i+v_iv^i+u_{ij}u^{ij})\, dx+2\int_{\partial\Omega}H(w_in^i)^2\, ds].
\end{equation*}
Differentiating in time, we obtain
$$\dot E(t)=\int_{\Omega}(\dot{w}_iw^i+\dot{v}_iv^i+\dot{u}_{ij}u^{ij})\, dx+
2\int_{\partial\Omega}H(\dot{w}_in^i)(w_in^i)\, ds.$$
By the system \eqref{hyps}, the derivatives in time can be substituted for spatial derivatives, and so
$$\dot E(t)=\int_{\Omega}[v_iw^i+(\partial^ju_{ij})v^i+f_iv^i+(\partial_jv_i)u^{ij}\, dx+
2\int_{\partial\Omega}H(v_in^i)(w_in^i)\, ds.$$
Integrating by parts, it follows that
$$\dot E(t)=\int_{\Omega}(v_iw^i+f_iv^i)\, dx +\int_{\partial\Omega}v^in^ju_{ij}\ ds+
2\int_{\partial\Omega}H(v_in^i)(w_in^i)\, ds.$$
By the orthogonal decomposition of $v^i=n^in^lv_l+\tau^{il}v_l$, we obtain $v^in^ju_{ij}=n^in^ju_{ij}n^lv_l+\tau^{il}v_ln^ju_{ij}$.
In view of the second boundary condition in \eqref{modelbcH}, the second term of the right-hand side vanishes. Thus,
$v^in^ju_{ij}=n^in^ju_{ij}n^lv_l$, and so
$$\dot E(t)=\int_{\Omega}(v_iw^i+f_iv^i)\, dx +\int_{\partial\Omega}(n^in^ju_{ij}+2Hw_in^i)(n^lv_l)\, ds.$$
Due to the second boundary condition in \eqref{modelbcH}, the boundary integral vanishes. Therefore,
$$\dot E(t)=\int_{\Omega}(v_iw^i+f_iv^i)\, dx.$$
Now, since $H\geq 0$ ($a.e.$), it is easy to see that
$$\dot E(t)\leq E(t)+\theta (t),$$
where $\theta (t)=(\int_{\Omega}f_if^i\, dx)/2$ for all positive time $t>0$.
By the Gronwall inequality, $E(t)\leq e^t[E(0)+\int_0^te^{-s}\theta (s)\, ds]$. 
This energy estimate is a key ingredient in proving well-posedness by Galerkin approximations.

The fact that the solution satisfies the constraint \eqref{hypsC} for all time follows
from Theorem~\ref{thm:mpconstraintH}.
\end{proof}

\section{Einstein-Christoffel Formulation}

\subsection{Maximal Nonnegative Constraint-Preserving Boundary Conditions}

In this section, we provide maximal nonnegative boundary conditions for 
the linearized EC system which are constraint-preserving in the sense that 
the analogue of Theorem~\ref{thm:constraint0} is true for the initial--boundary value
problem.  This will then imply the analogue of
Theorem~\ref{thm:equiv0}.  We assume that $\Omega$ is a polyhedral domain (see Figure~\ref{polyhedral_domain}) and
use the notations introduced in Subsection~\ref{secmodpoly}.

First we consider the following boundary conditions on the face:
\begin{equation}\label{set2a}
n^im^jK_{ij}=n^il^jK_{ij}=n^kn^in^jf_{kij}=n^km^im^jf_{kij}=n^kl^il^jf_{kij}=
n^km^il^jf_{kij}=0.
\end{equation}
These can be written as well:
\begin{equation}\label{set2b}
n^i\tau^{jk}K_{ij}=0, \quad
n^kn^in^jf_{kij}=0, \quad
n^k\tau^{il}\tau^{jm}f_{kij}=0,
\end{equation}
and so do not depend on the choice of basis for the tangent space.
We begin by showing that these boundary conditions are maximal nonnegative
for the hyperbolic system \eqref{K}, \eqref{F}, and so, according to
the classical theory of \cite{F} and \cite{LP}, the initial--boundary 
value problem is well-posed.

Let $V$ denote the vector space of pairs of constant tensors
$(K_{ij},f_{kij})$ both symmetric with respect to the indices $i$ and
$j$.  Thus $\dim V=24$.
The boundary operator $A_n$ associated to the evolution equations
\eqref{K}, \eqref{F} is the symmetric linear operator $V\to V$ given by
\begin{equation}\label{An}
\tilde K_{ij}=n^kf_{kij}, \quad 
\tilde f_{kij}=n_kK_{ij}.
\end{equation}
A subspace $N$ of $V$ is nonnegative for $A_n$ if
\begin{equation}\label{ip}
K_{ij}\tilde K^{ij} + f_{kij}\tilde f^{kij} \ge 0
\end{equation}
whenever $(K_{ij},f_{kij})\in N$ and $(\tilde K_{ij},\tilde f_{kij})$
is defined by \eqref{An}.  The subspace is maximal nonnegative if
also no larger subspace has this property.  
The eingenvalues $\lambda$ of $A_n$, together with the corresponding eigenvectors $(K_{ij},\, f_{kij})$, are
the following: 

$\lambda=0$ with multiplicity 12 and eingenvectors: $(0,\, m_kn_in_j)$, $(0,\, m_kn_{(i}m_{j)})$,
$(0,\, m_kn_{(i}l_{j)})$, $(0,\, m_km_{i}m_{j})$, $(0,\, m_km_{(i}l_{j)})$, $(0,\, m_kl_{i}l_{j})$, 
$(0,\, l_kn_in_j)$, $(0,\, l_kn_{(i}m_{j)})$,
$(0,\, l_kn_{(i}l_{j)})$, $(0,\, l_km_{i}m_{j})$, $(0,\, l_km_{(i}l_{j)})$, $(0,\, l_kl_{i}l_{j})$,

$\lambda=-1$ with multiplicity six and eingenvectors: $(n_in_j,\, -n_kn_in_j)$, $(m_im_j,\, -n_km_im_j)$,
$(l_il_j,\, -n_kl_il_j)$, $(n_{(i}m_{j)},\, -n_kn_{(i}m_{j)})$, $(n_{(i}l_{j)},\, -n_kn_{(i}l_{j)})$,
$(l_{(i}m_{j)},\, -n_kl_{(i}m_{j)})$,

$\lambda=1$ with multiplicity six and eingenvectors: $(n_in_j,\, n_kn_in_j)$, $(m_im_j,\, n_km_im_j)$,
$(l_il_j,\, n_kl_il_j)$, $(n_{(i}m_{j)},\, n_kn_{(i}m_{j)})$, $(n_{(i}l_{j)},\, n_kn_{(i}l_{j)})$,
$(l_{(i}m_{j)},\, n_kl_{(i}m_{j)})$.

Since $A_n$ has six
positive, 12 zero, and six negative eigenvalues, a nonnegative subspace
is maximal nonnegative if and only if it has dimension $18$.  Our
claim is that the subspace $N$ defined by \eqref{set2a} is maximal
nonnegative.  The dimension is clearly $18$.  In view of \eqref{An},
the verification of \eqref{ip} reduces to showing that
$n^kf_{kij}K^{ij}\ge0$
whenever \eqref{set2a} holds.  In fact, $n^kf_{kij}K^{ij}=0$,
that is, $n^kf_{kij}$ and $K_{ij}$ are orthogonal (when \eqref{set2a}
holds).
To see this, we use orthogonal expansions of each based
on the normal and tangential components:
\begin{align}\label{expand1}
K_{ij}&=n^ln_in^mn_jK_{lm} + n^ln_i\tau_j^mK_{lm}
+\tau_i^ln^mn_jK_{lm}+\tau_i^l\tau_j^mK_{lm},
\\\label{expand2}
n^kf_{kij}&=n^ln_in^mn_jn^kf_{klm} + n^ln_i\tau_j^mn^kf_{klm}
+\tau_i^ln^mn_jn^kf_{klm}+\tau_i^l\tau_j^mn^kf_{klm}.
\end{align}
In view of the boundary conditions (in the form \eqref{set2b}),
the two inner terms on the right-hand side of \eqref{expand1}
and the two outer terms on the right-hand side of \eqref{expand2}
vanish, 
and so the orthogonality is evident.

Next we show that the boundary conditions are constraint-preserving.
This is based on the following lemma.
\begin{lem}\label{lem}
Suppose that $\alpha$ and $\beta^i$ vanish.
Let $K_{ij}$, $f_{kij}$ be a solution to the homogeneous hyperbolic system
\eqref{K}, \eqref{F} and suppose that the boundary conditions \eqref{set2a}
are satisfied on some face of $\partial\Omega$.  Let $C_j$ be defined by
\eqref{Cj}.  Then 
\begin{equation}\label{bndid}
\dot C_j n^l\partial_l C^j=0
\end{equation}
on the face.
\end{lem}
\begin{proof}
 In fact we shall
show that $n^jC_j=0$ (so also $n^j\dot C_j=0$) and $\tau_j^p n^l\partial_l C^j=0$, which,
by \eqref{dotid} implies \eqref{bndid}.  First note that
\begin{equation*}
C_j=(\delta_j^m\delta^{ik}-\delta_j^k\delta^{im})\partial_k K_{im}
=(\delta_j^mn^in^k+\delta_j^m\tau^{ik}-\delta_j^k\delta^{im})\partial_k
K_{im},
\end{equation*}
where we have used the first identity in \eqref{delid}.  Contracting
with $n^j$ gives
\begin{align*}
n^jC_j&=(n^mn^in^k+n^m\tau^{ik}-n^k\delta^{im})\partial_k K_{im}
\\
&=-n^mn^in^k\dot f_{kim}+\tau^{il}\tau_l^kn^m\partial_k K_{im}
+n^k \delta^{im}\dot f_{kim},
\end{align*}
where now we have used the equation \eqref{F} (with $\beta_i=0$)
for the first and last term
and the second identity in \eqref{delid} for the middle term.
From the boundary conditions we know that
$n^mn^in^kf_{kim}=0$, and so the first term on the right-hand side
vanishes.  Similarly, we know that $\tau^{il}n^mK_{im}=0$ on the
boundary face, and so the second term vanishes as well (since the
differential operator $\tau_l^k\partial_k$ is purely tangential).
Finally, $n^k\delta^{im}f_{kim}=n^k(n^in^m+l^il^m+m^im^m)f_{kim}=0$,
and so the third term vanishes.
We have established that $n^jC_j=0$ holds on the face.

To show that $\tau_j^p n^l\partial_l C^j=0$ on the face, we start with the identity
\begin{equation*}
\tau_j^p n^l\delta^{mj}\delta^{ik}=\tau^{pm}(n^in^k+\tau^{ik})n^l
=\tau^{pm}n^i(\delta^{kl}-\tau^{kl})+\tau^{pm}\tau^{ik}n^l.
\end{equation*}
Similarly
\begin{equation*}
\tau_j^pn^l\delta^{kj}\delta^{im} =
\tau^{pk}n^ln^in^m+\tau^{pk}\tau^{im}n^l.
\end{equation*}
Therefore,
\begin{align*}
\tau_j^p n^l\partial_l C^j
&=\tau^p_jn^l
\partial_l(\delta^{mj}\delta^{ik}-\delta^{kj}\delta^{im})\partial_k K_{im}
\\
&=(\tau^{pm}n^i\delta^{kl}-\tau^{pm}n^i\tau^{kl}+\tau^{pm}\tau^{ik}n^l
-\tau^{pk}n^ln^in^m-\tau^{pk}\tau^{im}n^l)\partial_k\partial_l K_{im}.
\end{align*}
For the last three terms, we again use \eqref{F} to replace
$\partial_l K_{im}$ with $-\dot f_{lim}$ and argue as before to
see that these terms vanish.  For the first term we notice
that $\delta^{kl}\partial_k\partial_l K_{im}=\partial^k\partial_kK_{im}
=\ddot K_{im}$ (from \eqref{K} and \eqref{F} with vanishing
$\alpha$ and $\beta^i$).  Since
$\tau^{pm}n^iK_{im}$ vanishes on the boundary, this term vanishes.
Finally we recognize that the second term is the tangential Laplacian,
$\tau^{kl}\partial_k\partial_l$ applied to the quantity
$n^i\tau^{pm} K_{im}$, which vanishes.  This concludes the proof of
\eqref{bndid}.
\end{proof}

The next theorem asserts that the boundary conditions are
constraint-preserving.
\begin{thm}\label{thm:constraint}
Let $\Omega$ be a polyhedral domain. Given initial data
$K_{ij}(0)$, $f_{kij}(0)$ on $\Omega$ satisfying the constraints
\eqref{Cj} and \eqref{Fc}, define $K_{ij}$ and $f_{kij}$ for
positive time by the evolution equations \eqref{K}, \eqref{F}
and the boundary conditions \eqref{set2a}.  Then the constraints
\eqref{Cj} and \eqref{Fc} are satisfied for all time.
\end{thm}
\begin{proof}
Exactly as for Theorem~\ref{thm:constraint0} we find that $C_j$
satisfies the wave equation \eqref{w} and both $C_j$ and $\dot C_j$
vanish at the initial time.  Define the usual energy
\begin{equation*}
E(t)=\frac12 \int_{\Omega}(\dot C_j\dot
C^j+\partial^lC_j\partial_lC^j)\,dx.
\end{equation*}
Clearly $E(0)=0$.  From \eqref{w} and integration by parts
\begin{equation}\label{dtE}
\dot E=\int_{\partial\Omega}\dot C_jn^l\partial_lC^j d\sigma.
\end{equation}
Therefore, if $\alpha=0$ and $\beta^i=0$, we can invoke
Lemma~\ref{lem}, and conclude that $E$ is constant in time.
Hence $E$ vanishes identically.  Thus $C_j$ is constant, and, since
it vanishes at time $0$, it vanishes for all time.  This establishes
the theorem under the additional assumption that $\alpha$ and $\beta^i$
vanish.

To extend to the case of general $\alpha$ and $\beta^i$ we use
Duhamel's principle.  Let $S(t)$ denote the solution operator
associated to the homogeneous boundary value problem.  That is,
given functions $\kappa_{ij}(0)$, $\phi_{kij}(0)$ on $\Omega$,
define $S(t)(\kappa_{ij}(0),\phi_{kij}(0))=
(\kappa_{ij}(t),\phi_{kij}(t))$ where $\kappa_{ij}$, $\phi_{kij}$ is
the solution to the homogeneous evolution equations
\begin{equation*}
\dot\kappa_{ij}=-\partial^k\phi_{kij},
\quad \dot\phi_{kij}=-\partial_k\kappa_{ij},
\end{equation*}
satisfying the boundary conditions and assuming the given initial
values.  Then Duhamel's principle represents the solution
$K_{ij}$, $f_{kij}$ of the inhomogeneous initial-boundary value
problem \eqref{K}, \eqref{F}, \eqref{set2a} as
\begin{equation}\label{duhamel}
(K_{ij}(t),f_{kij}(t))=S(t)(K_{ij}(0),f_{kij}(0))
+\int_0^t S(t-s)(-\partial_i\partial_j\alpha(s),L_{kij}(s))\,ds.
\end{equation}
Now it is easy to check that the momentum constraint \eqref{Cj} is
satisfied when $K_{ij}$ is replaced by $\partial_i\partial_j\alpha(s)$
(for any smooth function $\alpha$), and the constraint \eqref{Fc}
is satisfied when $f_{kij}$ is replaced by $L_{kij}(s)$ defined by 
\eqref{defL} (for any smooth function $\beta^i$).  Hence the integrand
in \eqref{duhamel} satisfies the constraints by the result for the homogeneous case,
as does the first term on the right-hand side, and thus the constraints
are indeed satisfied by $K_{ij}$, $f_{kij}$.
\end{proof}

The analogue of Theorem~\ref{thm:equiv0} for the initial--boundary
value problem follows from the preceding theorem exactly as before.
\begin{thm}\label{thm:equiv}
Let $\Omega$ be a polyhedral domain.
Suppose that initial data $g_{ij}(0)$ and $K_{ij}(0)$ are given
satisfying the Hamiltonian constraint \eqref{C} and momentum constraint
\eqref{Cj}, respectively, and that initial data $f_{kij}(0)$ is defined
by \eqref{finit}.  Then the unique solution of the linearized EC
initial--boundary value problem \eqref{G}, \eqref{K}, \eqref{F},
together with the boundary conditions \eqref{set2a}
satisfies the linearized ADM system
\eqref{G}--\eqref{Cj} in $\Omega$.
\end{thm}

We close by noting a second set of boundary
conditions which are maximal nonnegative and constraint-preserving.
These are
\begin{equation}\label{set1a}
n^in^jK_{ij}=m^im^jK_{ij}=l^il^jK_{ij}=m^il^jK_{ij}=n^kn^im^jf_{kij}=
n^kn^il^jf_{kij}=0,
\end{equation}
or, equivalently,
\begin{equation*}
n^in^jK_{ij}=0, \quad
\tau^{il}\tau^{jm}K_{ij}=0, \quad
n^kn^i\tau^{jl}f_{kij}=0.
\end{equation*}
Now when we make an orthogonal expansion as in \eqref{expand1},
\eqref{expand2}, the outer terms on the right-hand side of the
first equation and the inner terms on the right-hand side of the
second equation vanish (it was the reverse before), so we
again have the necessary orthogonality to demonstrate that
the boundary conditions are maximal nonnegative.  Similarly, to
prove the analogue of Lemma~\ref{lem}, for these boundary
conditions we show that the tangential component of $\dot C_j$ vanishes
and the normal component of $n^l\partial_l C^j$ vanishes (it was
the reverse before).  Otherwise the analysis is essentially the
same as for the boundary conditions \eqref{set2a}.

\subsection{Extended EC System}\label{Ext_EC}

In this section we indicate an extended initial boundary value
problem whose solution solves the linearized ADM system 
\eqref{G}--\eqref{Cj} in $\Omega$. This approach could present advantages
from the numerical point of view since the momentum constraint
is ``built-in'', and so controlled for all time.
The new system consists of \eqref{G}, \eqref{F},
and two new sets of equations corresponding to \eqref{K} 
\begin{equation}\label{KE}
\dot{K}_{ij}=-\partial^kf_{kij}+\frac{1}{2} (\partial_ip_j+\partial_jp_i)-
\partial^kp_k\delta_{ij}-\partial_i\partial_j\alpha
\end{equation}
and to
a new three dimensional vector field $p_i$ defined by
\begin{equation}\label{p}
\dot{p}_i=\partial^lK_{li}-\partial_iK_l^l,
\end{equation}
respectively.
Observe that the additional terms that appear on the right-hand side
of \eqref{KE} compared with \eqref{K} are nothing but the negative components
of the formal adjoint of the momentum constraint operator applied to $p_i$.

Let $\tilde{V}$ be the vector space of quadruples of constant tensors
$(g_{ij},K_{ij},f_{kij},p_k)$ symmetric with respect to the indices
$i$ and $j$. Thus $\dim \tilde{V}=33$. The boundary operator 
$\tilde{A}_n :\tilde{V}\to\tilde{V}$ in this case is given by
\begin{equation}\label{Andef}
\tilde{g}_{ij}=0, \quad
\tilde{K}_{ij}=n^kf_{kij}-\frac{1}{2}(n_ip_j+n_jp_i)+n^kp_k\delta_{ij},
\quad
\tilde{f}_{kij}=n_kK_{ij}, \quad
\tilde{p}_i=-n^lK_{il}+n_iK_l^l.
\end{equation}
The boundary operator $\tilde{A}_n$ associated to the evolution equations
\eqref{G}, \eqref{KE}, \eqref{F}, and \eqref{p} has six positive, 21
zero, and six negative eigenvalues. Therefore, a nonnegative subspace
is maximal nonnegative if and only if it has
dimension 27. We claim that the following boundary conditions are
maximal nonnegative for \eqref{G}, \eqref{KE}, \eqref{F}, and \eqref{p}
\begin{equation}\label{set2ae}
n^im^jK_{ij}=n^il^jK_{ij}=n^kn^in^jf_{kij}=n^k(m^im^jf_{kij}+p_k)=
n^k(l^il^jf_{kij}+p_k)=n^km^il^jf_{kij}=0.
\end{equation}
These can be written as well:
\begin{equation}\label{set2be}
n^i\tau^{jk}K_{ij}=0, \quad
n^kn^in^jf_{kij}=0, \quad
n^k(\tau^{il}\tau^{jm}f_{kij}+\tau^{lm}p_k)=0,
\end{equation}
and so do not depend on the choice of basis for the tangent space.

Let us prove the claim that the subspace $\tilde{N}$ defined by
\eqref{set2ae} is maximal nonnegative. Obviously,
$\dim\tilde{N}=27$. Hence, it remains to be proven that $\tilde{N}$ is
also nonnegative. In view of \eqref{Andef}, the verification of
nonnegativity of $\tilde{N}$ reduces to showing that
\begin{equation}\label{positivity}
n^kf_{kij}K^{ij}-n^ip^jK_{ij}+n^kp_kK_l^l\geq 0
\end{equation}
whenever \eqref{set2ae} holds. In fact, we can prove that the
left-hand side of \eqref{positivity} vanishes pending 
\eqref{set2ae} holds. From the boundary conditions (in the
form \eqref{set2be}) and the orthogonal expansions \eqref{expand1} and
\eqref{expand2} of $K_{ij}$ and $f_{kij}$, respectively, the first
term on the right-hand side of \eqref{positivity} reduces to
$n^k\tau^{il}\tau^{jm}f_{kij}K_{lm}=-n^kp_k\tau^{lm}K_{lm}$.
Then, combining the first and third terms of the left-hand side of 
\eqref{positivity} gives
$-n^kp_k\tau^{ij}K_{ij}+n^kp_k\delta^{ij}K_{ij}=n^kp_kn^in^jK_{ij}$.
Finally, by using the orthogonal decomposition
$p^j=n^kp_kn^j+\tau^{kj}p_k$ and the first part of the boundary 
conditions \eqref{set2be}
the second term of the left-hand side of \eqref{positivity} is
$-n^kp_kn^in^jK_{ij}-p_kn^i\tau^{kj}K_{ij}=-n^kp_kn^in^jK_{ij}$, which
is precisely the negative sum of the first and third terms of the
left-hand side of \eqref{positivity}. This concludes the proof of 
\eqref{positivity}.  

\begin{thm}\label{thm:extended}
Let $\Omega$ be a polyhedral domain. Suppose that the initial data
$g_{ij}(0)$ and $K_{ij}(0)$ are given satisfying the Hamiltonian \eqref{C}
and momentum constraints \eqref{Cj}, respectively, $f_{kij}(0)$ is 
defined by \eqref{deff}, and $p_i(0)=0$. Then the unique solution
$(g_{ij},K_{ij},f_{kij},p_i)$ of the initial boundary value problem
\eqref{G}, \eqref{KE}, \eqref{F}, and \eqref{p}, together with the
boundary conditions \eqref{set2ae}, satisfies the properties
$p_i=0$ for all time, and $(g_{ij},K_{ij})$ solves the linearized
ADM system \eqref{G}--\eqref{Cj} in $\Omega$.
\end{thm}

\begin{proof}
Observe that the solution of the initial boundary value problem
\eqref{G}, \eqref{K}, \eqref{F}, and \eqref{set2a} (boundary
conditions), together with $p_i=0$ for all time, is the unique
solution of the initial boundary value problem \eqref{G}, \eqref{KE}, 
\eqref{F}, and \eqref{p}, together with the boundary conditions \eqref{set2ae}.
The conclusion follows from Theorem~\ref{thm:equiv}.
\end{proof}

We close by indicating a second set of maximal nonnegative boundary
conditions (corresponding to \eqref{set1a}) for \eqref{G}, \eqref{KE}, 
\eqref{F}, and \eqref{p} for which Theorem~\ref{thm:extended} holds as
well. These are \begin{equation}\label{set1ae}
n^in^jK_{ij}=m^im^jK_{ij}=m^il^jK_{ij}=l^il^jK_{ij}=n^kn^im^jf_{kij}-m^kp_k=
n^kn^il^jf_{kij}-l^kp_k=0,
\end{equation}
or, equivalently,
\begin{equation}\label{set1be}
n^in^jK_{ij}=0, \quad
\tau^{il}\tau^{jm}K_{ij}=0, \quad
n^kn^i\tau^{jl}f_{kij}-\tau^{kl}p_k=0.
\end{equation}

\subsection{Inhomogeneous Boundary Conditions}

In this subsection we provide well-posed constraint-preserving
{\it inhomogeneous} boundary conditions for \eqref{G}, \eqref{K}, and \eqref{F}
corresponding to the two sets of boundary conditions \eqref{set2a}, and
\eqref{set1a}, respectively. The first set of inhomogeneous boundary
conditions corresponds to \eqref{set2a} and can be written in the
following form
\begin{equation}\label{set2ai}
n^im^j\tilde{K}_{ij}=n^il^j\tilde{K}_{ij}=n^kn^in^j\tilde{f}_{kij}=n^km^im^j\tilde{f}_{kij}=
n^kl^il^j\tilde{f}_{kij}=
n^km^il^j\tilde{f}_{kij}=0,
\end{equation}
where $\tilde{K}_{ij}=K_{ij}-\kappa_{ij}$,
$\tilde{f}_{kij}=f_{kij}-F_{kij}$, with $\kappa_{ij}$ and $F_{kij}$
given in $\overline{\Omega}$ for all time and satisfying the constraints \eqref{Cj} and \eqref{Fc},
respectively. The matter of choosing $\kappa_{ij}$ and $F_{kij}$ is
deferred to the Appendix A.

The analogue of Theorem~\ref{thm:constraint} for the inhomogeneous boundary
conditions \eqref{set2ai} is true.

\begin{thm}\label{thm:constrainti}
Let $\Omega$ be a polyhedral domain. Given initial data
$K_{ij}(0)$, $f_{kij}(0)$ on $\Omega$ satisfying the constraints
\eqref{Cj} and \eqref{Fc}, define $K_{ij}$ and $f_{kij}$ for
positive time by the evolution equations \eqref{K}, \eqref{F}
and the boundary conditions \eqref{set2ai}.  Then the constraints
\eqref{Cj} and \eqref{Fc} are satisfied for all time.
\end{thm}

\begin{proof}
Observe that $\tilde{K}_{ij}$ and $\tilde{f}_{kij}$ satisfy \eqref{K}
and \eqref{F} with the forcing terms replaced by
$-\partial_i\partial_j\alpha-\partial^kF_{kij}-\dot{\kappa}_{ij}$ and
$L_{kij}-\partial_k\kappa_{ij}-\dot{F}_{kij}$, respectively.
Exactly as in Theorem~\ref{thm:constraint}, it follows that
$\tilde{K}_{ij}$ and $\tilde{f}_{kij}$ satisfy \eqref{Cj}
and \eqref{Fc}, respectively, for all time. Thus, $K_{ij}$ and
$f_{kij}$ satisfy \eqref{Cj} and \eqref{Fc}, respectively, for all time.
\end{proof}

The analogue of Theorem~\ref{thm:equiv} for the case of the
inhomogeneous boundary conditions \eqref{set2ai} follows from
the preceding theorem by using the same arguments as in the proof
of Theorem~\ref{thm:equiv0}.

\begin{thm}\label{thm:equivi}
Let $\Omega$ be a polyhedral domain.
Suppose that initial data $g_{ij}(0)$ and $K_{ij}(0)$ are given
satisfying the Hamiltonian constraint \eqref{C} and momentum constraint
\eqref{Cj}, respectively, and that initial data $f_{kij}(0)$ is defined
by \eqref{finit}.  Then the unique solution of the linearized EC
initial--boundary value problem \eqref{G}, \eqref{K}, \eqref{F},
together with the {\it inhomogeneous} boundary conditions \eqref{set2ai}
satisfies the linearized ADM system
\eqref{G}--\eqref{Cj} in $\Omega$.
\end{thm}

Note that there is a second set of inhomogeneous
boundary conditions corresponding to \eqref{set1a} for which 
Theorem~\ref{thm:constrainti} and Theorem~\ref{thm:equivi} remain
valid. These are
\begin{equation}\label{set1ai}
n^in^j\tilde{K}_{ij}=m^im^j\tilde{K}_{ij}=l^il^j\tilde{K}_{ij}=
m^il^j\tilde{K}_{ij}=n^kn^im^j\tilde{f}_{kij}=n^kn^il^j\tilde{f}_{kij}=0,
\end{equation}
where again $\tilde{K}_{ij}=K_{ij}-\kappa_{ij}$,
$\tilde{f}_{kij}=f_{kij}-F_{kij}$, with $\kappa_{ij}$ and $F_{kij}$
given and satisfying the constraints \eqref{Cj} and \eqref{Fc},
respectively.

Similar considerations can be made for the extended system introduced
in the previous section. There are two sets of {\it inhomogeneous} 
boundary conditions for which the extended system produces solutions 
of the linearized ADM system \eqref{G}--\eqref{Cj} on a polyhedral domain $\Omega$.
These are
\begin{equation}\label{set2aei}
n^im^j\tilde{K}_{ij}=n^il^j\tilde{K}_{ij}=n^kn^in^j\tilde{f}_{kij}=
n^k(m^im^j\tilde{f}_{kij}+p_k)=n^k(l^il^j\tilde{f}_{kij}+p_k)=
n^km^il^j\tilde{f}_{kij}=0
\end{equation}
and, respectively,
\begin{equation}\label{set1aei}
n^in^j\tilde{K}_{ij}=m^im^j\tilde{K}_{ij}=m^il^j\tilde{K}_{ij}=
l^il^j\tilde{K}_{ij}=n^kn^im^j\tilde{f}_{kij}-m^kp_k=
n^kn^il^j\tilde{f}_{kij}-l^kp_k=0,
\end{equation}
where $\tilde{K}_{ij}$ and $\tilde{f}_{kij}$ are
defined as before.

The next theorem is an extension of Theorem~\ref{thm:extended} to the
case of inhomogeneous boundary conditions.

\begin{thm}\label{thm:extendedi}
Let $\Omega$ be a polyhedral domain. Suppose that the initial data
$g_{ij}(0)$ and $K_{ij}(0)$ are given satisfying the Hamiltonian \eqref{C}
and momentum constraints \eqref{Cj}, respectively, $f_{kij}(0)$ is 
defined by \eqref{deff}, and $p_i(0)=0$. Then the unique solution
$(g_{ij},K_{ij},f_{kij},p_i)$ of the initial boundary value problem
\eqref{G}, \eqref{KE}, \eqref{F}, and \eqref{p}, together with the
{\it inhomogeneous} boundary conditions \eqref{set2aei} (or
\eqref{set1aei}), satisfies the properties
$p_i=0$ for all time, and $(g_{ij},K_{ij})$ solves the linearized
ADM system \eqref{G}--\eqref{Cj} in $\Omega$.
\end{thm}

\begin{proof}
Note that the solution of the initial boundary value problem
\eqref{G}, \eqref{K}, \eqref{F}, and \eqref{set2ai} (or
\eqref{set1ai}, respectively), together
with $p_i=0$ for all time, is the unique solution of the initial
boundary value problem \eqref{G}, \eqref{KE}, \eqref{F}, and
\eqref{p},
together with the boundary conditions \eqref{set2aei} (or
\eqref{set1aei}, respectively). The conclusion follows from
Theorem~\ref{thm:equivi}.
\end{proof}

\section{Alekseenko-Arnold Formulation}

\subsection{Maximal Nonnegative Constraint-Preserving Boundary Conditions}

The equivalence of \eqref{AAK}, \eqref{AAF} and the linearized ADM system
has been studied in the second section of \cite{AA} for the case of 
pure initial value problem with the result that 
for given initial data satisfying the
constraints, the unique solution of the linearized AA evolution
equations satisfies the linearized ADM system, and so the linearized
ADM system and the linearized AA system are equivalent (see \cite{AA}, Theorem 1.).
Our interest is in the case when the spatial domain is bounded and 
appropriate boundary conditions are imposed.
In this subsection,  we provide maximal nonnegative
boundary conditions for the linearized AA system which are
constraint-preserving. This will then imply the analogue of the
equivalence result proven in \cite{AA} (Theorem 1.) for the case of
bounded domains. The ideas and techniques used in this section
have many points in common to those used in Section 4.3
for the case of EC formulation.

Assume that $\Omega$ is a polyhedral domain.
Consider an arbitrary face of $\partial\Omega$ and let $n^i$ denote its
exterior unit normal.  Denote by $m^i$ and $l^i$ two additional vectors
which together $n^i$ form an orthonormal basis (see Figure~\ref{polyhedral_domain}).  

First we consider the following boundary conditions on the face:
\begin{equation}\label{AAset2a}
n^im^j\kappa_{ij}=n^il^j\kappa_{ij}=n^km^im^j\lambda_{kij}=n^kl^il^j\lambda_{kij}
=n^km^il^j\lambda_{kij}=0.
\end{equation}
These are equivalent to:
\begin{equation}\label{AAset2b}
n^i\tau^{jk}\kappa_{ij}=0, \quad
n^k\tau^{il}\tau^{jm}\lambda_{kij}=0,
\end{equation}
where $\tau_i^j:=m_im^j+l_il^j$ is the projection operator to $n^i$,
and so do not depend on the choice of basis for the tangent space.
We begin by showing that these boundary conditions are maximal nonnegative
for the hyperbolic system \eqref{AAK}, \eqref{AAF}, and so, according to
the classical theory of \cite{F} and \cite{LP}, the initial boundary 
value problem is well-posed.

Let $V$ denote the vector space of pairs of constant tensors
$(\kappa_{ij},\lambda_{kij})$. Here $\kappa_{ij}$ is symmetric with
respect to the indices $i$ and $j$, and $\lambda_{kij}$ is a
third-order constant tensor which is antisymmetric with respect to the 
first two indices and satisfies the cyclic identity 
$\lambda_{kij}+\lambda_{jki}+\lambda_{ijk}=0$. Thus $\dim V=14$.
The boundary operator $A_n$ associated to the evolution equations
\eqref{K}, \eqref{F} is the symmetric linear operator $V\to V$ given by
\begin{equation}\label{AAAn}
\tilde \kappa_{ij}=-n^k\lambda_{k(ij)}, \quad 
\tilde \lambda_{kij}=-n_{[k}\kappa_{i]j}.
\end{equation}
A subspace $N$ of $V$ is called nonnegative for $A_n$ if
\begin{equation}\label{AAip}
\kappa_{ij}\tilde \kappa^{ij} + \lambda_{kij}\tilde \lambda^{kij} \ge 0
\end{equation}
whenever $(\kappa_{ij},\lambda_{kij})\in N$ and $(\tilde \kappa_{ij},
\tilde \lambda_{kij})$ is defined by \eqref{AAAn}.  
The subspace is maximal nonnegative if
also no larger subspace has this property.  Since $A_n$ has five
positive, four zero, and five negative eigenvalues, a nonnegative subspace
is maximal nonnegative if and only if it has dimension nine.  Our
claim is that the subspace $N$ defined by \eqref{AAset2a} is maximal
nonnegative.  The dimension is clearly nine.  In view of \eqref{AAAn},
the verification of \eqref{AAip} reduces to showing that
$n^k\lambda_{kij}\kappa^{ij}\leq 0$
whenever \eqref{set2a} holds.  In fact, $n^k\lambda_{kij}\kappa^{ij}=0$,
that is, $n^k\lambda_{kij}$ and $\kappa_{ij}$ are orthogonal 
(when \eqref{AAset2a} holds).
To see this, we use orthogonal expansions of each based
on the normal and tangential components:
\begin{align}\label{AAexpand1}
\kappa_{ij}&=n^ln_in^mn_j\kappa_{lm} + n^ln_i\tau_j^m\kappa_{lm}
+\tau_i^ln^mn_j\kappa_{lm}+\tau_i^l\tau_j^m\kappa_{lm},
\\\label{AAexpand2}
n^k\lambda_{kij}&=n^ln_in^mn_jn^k\lambda_{klm} + n^ln_i\tau_j^mn^k\lambda_{klm}
+\tau_i^ln^mn_jn^k\lambda_{klm}+\tau_i^l\tau_j^mn^k\lambda_{klm}.
\end{align}
In view of the boundary conditions (in the form \eqref{AAset2b}),
the two inner terms on the right-hand side of \eqref{AAexpand1}
and the last term of the right-hand side of \eqref{AAexpand2}
vanish. Also, the first two terms of the right-hand side of \eqref{AAexpand2}
vanish due to the antisymmetry of $\lambda_{kij}$ with respect to
the first two indices. Thus, the orthogonality is evident.

Next we show that the boundary conditions are constraint-preserving.
This is based on the following lemma.

\begin{lem}\label{AAlem}
Suppose that $\alpha$ and $\beta^i$ vanish.
Let $\kappa_{ij}$, $\lambda_{kij}$ be a solution to the 
homogeneous hyperbolic system \eqref{AAK}, \eqref{AAF} and suppose 
that the boundary conditions \eqref{AAset2a}
are satisfied on some face of $\partial\Omega$.  Let $C_j$ be defined by
\eqref{LCj}.  Then 
\begin{equation}\label{AAbndid}
\dot C_j n^l\partial_l C^j+n^j\dot C_j\partial_lC^l=0
\end{equation}
on the face.
\end{lem}
\begin{proof}
In fact we shall
show that $n^jC_j=0$ (so also $n^j\dot C_j=0$) and $\tau_j^p n^l\partial_l C^j=0$, which,
by \eqref{dotid} implies \eqref{AAbndid}.  First note that
\begin{equation*}
C_j=(\delta_j^m\delta^{ik}-\delta_j^k\delta^{im})\partial_k \kappa_{im}
\end{equation*}
Contracting with $n^j$ and using the first identity of \eqref{delid}  
give
\begin{align*}
n^jC_j&=(n^m\delta^{ik}-n^k\delta^{im})\partial_k\kappa_{im}
\\
&=[n^m(n^in^k+\tau^{ik})-n^k(n^in^m+\tau^{im})]\partial_k\kappa_{im}
\\
&=(n^m\tau^{ik}-n^k\tau^{im})\partial_k\kappa_{im}
\\
&=n^m\tau^{ik}\partial_k\kappa_{im}-n^k\tau^{im}(\sqrt{2}\dot\lambda_{kim}+
\partial_i\kappa_{km}),
\end{align*}
where now we have used the equation \eqref{AAF} (with $\beta_i=0$)
for the last term.
From the boundary conditions we know that
$n^m\tau^{ik}\partial_k\kappa_{im}=0$, and so the first term on the 
right-hand side vanishes (since the
differential operator $\tau^{ik}\partial_k$ is purely tangential).  Similarly,
$n^k\tau^{im}\lambda_{kim}=n^k\tau^{im}\partial_i\kappa_{km}=0$ 
on the boundary face, and so the second term vanishes as well.
We have established that $n^jC_j=0$ holds on the face.

To show that $\tau_j^p n^l\partial_l C^j=0$ on the face, we start with the identity
\begin{equation*}
\tau_j^p n^l\delta^{mj}\delta^{ik}=\tau^{pm}(n^in^k+\tau^{ik})n^l
=\tau^{pm}n^i(\delta^{kl}-\tau^{kl})+\tau^{pm}\tau^{ik}n^l.
\end{equation*}
Similarly
\begin{equation*}
\tau_j^pn^l\delta^{kj}\delta^{im} =
\tau^{pk}n^ln^in^m+\tau^{pk}\tau^{im}n^l.
\end{equation*}
Therefore,
\begin{align*}
\tau_j^p n^l\partial_l C^j
&=\tau^p_jn^l
\partial_l(\delta^{mj}\delta^{ik}-\delta^{kj}\delta^{im})\partial_k \kappa_{im}
\\
&=(\tau^{pm}n^i\delta^{kl}-\tau^{pm}n^i\tau^{kl}+\tau^{pm}\tau^{ik}n^l
-\tau^{pk}n^ln^in^m-\tau^{pk}\tau^{im}n^l)\partial_k\partial_l \kappa_{im}.
\end{align*}

The second term of the right-hand side vanishes since it 
involves the tangential Laplacian $\tau^{kl}\partial_k\partial_l$ 
applied to the quantity $\tau^{pm}n^i\kappa_{im}$, which vanishes. 
For the third and last terms, we again use \eqref{AAF} to replace
$\partial_l\kappa_{im}$ with
$\sqrt{2}\dot\lambda_{lim}+\partial_i\kappa_{lm}$ and argue as before
to see that the resulting terms vanish.
For the first term we notice
that $\delta^{kl}\partial_k\partial_l \kappa_{im}=\partial^k\partial_k\kappa_{im}
=\ddot \kappa_{im}+\partial_{(i}C_{m)}-\partial_i\partial_m\kappa_l^l$ 
(from \eqref{AAK} and \eqref{AAF} with vanishing $\alpha$ and
$\beta^i$). So, $\tau^{pm}n^i\delta^{kl}\partial_k\partial_l
\kappa_{im}=\tau^{pm}n^i\ddot\kappa_{im}+\frac12\tau^{pm}n^i\partial_iC_m+\frac12
\tau^{pm}n^i\partial_mC_i-\tau^{pm}n^i\partial_i\partial_m\kappa_l^l$.
From
$\tau^{pm}n^i\kappa_{im}=0$ and $n^iC_i=0$ on the face, this identity
reduces to $\tau^{pm}n^i\delta^{kl}\partial_k\partial_l
\kappa_{im}=\frac12\tau^{pm}n^i\partial_iC_m-\tau^{pm}n^i\partial_i\partial_m\kappa_l^l$.

Finally,
$\tau^{pk}n^ln^in^m\partial_k\partial_l\kappa_{im}=\tau^{pk}(\delta^{li}-
\tau^{li})n^m\partial_k\partial_l\kappa_{im}=\tau^{pk}n^m\partial_k\partial^l\kappa_{lm}
-\tau^{pk}\tau^{li}n^m\partial_k\partial_l\kappa_{im}=
\tau^{pk}n^m\partial_kC_m-\tau^{pk}n^m\partial_k\partial_m\kappa_l^l-\tau^{pk}\tau^{li}
n^m\partial_k\partial_l\kappa_{im}$. Again, from $n^mC_m=0$ on the face and
the tangentiality of $\tau^{pk}\partial_k$, the first term
vanishes. Also, from the boundary conditions and the tangentiality of
$\tau^{pk}\tau^{li}\partial_k\partial_l$, the last term vanishes. So,
$\tau^{pk}n^ln^in^m\partial_k\partial_l\kappa_{im}=-\tau^{pk}n^m\partial_k\partial_m\kappa_l^l$.

From above,
$\tau_j^pn^l\partial_lC^j=\frac12\tau^{pm}n^i\partial_iC_m$, which
implies $\tau_j^pn^l\partial_lC^j=0$ on the face. This concludes the proof of
\eqref{AAbndid}.
\end{proof}

The next theorem asserts that the boundary conditions are
constraint-preserving.
\begin{thm}\label{thm:AAconstraint}
Let $\Omega$ be a polyhedral domain. Given
$\gamma_{ij}(0)$, $\kappa_{ij}(0)$ on $\Omega$ satisfying the constraints
\eqref{C} and \eqref{Cj}, define $\lambda_{kij}(0)$ by \eqref{AAfinit}.
Having $\kappa_{ij}(0)$ and $\lambda_{kij}(0)$, define $\kappa_{ij}$
and $\lambda_{kij}$ for positive time by
the evolution equations \eqref{AAK}, \eqref{AAF}, and the boundary
conditions \eqref{AAset2a}. Then the constraints \eqref{LCj} are satisfied
for all time.
\end{thm}
\begin{proof}
Exactly as in the proof of Theorem 1. of  \cite{AA}, we find that
$C_j$ satisfies 
satisfies the elastic  wave equation
\begin{equation}\label{AAwave}
\ddot C_j=\partial^l\partial_{(l}C_{j)}
\end{equation} 
and both $C_j$ and $\dot C_j$ vanish at the initial time.  Define the usual energy
\begin{equation*}
E(t)=\frac12 \int_{\Omega}\{ \dot C_j\dot
C^j+\frac12[\partial^lC_j\partial_lC^j+(\partial_lC^l)^2]\} \,dx.
\end{equation*}
Clearly $E(0)=0$.  From \eqref{AAwave} and integration by parts
\begin{equation}\label{AAdtE}
\dot E=\int_{\partial\Omega}(\dot C_jn^l\partial_lC^j+n^j\dot C_j\partial_lC^l) d\sigma.
\end{equation}
Therefore, if $\alpha=0$ and $\beta^i=0$, we can invoke
Lemma~\ref{AAlem}, and conclude that $E$ is constant in time.
Hence $E$ vanishes identically.  Thus $C_j$ is constant, and, since
it vanishes at time $0$, it vanishes for all time.  This establishes
the theorem under the additional assumption that $\alpha$ and $\beta^i$
vanish.

To extend to the case of general $\alpha$ and $\beta^i$ we use
Duhamel's principle.  Let $S(t)$ denote the solution operator
associated to the homogeneous boundary value problem.  That is,
given functions $\theta_{ij}(0)$, $\phi_{kij}(0)$ on $\Omega$,
define $S(t)(\theta_{ij}(0),\phi_{kij}(0))=
(\theta_{ij}(t),\phi_{kij}(t))$ where $\theta_{ij}$, $\phi_{kij}$ is
the solution to the homogeneous evolution equations
\begin{equation*}
\frac{1}{\sqrt{2}}\dot\theta_{ij}=\partial^k\phi_{k(ij)},
\quad \frac{1}{\sqrt{2}}\dot\phi_{kij}=\partial_{[k}\theta_{i]j},
\end{equation*}
satisfying the boundary conditions and assuming the given initial
values.  Then Duhamel's principle represents the solution
$\kappa_{ij}$, $\lambda_{kij}$ of the inhomogeneous initial boundary value
problem \eqref{AAK}, \eqref{AAF}, \eqref{AAset2a} as
\begin{equation}\label{AAduhamel}
(\kappa_{ij}(t),\lambda_{kij}(t))=S(t)(\kappa_{ij}(0),\lambda_{kij}(0))
+\int_0^t S(t-s)(-\frac{1}{\sqrt{2}}\partial_i\partial_j\alpha(s),\eta_{kij}(s))\,ds.
\end{equation}
Now it is easy to check that the momentum constraint \eqref{LCj} is
satisfied when $\kappa_{ij}$ is replaced by 
$-\frac{1}{\sqrt{2}}\partial_i\partial_j\alpha(s)$
(for any smooth function $\alpha$), and, from \eqref{AAK},
$\dot
C_j(0)=\sqrt{2}\partial^k(\partial^l\eta_{k(lj)}-\partial_j\eta_{kl}^{\phantom{kl}l})=0$
for any smooth shift vector $\beta^i$. Hence the integrand
in \eqref{AAduhamel} satisfies the constraints by the result for the homogeneous case,
as does the first term on the right-hand side, and thus the constraints
are indeed satisfied by $\kappa_{ij}$, $\lambda_{kij}$.
\end{proof}

The analogue of Theorem 1. of \cite{AA} for the initial boundary
value problem follows from the preceding theorem exactly as there.
\begin{thm}\label{thm:AAequiv}
Let $\Omega$ be a polyhedral domain.
Suppose that initial data $\gamma_{ij}(0)$ and $\kappa_{ij}(0)$ are given
satisfying the Hamiltonian constraint \eqref{LC} and momentum constraint
\eqref{LCj}, respectively, and that initial data $\lambda_{kij}(0)$ is defined
by \eqref{AAfinit}.  Then the unique solution of the linearized AA
initial boundary value problem \eqref{AAK}, \eqref{AAF}, and $\gamma$
defined by \eqref{AAgamma} together with the boundary conditions \eqref{AAset2a}
satisfies the linearized ADM system
\eqref{LG}--\eqref{LCj} in $\Omega$.
\end{thm}

We close by noting a second set of boundary
conditions which are maximal nonnegative and constraint-preserving.
These are
\begin{equation}\label{AAset1a}
n^in^j\kappa_{ij}=m^im^j\kappa_{ij}=l^il^j\kappa_{ij}=m^kn^in^j\lambda_{kij}=
l^kn^in^j\lambda_{kij}=0.
\end{equation}
Now when we make an orthogonal expansion as in \eqref{AAexpand1},
\eqref{AAexpand2}, the first term on the right-hand side of the
first equation and the first two terms on the right-hand side of the
second equation vanish. The necessary orthogonality to demonstrate that
the boundary conditions are maximal nonnegative follows from the
boundary conditions and the antisymmetry of $\lambda_{kij}$ with respect to its
first two indices.  Similarly, to
prove the analogue of Lemma~\ref{AAlem} for these boundary
conditions, we show that the tangential component of $\dot C_j$ vanishes
and the normal component of $n^l\partial_l C^j$ vanishes (it was
the reverse before). The analysis is essentially the
same as for the boundary conditions \eqref{AAset2a}. However, we need to
do some more work here since the vanishing of both tangential
component of $\dot C_j$ and normal component of $n^l\partial_l C^j$ is
not enough for proving \eqref{AAbndid}. We prove that 
$\partial_jC^j$ vanishes on the face. From \eqref{AAF}, observe that
$\dot \lambda_l^{\phantom{l}jl}=C^j/\sqrt{2}$. So, from \eqref{AAK} and
the
antisymmetry of $\lambda_{kij}$ in $k$ and $i$, 
$\partial_jC^j=\sqrt{2}\partial_j\dot
\lambda_l^{\phantom{l}jl}=-\sqrt{2}\partial_j\dot
\lambda^{jl}_{\phantom{jl}l}=-\ddot\kappa_l^l=-\delta^{ij}\ddot
  \kappa_{ij}=-(n^in^j+m^im^j+l^il^j)\ddot\kappa_{ij}$. From
  \eqref{AAset1a}, it follows that the divergence of $C^j$ vanishes on
  the face. Otherwise the  approach follows exactly the ideas and
  techniques used for the boundary conditions \eqref{AAset2a}.

\subsection{Extended AA System}\label{Ext_AA}

By using similar ideas and techniques as in Subsection~\ref{Ext_EC}, we indicate
an extended initial boundary value problem corresponding to the AA formulation
whose solution solves the linearized ADM system 
\eqref{LG}--\eqref{LCj} in $\Omega$.
The new system consists of \eqref{LG} for $\gamma_{ij}$, \eqref{AAF},
and two new sets of equations, one replacing \eqref{AAK} 
\begin{equation}\label{AAKE}
\frac{1}{\sqrt{2}}\dot\kappa_{ij}=\partial^k\lambda_{k(ij)}
+\frac{1}{2}(\partial_ip_j+\partial_jp_i)-\partial^kp_k\delta_{ij}
-\frac{1}{\sqrt{2}}\partial_i\partial_j\alpha,
\end{equation}
and the other one corresponding to a new three dimensional vector field $p_i$ 
defined by 
\begin{equation}\label{AAp}
\dot{p}_i=\partial^l\kappa_{li}-\partial_i\kappa_l^l.
\end{equation}
Observe that the additional terms that appear on the right-hand side
of \eqref{AAKE} compared with \eqref{AAK} are precisely the negative components
of the formal adjoint of the momentum constraint operator \eqref{LCj} applied to $p_i$.

Let $\tilde{V}$ be the vector space of quadruples of constant tensors
$(\gamma_{ij},\kappa_{ij},\lambda_{kij},p_k)$, where $\gamma_{ij}$, $\kappa_{ij}$ are 
symmetric with respect to the indices
$i$ and $j$ and $\lambda_{kij}$ belong to the 8-dimensional space $\T$ defined
in Subsection 3.4.3. Thus $\dim \tilde{V}=23$. The boundary operator 
$\tilde{A}_n :\tilde{V}\to\tilde{V}$ in this case is given by
\begin{equation}\label{AAndef}
\tilde{\gamma}_{ij}=0, \quad
\tilde{\kappa}_{ij}=-n^k\lambda_{k(ij)}-\frac{1}{2}(n_ip_j+n_jp_i)+n^kp_k\delta_{ij},
\quad
\tilde{\lambda}_{kij}=-n_{[k}\kappa_{i]j}, \quad
\tilde{p}_i=-n^l\kappa_{il}+n_i\kappa_l^{\phantom{l}l}.
\end{equation}
The boundary operator $\tilde{A}_n$ associated to the evolution equations
\eqref{LG}, \eqref{AAKE}, \eqref{AAF}, and \eqref{AAp} has five positive, 13
zero, and five negative eigenvalues. Therefore, a nonnegative subspace
is maximal nonnegative if and only if it has
dimension 18. We claim that the following boundary conditions are
maximal nonnegative for \eqref{LG}, \eqref{AAKE}, \eqref{AAF}, and \eqref{AAp}
\begin{equation}\label{AAset2ae}
n^im^j\kappa_{ij}=n^il^j\kappa_{ij}=n^k(m^im^j\lambda_{kij}-p_k)=
n^k(l^il^j\lambda_{kij}-p_k)=n^km^il^j\lambda_{kij}=0.
\end{equation}
These can be written as well:
\begin{equation}\label{AAset2be}
n^i\tau^{jk}\kappa_{ij}=0, \quad
n^k(\tau^{il}\tau^{jm}\lambda_{kij}-\tau^{lm}p_k)=0,
\end{equation}
and so do not depend on the choice of basis for the tangent space.

Let us prove the claim that the subspace $\tilde{N}$ defined by
\eqref{AAset2ae} is maximal nonnegative. Obviously,
$\dim\tilde{N}=18$. Hence, it remains to be proven that $\tilde{N}$ is
also nonnegative. In view of \eqref{AAndef}, the verification of
nonnegativity of $\tilde{N}$ reduces to showing that
\begin{equation}\label{AApositivity}
\kappa_{ij}\tilde{\kappa}^{ij}+\lambda_{kij}\tilde{\lambda}^{kij}+p_i\tilde{p}^i\geq 0
\end{equation}
whenever \eqref{AAset2ae} holds. In fact, we can prove that the
left-hand side of \eqref{AApositivity} vanishes pending that 
\eqref{AAset2ae} holds:
\begin{equation}
\begin{gathered}
\kappa_{ij}\tilde{\kappa}^{ij}+\lambda_{kij}\tilde{\lambda}^{kij}+p_i\tilde{p}^i=\\
=-2n^k\kappa^{ij}\lambda_{kij}-2n^ip^j\kappa_{ij}+2n^ip_i\kappa_l^l\\
=-2(n^ln^in^mn^j\kappa_{lm}+\tau^{li}\tau^{mj}\kappa_{lm})(\tau_i^ln^mn_jn^k\lambda_{klm}
+\tau_i^l\tau_j^mn^k\lambda_{klm})-2n^ip^j\kappa_{ij}+2n^ip_i\kappa_l^l\\
=-2n^k\tau^{rl}\tau{pm}\kappa_{rp}\lambda_{klm}-2n^ip^j\kappa_{ij}+2n^ip_i\kappa_l^l\\
=-2n^kp_k\tau^{rp}\kappa_{rp}-2n^ip^j\kappa_{ij}+2n^ip_i\kappa_l^l\\
=-2n^kp_k(\delta^{rp}-n^rn^p)\kappa_{rp}-2n^ip^j\kappa_{ij}+2n^ip_i\kappa_l^l\\
=2n^kp_kn^rn^p\kappa_{rp}-2n^ip^j\kappa_{ij}\\
=2n^kp_kn^rn^p\kappa_{rp}-2n^i(n^kp_kn^j+\tau^{kj}p_k)\kappa_{ij}\\
=-2n^i\tau^{kj}\kappa_{ij}p_k=0.
\end{gathered}
\end{equation}
This concludes the proof of \eqref{AApositivity}.  

\begin{thm}\label{thm:AAextended}
Let $\Omega$ be a polyhedral domain. Suppose that the initial data
$\gamma_{ij}(0)$ and $\kappa_{ij}(0)$ are given satisfying the Hamiltonian \eqref{LC}
and momentum constraints \eqref{LCj}, respectively, $\lambda_{kij}(0)$ is 
defined by \eqref{AAfinit}, and $p_i(0)=0$. Then the unique solution
$(\gamma_{ij},\kappa_{ij},\lambda_{kij},p_i)$ of the initial boundary value problem
\eqref{LG}, \eqref{AAKE}, \eqref{AAF}, and \eqref{AAp}, together with the
boundary conditions \eqref{AAset2ae}, satisfies the properties
$p_i=0$ for all time, and $(\gamma_{ij},\kappa_{ij})$ solves the linearized
ADM system \eqref{LG}--\eqref{LCj} in $\Omega$.
\end{thm}

\begin{proof}
Observe that the solution of the initial boundary value problem
\eqref{LG}, \eqref{AAK}, \eqref{AAF}, and \eqref{AAset2a} (boundary
conditions), together with $p_i=0$ for all time, is the unique
solution of the initial boundary value problem \eqref{LG}, \eqref{AAKE}, 
\eqref{AAF}, and \eqref{AAp}, together with the boundary conditions \eqref{AAset2ae}.
The conclusion follows from Theorem~\ref{thm:AAequiv}.
\end{proof}

We close by indicating a second set of maximal nonnegative boundary
conditions (corresponding to \eqref{AAset1a}) for \eqref{LG}, \eqref{AAKE}, 
\eqref{AAF}, and \eqref{AAp} for which Theorem~\ref{thm:AAextended} holds as
well. These are \begin{equation}\label{AAset1ae}
n^in^j\kappa_{ij}=m^im^j\kappa_{ij}=l^il^j\kappa_{ij}=m^kn^in^j\lambda_{kij}+
m^kp_k=l^kn^in^j\lambda_{kij}+l^kp_k=0.
\end{equation}

\section{Arnold Formulation}
Throughout this section, we will use the notations and results introduced in
Subsection 3.4.4. In this part we unveil the technique that led us to the finding
of maximal nonnegative constraint preserving boundary conditions for the three
formulations, EC, AA, and A, respectively, analyzed in this thesis from this
point of view. 
\subsection{Maximal Nonnegative Constraint Preserving Boundary Conditions}
\begin{figure} 
 \centering 
 \includegraphics[width=5cm,height=6cm]{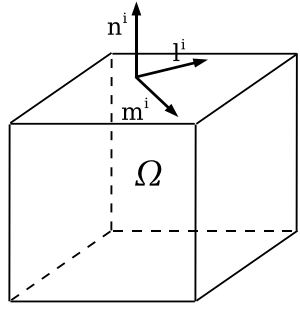} 
 \caption{A parallelipipedic domain.} 
 \label{3D_brick} 
\end{figure}

In this subsection we exhibit in detail our technique for finding maximal nonnegative
constraint preserving boundary conditions for Arnold's formulation. Everything that
follows can be done for general polyhedral domains, but for the sake of simplicity
let $\Omega$ be a parallelepiped (a box) with faces parallel to the 
coordinate planes. Again, by $m^i$ and $l^i$ we denote two additional vectors
which together $n^i$ form an orthonormal basis (see Figure~\ref{3D_brick}).

Denote by $\ub v=\ub M\ubb\nu$, and $\ub w=\ub M\ubb\mu$. Then, from
\eqref{Mcurls}, we have that $\ub v$ and $\ub w$ are solutions of the
following problem
\begin{eqnarray}\label{eq:v}
\dot{\ub v}=-\frac12\ubcurl{\ub w},\\ \label{eq:w}
\dot{\ub w}=\frac12\ubcurl{\ub v},\\ \label{eq:i}
\ub v (0)=\ub w (0)=0.
\end{eqnarray}

Observe that \eqref{eq:v}, \eqref{eq:w} is a first order symmetric hyperbolic
system. Its boundary matrix has the following eigenvalues (given with the corresponding
eigenvectors):

$\lambda_1=0:\ (\ub{n},\ 0)^T,\ (0,\ \ub{n})^T$, 

$\lambda_2=-\frac12:\ (\ub{m},\ \ub{l})^T,\ (\ub{l},\ -\ub{m})^T$,

$\lambda_3=\frac12:\ (\ub{l},\ \ub{m})^T\ (-\ub{m},\ \ub{l})^T$,

where $\ub n$ is the exterior unit normal and $\ub m$, $\ub l$ are chosen so that
$\ub m\perp \ub n$ and $\ub l=\ub n\times\ub m$ (see Figure~\ref{3D_brick}). 

If we consider maximal nonnegative boundary conditions for \eqref{eq:v}--\eqref{eq:i},
then the resulting initial boundary value problem has the unique solution 
$\ub{v}=0,\ \ub{w}=0$.

From Proposition~\ref{MaxNNeg}, the general form of maximal nonnegative boundary conditions on the face having 
as exterior unit normal vector $\ub{n}=(1,0,0)^T$ is given by
\begin{gather}\label{eq:bc1}
(1-\beta )v_2+\alpha v_3+\alpha w_2 +(1+\beta )w_3=0,\\ \label{eq:bc2}
-\delta v_2+(1+\gamma )v_3+(\gamma -1)w_2+\delta w_3=0,
\end{gather}  
with
\[ ||\left(\begin{array}{cc}
\alpha & \beta \\
\gamma & \delta
\end{array}\right) ||\leq 1 .\]
From 
\begin{gather*}
v_2=\partial_x\nu_{12}+\partial_y(-\nu_{11}-\nu_{33})+\partial_z\nu_{23},\\
v_3=\partial_x\nu_{13}+\partial_y\nu_{23}+\partial_z(-\nu_{11}-\nu_{22}),\\
w_2=\partial_x\mu_{12}+\partial_y(-\mu_{11}-\mu_{33})+\partial_z\mu_{23},\\
w_3=\partial_x\mu_{13}+\partial_y\mu_{23}+\partial_z(-\mu_{11}-\mu_{22}),
\end{gather*}
the conditions \eqref{eq:bc1}--\eqref{eq:bc2} become
\begin{equation}\label{eq:dc1}
[(1-\beta)\partial_x\nu_{12}+\alpha\partial_x\nu_{13}+\alpha\partial_x\mu_{12}
+(1+\beta)\partial_x\mu_{13}]+\end{equation}
$$\partial_y[-(1-\beta)\nu_{11}-(1-\beta)\nu_{33}-
-\alpha \mu_{11}-\alpha\mu_{33}+(1+\beta)\mu_{23}]+$$ $$\partial_z[(1-\beta)\nu_{23}-
\alpha\nu_{11}-\alpha\nu_{22}+\alpha\mu_{23}-(1+\beta)\mu_{11}-(1+\beta)\mu_{22}]=0,$$
\begin{equation}\label{eq:dc2}
[-\delta\partial_x\nu_{12}+(\gamma+1)\partial_x\nu_{13}+(\gamma-1)\partial_x\mu_{12}
+\delta\partial_x\mu_{13}]+\end{equation}
$$\partial_y[\delta\nu_{11}+\delta\nu_{33}+
(\gamma+1)\nu_{23}+(1-\gamma)\mu_{11}+(1-\gamma)\mu_{33}+\delta\mu_{23}]+$$
$$\partial_z[-\delta\nu_{23}-(\gamma+1)\nu_{11}-(\gamma+1)\nu_{22}+(\gamma-1)\mu_{23}-
\delta\mu_{11}-\delta\mu_{22}]=0$$

With no loss of generality (via Duhamel's principle), we can assume that the forcing
terms in \eqref{eq:n},\eqref{eq:m} vanish.

From the system \eqref{eq:n}--\eqref{eq:m}, it follows that
\begin{gather}\label{eq:n13}
\partial_x\nu_{13}=\partial_y\nu_{23}+\partial_z(\nu_{11}-\nu_{22})-2\dot{\mu}_{12},\\
\partial_x\nu_{12}=\partial_y(\nu_{11}-\nu_{33})+\partial_z\nu_{23}+2\dot{\mu}_{13},\\
\partial_x\mu_{13}=\partial_y\mu_{23}+\partial_z(\mu_{11}-\mu_{22})+2\dot{\nu}_{12},\\ \label{eq:m12}
\partial_x\mu_{12}=\partial_y(\mu_{11}-\mu_{33})+\partial_z\mu_{23}-2\dot{\nu}_{13}.
\end{gather}

By introducing \eqref{eq:n13}--\eqref{eq:m12} into \eqref{eq:dc1}--\eqref{eq:dc2}, we
get
\begin{equation}\label{eq:py1}
\partial_t[(1-\beta)\mu_{13}-\alpha\mu_{12}-\alpha\nu_{13}+(1+\beta)\nu_{12}]+
\end{equation}
$$\partial_y[(\beta -1)\nu_{33}+\alpha\nu_{23}-\alpha\mu_{33}+(1+\beta)\mu_{23}]+$$
$$\partial_z[-\alpha\nu_{22}+(1-\beta)\nu_{23}-(1+\beta)\mu_{22}+\alpha\mu_{23}]=0,$$

\begin{equation}\label{eq:py2}
\partial_t[-\delta\mu_{13}-(\gamma+1)\mu_{12}-(\gamma-1)\nu_{13}+
\delta\nu_{12}]+
\end{equation}
$$\partial_y[\delta\nu_{33}+(\gamma +1)\nu_{23}+(1-\gamma)\mu_{33}+\delta\mu_{23}]+$$
$$\partial_z[-(\gamma+1)\nu_{22}-\delta\nu_{23}-\delta\mu_{22}-(1-\gamma)\mu_{23}]=0,$$
on the part of the boundary having $(1,0,0)^T$ as the unit exterior normal vector.

Next, observe that we can get from the system \eqref{eq:n}--\eqref{eq:m} four relations
that do not involve the transverse derivative $\partial_x$
\begin{gather}\label{eq:1}
\partial_t\nu_{11}+\partial_y\mu_{13}-\partial_z\mu_{12}=0,\\ \label{eq:2}
\partial_t(\nu_{22}+\nu_{33})-\partial_y\mu_{13}+\partial_z\mu_{12}=0,\\ \label{eq:3}
\partial_t\mu_{11}-\partial_y\nu_{13}+\partial_z\nu_{12}=0,\\ \label{eq:4}
\partial_t(\mu_{22}+\mu_{33})+\partial_y\nu_{13}-\partial_z\nu_{12}=0.
\end{gather}

We will not use the relations \eqref{eq:1} and \eqref{eq:3} since they
seem to introduce only complications in what follows.

By adding combinations of \eqref{eq:2} and \eqref{eq:4} to both \eqref{eq:py1}
and \eqref{eq:py2}, we get
\begin{equation}\label{eq:pt1}
\partial_t[(1-\beta)\mu_{13}-\alpha\mu_{12}-\alpha\nu_{13}+(1+\beta)\nu_{12}
+m\nu_{22}+m\nu_{33}+n\mu_{22}+n\mu_{33}]+
\end{equation}
$$\partial_y[(\beta -1)\nu_{33}+\alpha\nu_{23}-\alpha\mu_{33}+(1+\beta)\mu_{23}-
m\mu_{13}+n\nu_{13}]+$$
$$\partial_z[-\alpha\nu_{22}+(1-\beta)\nu_{23}-(1+\beta)\mu_{22}+\alpha\mu_{23}+
m\mu_{12}-n\nu_{12}]=0$$
and
\begin{equation}\label{eq:pt2}
\partial_t[-\delta\mu_{13}-(\gamma+1)\mu_{12}-(\gamma-1)\nu_{13}+
\delta\nu_{12}+p\nu_{22}+p\nu_{33}+q\mu_{22}+q\mu_{33}]+
\end{equation}
$$\partial_y[\delta\nu_{33}+(\gamma +1)\nu_{23}+(1-\gamma)\mu_{33}+\delta\mu_{23}
-p\mu_{13}+q\nu_{13}]+$$
$$\partial_z[-(\gamma+1)\nu_{22}-\delta\nu_{23}-\delta\mu_{22}-(1-\gamma)\mu_{23}
+p\mu_{12}-q\nu_{12}]=0$$
on the part of the boundary having $(1,0,0)^T$ as the unit exterior normal vector.

It is easy to see now that \eqref{eq:py1}--\eqref{eq:py2} are satisfied if the
following conditions hold:
\begin{gather}\label{eq:b1}
m\nu_{22}+m\nu_{33}-\alpha\nu_{13}+(1+\beta)\nu_{12}+n\mu_{22}+n\mu_{33}+
(1-\beta )\mu_{13}-\alpha\mu_{12}=0,\\ \label{eq:b2}
(\beta -1)\nu_{33}+\alpha\nu_{23}+n\nu_{13}-\alpha\mu_{33}+(1+\beta)\mu_{23}-
m\mu_{13}=0,\\ \label{eq:b3}
\alpha\nu_{22}+(\beta-1)\nu_{23}+n\nu_{12}+(1+\beta)\mu_{22}-\alpha\mu_{23}-m\mu_{12}=0,\\ \label{eq:b4}
p\nu_{22}+p\nu_{33}+(1-\gamma)\nu_{13}+\delta\nu_{12}+q\mu_{22}+q\mu_{33}-
\delta\mu_{13}-(\gamma+1)\mu_{12}=0,\\ \label{eq:b5}
\delta\nu_{33}+(\gamma+1)\nu_{23}+q\nu_{13}+(1-\gamma)\mu_{33}+\delta\mu_{23}
-p\mu_{13}=0,\\ \label{eq:b6}
(\gamma+1)\nu_{22}+\delta\nu_{23}+q\nu_{12}+\delta\mu_{22}+(1-\gamma)\mu_{23}
-p\mu_{12}=0.
\end{gather}
The matrix $A$ corresponding to this linear homogeneous system is:
\[ A=\left(\begin{array}{cccccccccc}
m&m&0&-\alpha &1+\beta &n&n&0&1-\beta&-\alpha\\
0&\beta -1&\alpha&n&0&0&-\alpha&1+\beta&-m&0\\
\alpha&0&\beta-1&0&n&1+\beta&0&-\alpha&0&-m\\
p&p&0&1-\gamma&\delta&q&q&0&-\delta&-1-\gamma\\
0&\delta&1+\gamma&q&0&0&1-\gamma&\delta&-p&0\\
1+\gamma&0&\delta&0&q&\delta&0&1-\gamma&0&-p
\end{array}\right) .\]

Before we go any further, let us characterize the maximal nonnegative
boundary conditions for \eqref{eq:n}--\eqref{eq:m}.

The boundary matrix corresponding to this symmetric hyperbolic system
has the following eigenvalues, given with the corresponding eigenvectors:
\[ \lambda =0:\ \left(\begin{array}{c}
\ub{n}\odot\ub{n}\\
\ubb{0}\end{array}\right) ,\ 
\left(\begin{array}{c}
\ubb{0}\\
\ub{n}\odot\ub{n}\end{array}\right) ,\ 
\left(\begin{array}{c}
\ub{m}\odot\ub{m}+\ub{l}\odot\ub{l}\\
\ubb{0}\end{array}\right) ,\ 
\left(\begin{array}{c}
\ubb{0}\\
\ub{m}\odot\ub{m}+\ub{l}\odot\ub{l}
\end{array}\right) , \]

\[ \lambda=\frac{1}{2}:\ 
\left(\begin{array}{c}
\ub{l}\odot\ub{n}\\
\ub{m}\odot\ub{n}\end{array}\right) ,\ 
\left(\begin{array}{c}
\ub{m}\odot\ub{n}\\
-\ub{n}\odot\ub{l}\end{array}\right), \]

\[ \lambda=1:\ 
\left(\begin{array}{c}
\frac{1}{2}(\ub{l}\odot\ub{l}-\ub{m}\odot\ub{m})\\
\ub{m}\odot\ub{l}\end{array}\right) ,\ 
\left(\begin{array}{c}
\ub{m}\odot\ub{l}\\
\frac{1}{2}(\ub{m}\odot\ub{m}-\ub{l}\odot\ub{l})\end{array}\right) ,\]

\[ \lambda=-\frac{1}{2}:\ 
\left(\begin{array}{c}
\ub{m}\odot\ub{n}\\
\ub{l}\odot\ub{n}\end{array}\right) ,\ 
\left(\begin{array}{c}
\ub{n}\odot\ub{l}\\
-\ub{m}\odot\ub{n}\end{array}\right) ,\]

\[ \lambda=-1:\
\left(\begin{array}{c}
\ub{m}\odot\ub{l}\\
\frac{1}{2}(\ub{l}\odot\ub{l}-\ub{m}\odot\ub{m})\end{array}\right) ,\ 
\left(\begin{array}{c}
\frac{1}{2}(\ub{m}\odot\ub{m}-\ub{l}\odot\ub{l} )\\
\ub{m}\odot\ub{l}\end{array}\right) .\]

From Proposition~\ref{MaxNNeg}, we know that if $M\in\R^{4\times 4}$ and the
norm of the matrix 
\[\left(\begin{array}{cccc}
1/\sqrt{2}&0&0&0\\
0&1/\sqrt{2}&0&0\\
0&0&1&0\\
0&0&0&1
\end{array}\right) M
\left(\begin{array}{cccc}
\sqrt{2}&0&0&0\\
0&\sqrt{2}&0&0\\
0&0&1&0\\
0&0&0&1
\end{array}\right)\]
is less or equal to $1$, then the following boundary conditions are
maximal nonnegative:
\begin{equation}\label{eq:mnnbc}
\left(\begin{array}{c}
(\ub{m}\odot\ub{n})\ubb\nu +(\ub{l}\odot\ub{n})\ubb\mu \\
(\ub{n}\odot\ub{l})\ubb\nu- (\ub{m}\odot\ub{n})\ubb\mu\\
(\ub{m}\odot\ub{l})\ubb\nu+\ubb\mu (\ub{l}\odot\ub{l}-\ub{m}\odot
\ub{m})/2\\
\ubb\nu (\ub{m}\odot\ub{m}-\ub{l}\odot\ub{l})/2+
(\ub{m}\odot\ub{l})\ubb\mu\end{array}\right) +M
\left(\begin{array}{c}
(\ub{l}\odot\ub{n})\ubb\nu +(\ub{m}\odot\ub{n})\ubb\mu\\
(\ub{m}\odot\ub{n})\ubb\nu -(\ub{n}\odot\ub{l})\ubb\mu\\
\ubb\nu (\ub{l}\odot\ub{l}-\ub{m}\odot\ub{m})/2+(\ub{m}\odot\ub{l})\ubb\mu\\
(\ub{m}\odot\ub{l})\ubb\nu+\ubb\mu (\ub{m}\odot\ub{m}-\ub{l}\odot
\ub{l})/2
\end{array}\right) =0.
\end{equation}
For $\ub{n}=(1,0,0)^T,\, \ub{m}=(0,1,0)^T$, and $\ub{l}=(0,0,1)^T$,
\eqref{eq:mnnbc} is equivalent to
\[ B\left(\begin{array}{c}
\nu_{22}\\
\vdots\\
\mu_{12}
\end{array}\right) =0, \]
where
\begin{equation}
B=\left(\begin{array}{ccccccc}
-m_{13}&m_{13}&2m_{14}&2m_{11}&2(1+m_{12})&m_{14}&
-m_{14}\\
-m_{23}&m_{23}&2m_{24}&2(1+m_{21})&2m_{22}&m_{24}&-m_{24}\\
-m_{33}&m_{33}&2(1+m_{34})&2m_{31}&2m_{32}&-1+m_{34}&1-m_{34}\\
1-m_{43}&-1+m_{43}&2m_{44}&2m_{41}&2m_{42}&m_{44}&-m_{44}
\end{array}\right.
\end{equation}
\begin{equation*}
\left. \begin{array}{ccc}
2m_{13}&2(1-m_{12})&2m_{11}\\
2m_{23}&-2m_{22}&2(-1+m_{21})\\
2m_{33}&-2m_{32}&2m_{31}\\
2(1+m_{43})&-2m_{42}&2m_{41}
\end{array}\right) .
\end{equation*}

Now, we want a match between the systems \eqref{eq:b1}--\eqref{eq:b6}
and \eqref{eq:mnnbc}. First of all, observe that the first two columns
of $B$ coincide except the sign and, since we want each row of $A$ to be
a linear combination of rows of $B$, this induces
$$\alpha =0;\ \beta =1;\ \gamma=-1;\ \delta=0;\ m=0;\ p=0.$$
Also, we have to worry about the rank of the matrix $A$, which
is a $6\times 10$--matrix. Obviously, the rank of $A$ cannot exceed four. 
Fortunately, we know that  the rank of $A$ equals the rank of $AA^T$ 
which is a $6\times 6$--matrix. Using this fact, we can see that 
$A$ has rank four for the above values of the other parameters if and only if
$n^2+q^2=4$ (otherwise $A$ has rank greater or equal to five). 
From here, we get that the rows of $B$ must be linear combinations of 
the rows of $A$ too and it turns out that $M$ must be

\[ M=\left(\begin{array}{cccc}
0&1&0&0\\
1&0&0&0\\
0&0&0&-1\\
0&0&1&0
\end{array}\right) .\]

So, the maximal nonnegative boundary conditions on the face are:
\begin{gather}
\nu_{12}=0,\\
\nu_{13}=0,\\
-\mu_{22}+\mu_{33}=0,\\
\mu_{23}=0.
\end{gather}

Same kind of conditions could be obtained for each face of $\Omega$. In fact, 
we can write the boundary conditions for all faces in
the following form:
\begin{gather}\label{1bc}
\left(\begin{array}{c}
\ub n\odot\ub m\\
\ubb 0\end{array}\right) :\left(\begin{array}{c}
\ubb\nu\\
\ubb\mu\end{array}\right) =0,\\ 
\left(\begin{array}{c}
\ub n\odot\ub l\\
\ubb 0\end{array}\right) :\left(\begin{array}{c}
\ubb\nu\\
\ubb\mu\end{array}\right) =0,\\
\left(\begin{array}{c}
\ubb 0\\
\ub l\odot\ub l -\ub m\odot\ub m \end{array}\right) :\left(\begin{array}{c}
\ubb\nu\\
\ubb\mu\end{array}\right) =0,\\ \label{4bc}
\left(\begin{array}{c}
\ubb 0\\
\ub m\odot\ub l\end{array}\right) :\left(\begin{array}{c}
\ubb\nu\\
\ubb\mu\end{array}\right) =0,
\end{gather} 
or
\begin{equation}\label{pbc}
\left(\begin{array}{c}
\ubb\nu\\
\ubb\mu\end{array}\right)_{|\partial\Omega}\in X,
\end{equation}
where
\[ X=\Span\{ \left(\begin{array}{c}
\ub n\odot\ub n\\
\ubb 0\end{array}\right),\,
\left(\begin{array}{c}
\ub m\odot\ub l\\
\ubb 0\end{array}\right),\,
\left(\begin{array}{c}
\ub m\odot\ub m\\
\ubb 0\end{array}\right),\,
\left(\begin{array}{c}
\ub m\odot\ub m-\ub l\odot\ub l\\
\ubb 0\end{array}\right),\,
\left(\begin{array}{c}
\ubb 0\\
\ub n\odot\ub n\end{array}\right),\, \]
\[\left(\begin{array}{c}
\ubb 0\\
\ub n\odot\ub m\end{array}\right),\,
\left(\begin{array}{c}
\ubb 0\\
\ub n\odot\ub l\end{array}\right),\,
\left(\begin{array}{c}
\ubb 0\\
\ub m\odot\ub m +\ub l\odot\ub l\end{array}\right)\}. \]
Observe that $X$ is a maximal nonnegative (in fact null) subspace for the boundary operator
\[ A_n=\left(\begin{array}{cc}
\ubb 0 & K_n\\
-K_n & \ubb 0
\end{array}\right) ,\]
where by definition
\[ K_n\ubb u =\frac{1}{2}[\Skw (\ub n )\ubb u -\ubb u\Skw (\ub n )],\]
for all $\ubb u\in \R^{3\times 3}_{sym}$.

Also, it is easy to see that the boundary conditions \eqref{1bc}--\eqref{4bc} can be
written as:
\begin{equation}\label{Abc_independent}
n^i\tau^{jk}\nu_{ij}=0, \quad
(2\tau^{ik}\tau^{jl}-\tau^{kl}\tau^{ij})\mu_{ij}=0,
\end{equation}
where $\tau^{ij}:=m^im^j+l^il^j$ is the projection operator orthogonal to $n^i$, and so,
there is no dependence on the particular choice of $m^i$ and $l^i$.

Define $$E(t)=\int_{\Omega}|\ub v |^2+|\ub w |^2dx.$$
Then,
$$\dot{E}(t)=\int_{\partial\Omega}\ub w (\ub n \times\ub v )d\sigma=
-\int_{\partial\Omega}\ub v (\ub n \times\ub w )d\sigma,$$
where $\ub n$ is the exterior unit normal vector.

By using \eqref{1bc}--\eqref{4bc}, it is easy to prove that $\dot{E}(t)=0$ and this
implies again that $\ub M\ubb\nu=0$ and $\ub M\ubb\mu=0$ for all time.

By the stucture of the main system of differential equations \eqref{eq:n}--\eqref{eq:m},
there is a second set of maximal nonnegative constraint preserving boundary conditions
\begin{equation}\label{Abc_independent_2}
n^i\tau^{jk}\mu_{ij}=0, \quad
(2\tau^{ik}\tau^{jl}-\tau^{kl}\tau^{ij})\nu_{ij}=0.
\end{equation}
Moreover, by our above analysis there are no other sets of maximal nonnegative constraint preserving
boundary conditions corresponding to the system \eqref{eq:n}--\eqref{eq:m}.

\subsection{Equivalent Unconstrained Initial Boundary Value Problem}

Consider the constrained initial boundary valu problem 
\eqref{a1h}--\eqref{a3h} together with maximal nonnegative boundary
conditions
\begin{equation}\label{eq:u}
u_{|\partial\Omega}\in Y,
\end{equation}
where $Y$ is a maximal nonnegative boundary space. Also, consider the
associated extended problem \eqref{un1}--\eqref{un2} with boundary
conditions
\begin{equation}\label{eq:uz2}
\left(\begin{array}{c}
u\\
z
\end{array}\right)_{|\partial\Omega}\in \overline{Y}=Y\times (B_nY)^{\perp},
\end{equation}
where $B_n(x)=-\Sigma_{j=1}^NB^jn_j$ is the boundary matrix corresponding to
the first order differential operator $B$ at $x\in \partial\Omega$.

From Corollary~\ref{cor:nonneg}, $\overline{Y}$ is nonnegative for the boundary
matrix $\overline{A}_n$ associated to \eqref{un1} if and only if $Y$ is 
nonnegative for the boundary matrix $A_n$ associated to \eqref{a1h}.

Furthermore, assume $\overline{Y}$ is maximal nonnegative for  $\overline{A}_n$.
Let $v$ be a solution of \eqref{a1h}--\eqref{a3h} and \eqref{eq:u}, and
$(u,z)^T$ be a solution of \eqref{un1}--\eqref{un2} and \eqref{eq:uz2}.

A formal calculation shows that
\[ \frac{1}{2}\frac{d}{dt}(\| u-v\|^2_{L^2(\Omega)}+\| z\|^2_{L^2(\Omega)})=
-\frac{1}{2}\int_{\partial\Omega}(u-v)^TA_n(u-v)\, d\sigma\leq 0 .\]

From here, it follows that $z=0$ and $u=v$. In other words, the initial 
boundary value problems  \eqref{a1h}--\eqref{a3h} with \eqref{eq:u} and 
\eqref{un1}--\eqref{un2} with \eqref{eq:uz2} are equivalent.

Returning to our problem, observe that the boundary matrix of the extended system
\eqref{A_ext_system} reads
\[ \overline{A}_n=\left(\begin{array}{cc}
A_n & B^T_n\\
B_n & 0
\end{array}\right),\]
where
\[ B_n=\left(\begin{array}{cc}
M_n & 0\\
0 & M_n
\end{array}\right),\]
with $M_n\ubb u =\ubb u\; \ub n -(\tr\ubb u )\;\ub n,\ \forall \ubb u\in 
\R^{3\times 3}_{sym}$.

Straighforward computations show that
\[ B_nX=\Span\{ \left(\begin{array}{c}
\ub n\\
\ub 0
\end{array}\right) ;\;
\left(\begin{array}{c}
\ub m\\
\ub 0
\end{array}\right) ;\; 
\left(\begin{array}{c}
\ub l\\
\ub 0
\end{array}\right);\;
\left(\begin{array}{c}
\ub 0\\
\ub n
\end{array}\right) \} , \]
and so,
\[ (B_nX)^{\perp}=\Span\{ \left(\begin{array}{c}
\ub 0\\
\ub m
\end{array}\right) ;\;
 \left(\begin{array}{c}
\ub 0\\
\ub l
\end{array}\right)\} .\]
Therefore, $\overline{X}=X\times (B_nX)^{\perp}$ has dimension ten and it is 
nonnegative for $\overline{A}_n$. Since $\overline{A}_n$ has exactly ten
nonnegative and eight strictly negative eigenvalues, it follows that $\overline{X}$ is
also maximal nonnegative.
Thus, the following boundary conditions are suitable for the extended system \eqref{A_ext_system}
\begin{gather}
\left(\begin{array}{c}
\ub n\odot\ub m\\
\ubb 0\end{array}\right) :\left(\begin{array}{c}
\ubb\nu\\
\ubb\mu\end{array}\right) =0,\\ 
\left(\begin{array}{c}
\ub n\odot\ub l\\
\ubb 0\end{array}\right) :\left(\begin{array}{c}
\ubb\nu\\
\ubb\mu\end{array}\right) =0,\\
\left(\begin{array}{c}
\ubb 0\\
\ub l\odot\ub l -\ub m\odot\ub m \end{array}\right) :\left(\begin{array}{c}
\ubb\nu\\
\ubb\mu\end{array}\right) =0,\\
\left(\begin{array}{c}
\ubb 0\\
\ub m\odot\ub l\end{array}\right) :\left(\begin{array}{c}
\ubb\nu\\
\ubb\mu\end{array}\right) =0,\\
\ub p=0,\\
\ub q\;\ub n=0,
\end{gather} 
or,
\begin{equation}\label{A_ext_bc1}
n^i\tau^{jk}\nu_{ij}=0, \quad
(2\tau^{ik}\tau^{jl}-\tau^{kl}\tau^{ij})\mu_{ij}=0, \quad
p_i=0, \quad
n^iq_i=0.
\end{equation}
Note the existence of a second set of boundary conditions for the extended system \eqref{A_ext_system}
related to \eqref{Abc_independent_2}:
\begin{equation}\label{A_ext_bc2}
n^i\tau^{jk}\mu_{ij}=0, \quad
(2\tau^{ik}\tau^{jl}-\tau^{kl}\tau^{ij})\nu_{ij}=0, \quad
p_i=0, \quad
n^iq_i=0.
\end{equation}

\begin{appendix}
\chapter{General Solution for the Linearized Momentum Constraints}
\label{ap:addi}
In order to apply the considerations of Subsection 4.2.3 to the EC
problem, it would be useful to solve the linearized momentum constraint equations 
\eqref{Cj} for symmetric matrix fields defined on the entire space.

Define the symbol $\epsilon_{ijk}$ by
\begin{equation}
\epsilon_{ijk}=\left\{\begin{array}{lll}
+1&\mbox{ if $(ijk)$ is either $(123)$, $(231)$, or $(321)$,}\\
-1&\mbox{ if $(ijk)$ is either $(321)$, $(213)$, or $(132),$}\\
0&\mbox{ otherwise. }
\end{array}\right.
\end{equation}
Note that
\begin{equation}\label{epsilon}
\epsilon_{ik}^{\hphantom{ik}m}\epsilon_{psm}=\delta_{ip}\delta_{ks}
-\delta_{is}\delta_{kp}.
\end{equation}

From $C_j(\kappa):=\partial^l\kappa_{jl}-\partial_j\kappa_l^l=
\partial^l(\kappa_{jl}-\kappa_s^s\delta_{jl})=0$, there exists a matrix
field $\sigma_{ij}$ so that
$\kappa_{jk}-\kappa_l^l\delta_{jk}=\epsilon_k^{\hphantom{k}il}\partial_i\sigma_{jl}$.
Therefore,
\begin{equation}\label{kappa}
\kappa_{jk}=\epsilon_k^{\hphantom{k}il}\partial_i\sigma_{jl}-
\frac12 \epsilon^{pil}\partial_i\sigma_{pl}\delta_{jk}.
\end{equation}
Since $\kappa_{ij}$ is symmetric, 
\begin{equation}\label{symmetry}
\epsilon_k^{\hphantom{k}il}
\partial_i\sigma_{jl}=\epsilon_j^{\hphantom{il}}\partial_i\sigma_{kl}.
\end{equation}
It is not difficult to see that \eqref{symmetry} is equivalent
to
\begin{equation}\label{sigma1}
\partial^l\sigma_{li}-\partial_i\sigma_s^s=0.
\end{equation}
The proof of this fact follows easily by multiplying \eqref{symmetry}
with $\epsilon^{kji}$, and using the identity \eqref{epsilon}.

From \eqref{sigma1} and $\partial^l\sigma_{li}-\partial_i\sigma_s^s=
\partial^l(\sigma_{li}-\sigma_s^s\delta_{li})$, there exists a matrix
field $\eta_{ij}$ such that
\begin{equation}\label{eta}
\sigma_{ij}-\sigma_s^s\delta_{ij}=\epsilon_i^{\hphantom{i}kl}
\partial_k\eta_{lj},
\end{equation}
and so,
\begin{equation}\label{sigma2}
\sigma_{ij}=\epsilon_i^{\hphantom{i}kl}\partial_k\eta_{lj}-\frac12 
\epsilon^{skl}\partial_k\eta_{ls}\delta_{ij}.
\end{equation}
By substituting \eqref{sigma2} into \eqref{kappa} we get the general
solution of the linearized momentum constraint \eqref{Cj}.

\chapter{The $4$--D Covariant Formulation}
\label{ap:more}

Assume that the metric $g_{\mu\nu}$ is a slight perturbation of the Minkowski
metric $\eta_{\mu\nu}$: $$g_{\mu\nu}=\eta_{\mu\nu}+h_{\mu\nu},\ | h_{\mu\nu}|\ll 1.$$

To first order in $h$, the Christoffel symbols are given by
\begin{equation}\label{eq:Gamma}
\Gamma^{\lambda}_{\mu\nu}=\frac{1}{2}\eta^{\lambda\rho}(h_{\rho\nu ,\mu}+h_{\rho\mu , \nu}-h_{\mu\nu ,\rho}),
\end{equation}
and so, to first order in $h$, the Ricci tensor is
\begin{equation}\label{eq:Ricci}
R_{\mu\nu}=\Gamma^{\lambda}_{\mu\nu ,\lambda}-\Gamma^{\lambda}_{\lambda\mu ,\nu}=
\frac{1}{2}(-\Box h_{\mu\nu}+h^{\lambda}_{\nu ,\lambda\mu}+h^{\lambda}_{\mu ,\lambda\nu}
-h^{\lambda}_{\lambda ,\mu\nu}),
\end{equation}
where $\Box$ is the d'Alembertian.

Therefore, the Einstein equations in first approximation read
\begin{equation}\label{eq:Einstein}
\Box h_{\mu\nu}-h^{\lambda}_{\nu ,\lambda\mu}-h^{\lambda}_{\mu ,\lambda\nu}
+h^{\lambda}_{\lambda ,\mu\nu}=-16\pi G S_{\mu\nu},
\end{equation}
where $$S_{\mu\nu}=T_{\mu\nu}-\frac{1}{2}\eta_{\mu\nu}T^{\lambda}_{\lambda}$$
is the source term.

Here $T_{\mu\nu}$ is taken to lowest order in $h_{\mu\nu}$ and satisfies the 
ordinary conservation conditions $T^{\mu}_{\nu ,\mu}=0$.

We can not expect a field equation such as \eqref{eq:Einstein} to provide
unique solutions because given any solution, we can always generate other
solutions by performing coordinate transformations. The most general coordinate
transformation that leaves the field weak is of the form
\begin{equation}\label{eq:coordinates}
x^{\mu}\longrightarrow x'^{\mu}=x^{\mu}+\epsilon^{\mu}(x),
\end{equation}
where $\epsilon^{\mu}_{,\nu}$ is at most of the same order of magnitude as 
$h_{\mu\nu}$.

From
\begin{gather}
g'^{\mu\nu}=\frac{\partial x'^{\mu}}{\partial x^{\lambda}}
\frac{\partial x'^{\nu}}{\partial x^{\rho}}g^{\lambda\rho},\\
g^{\mu\nu}=\eta^{\mu\nu}-h^{\mu\nu},\\
g'^{\mu\nu}=\eta^{\mu\nu}-h'^{\mu\nu},
\end{gather}
it follows that
\begin{equation}\label{eq:h'}
h'_{\mu\nu}=h_{\mu\nu}-\epsilon_{\mu ,\nu}-\epsilon_{\nu ,\mu},
\end{equation}
where $\epsilon_{\mu}=\eta_{\mu\nu}\epsilon^{\nu}$.

By direct computations, we can prove that \eqref{eq:h'} is also
a solution of \eqref{eq:Einstein}. To remove this non-uniqueness 
of solution, we have to
work in some particular gauge. It turns out that one of the most
convenient choice is the {\it harmonic gauge}, for which
\begin{equation}\label{eq:harmonic}
\Gamma^{\lambda}=g^{\mu\nu}\Gamma^{\lambda}_{\mu\nu}=0.
\end{equation}

Let us prove that there is always a coordinate system in which 
\eqref{eq:harmonic} holds.

Recall that the Christoffel symbols transforms as
\begin{equation}\label{eq:Christoffel}
\Gamma '^{\lambda}_{\mu\nu}=\frac{\partial x'^{\lambda}}{\partial x^{\rho}}
\frac{\partial x^{\tau}}{\partial x'^{\mu}}\frac{\partial x^{\sigma}}{\partial x'^{\nu}}
\Gamma^{\rho}_{\tau\sigma}-\frac{\partial x^{\rho}}{\partial x'^{\nu}}
\frac{\partial x^{\sigma}}{\partial x'^{\mu}}
\frac{\partial^2 x'^{\lambda}}{\partial x^{\rho}\partial x^{\sigma}}.
\end{equation}

Contracting this with $g'^{\mu\nu}$, we get
\begin{equation}
\Gamma '^{\lambda}=\frac{\partial x'^{\lambda}}{\partial x^{\rho}}\Gamma^{\rho}-
g^{\rho\sigma}\frac{\partial^2 x'^{\lambda}}{\partial x^{\rho}\partial x^{\sigma}}.
\end{equation}

Thus, if $\Gamma^{\lambda}$ does not vanish, we can always find a new coordinate system
$x'^{\nu}$ by solving the second--order partial differential equations
\begin{equation}\label{eq:PDE}
\frac{\partial x'^{\lambda}}{\partial x^{\rho}}\Gamma^{\rho}-g^{\rho\sigma}\frac{\partial^2 x'^{\lambda}}{\partial x^{\rho}\partial x^{\sigma}}=0,
\end{equation}
which gives $\Gamma '^{\lambda}=0$ in the $x'$--coordinate system.

So, without loss of generality, assume that the chosen gauge is harmonic. Then,
since $\Gamma^{\lambda}=0$, it follows that the coordinates are harmonic functions
\begin{equation}\label{eq:hc}
\Box x^{\mu}=(g^{\lambda k}x^{\mu}_{;\lambda})_{;k}=
g^{\lambda k}\frac{\partial^2 x^{\mu}}{\partial x^{\lambda}\partial x^{k}}-\Gamma^{\lambda}
\frac{\partial x^{\mu}}{\partial x^{\lambda}}=-\Gamma^{\mu}=0.
\end{equation}

This explains the term {\it harmonic gauge}.

For a harmonic gauge, the equation \eqref{eq:Einstein} is
\begin{equation}\label{eq:he}
\Box h_{\mu\nu}=-16\pi GS_{\mu\nu}.
\end{equation}

One solution of this equation is the {\it retarded potential}
\begin{equation}\label{eq:retarded}
h_{\mu\nu}(x,t)=4G\int\frac{S_{\mu\nu}(x',t-|x-x'|)}{|x-x'|}dx'.
\end{equation}

To this solution, we can add any solution of the corresponding homogeneous
equation
\begin{equation}\label{eq:homogeneous}
\Box h_{\mu\nu}=0,
\end{equation}
which also satisfies
\begin{equation}\label{eq:cond}
\frac{\partial}{\partial x^{\mu}}h^{\mu}_{\nu}=
\frac{1}{2}\frac{\partial}{\partial x^{\nu}}h^{\mu}_{\mu},
\end{equation}
which comes from $\Gamma^{\lambda}=0$ to first order.

Observe that \eqref{eq:retarded} automatically satisfies the harmonic
condition \eqref{eq:cond} (at least for a compactly supported source
$S_{\mu\nu}$), since 
$$\frac{\partial}{\partial x^{\mu}}S^{\mu}_{\nu}=
\frac{1}{2}\frac{\partial}{\partial x^{\nu}}S^{\mu}_{\mu}$$
that follows from the conservation law 
$$\frac{\partial}{\partial x^{\mu}}T^{\mu}_{\nu}=0$$
and $$S_{\mu\nu}=T_{\mu\nu}-\frac{1}{2}\eta_{\mu\nu}T^{\lambda}_{\lambda}.$$

{\bf Interpretation:} The solution \eqref{eq:retarded} is interpreted as the gravitational
radiation produced by the source $S_{\mu\nu}$, whereas any additional term (solution
of \eqref{eq:homogeneous}, and \eqref{eq:cond}) represents the gravitational radiation
coming from infinity. The term $t-|x-x'|$ in \eqref{eq:retarded} shows that the
gravitational effects propagate with the speed of light.

The drawback of this method is the fact that the gauge condition $\Gamma^{\lambda}=0$
can be imposed only locally \cite{CB}.

\end{appendix}

\end{document}